\def\<{\langle}
\def\>{\rangle}

\newcommand{\ra}{\;\raise1.0pt\hbox{$'$}\hskip-6pt\partial\;}
\newcommand{\lo}{\;\overline{\raise1.0pt\hbox{$'$}\hskip-6pt\partial}\;}

\newcommand{\degree}{^\circ}

\newcommand{\Abs}[1]{\begin{abstract} #1 \end{abstract}}

\newcommand{\mktt}{  }

\documentclass[twocolumn, times, twocolappendix]{aastex631}

\usepackage{graphicx, epsfig, natbib, color, times, bm, amsmath,
hyperref, multirow, harpoon, amssymb, mathtools, mathrsfs, xcolor,
academicons, booktabs, soul}

\usepackage[ruled,linesnumbered]{algorithm2e}

\usepackage{snapshot}
\usepackage{upgreek}
\usepackage{physics}
\usepackage{placeins}
\usepackage{subfigure}
\usepackage{CJK}

\begin{document}
\begin{CJK*}{UTF8}{gbsn}

\title{Evaluation of the single-component thermal dust emission model in CMB experiments}

\author{Hao Liu (刘浩)}
\affiliation{School of Physics and optoelectronics engineering, Anhui University, 111 Jiulong Road, Hefei, Anhui, China 230601.}

\author{Jia-Rui Li (李嘉睿)}
\affiliation{Department of Astronomy, School of Physical Sciences, University of Science and Technology of China, No.96, JinZhai Road, Baohe District, Hefei, Anhui 230026, China.}
\affiliation{CAS Key Laboratory for Researches in Galaxies and Cosmology, School of Astronomy and Space Science, University of Science and Technology of China, Hefei, Anhui, 230026, China.}
\affiliation{School of Physics and optoelectronics engineering, Anhui University, 111 Jiulong Road, Hefei, Anhui, China 230601.}

\author{Yi-Fu Cai (蔡一夫)}
\affiliation{Department of Astronomy, School of Physical Sciences, University of Science and Technology of China, No.96, JinZhai Road, Baohe District, Hefei, Anhui 230026, China.}
\affiliation{CAS Key Laboratory for Researches in Galaxies and Cosmology, School of Astronomy and Space Science, University of Science and Technology of China, Hefei, Anhui, 230026, China.}

\correspondingauthor{Jia-Rui Li}
\email{jr981025@mail.ustc.edu.cn}
\correspondingauthor{Yi-Fu Cai}
\email{yifucai@ustc.edu.cn}

\Abs{
It is well known that multiple Galactic thermal dust emission components may exist along the line of sight, but a single-component approximation is still widely used, since a full multi-component estimation requires a large number of frequency bands that are only available with future experiments. 
In light of this, we present a reliable, quantitative, and sensitive criterion to test the goodness of \emph{all} kinds of dust emission estimations. 
This can not only give a definite answer to the quality of current single-component approximations; 
but also help determine preconditions of future multi-component estimations. 
Upon the former, previous works usually depend on a more complicated model to improve the single-component dust emission; however, our method is free from any additional model, and is sensitive enough to directly discover a substantial discrepancy between the \textit{Planck} HFI data (100-857 GHz) and associated single-component dust emission estimations. 
This is the first time that the single-component estimation is ruled out by the data itself. 
For the latter, a similar procedure will be able to answer two important questions for estimating the complicated Galactic emissions: the number of necessary foreground components and their types. 
}


\mktt

\section{Introduction}
\label{sec:introduction}
\end{CJK*}

The discovery of the cosmic microwave background (CMB) in the last century \citep{1965ApJ...142..419P} and the precise measurements of it in subsequent decades have become one of the key pillars supporting the foundation of precision cosmology \citep{2004MSAIS...5..325B}. 
To date, three generations of space telescopes---COBE \citep{1992ApJ...397..420B}, WMAP \citep{2013ApJS..208...20B} and \textit{Planck} \citep{2010A&A...520A...1T}---have successively achieved increasingly higher-precision observations of the CMB across the full sky. 
Furthermore, numerous ground-based and space-based telescope projects targeting CMB anisotropies, such as POLARBEAR \citep{2010SPIE.7741E..1EA}, BICEP \citep{2014PhRvL.112x1101B}, CMB-S4 \citep{2019arXiv190704473A}, AliCPT \citep{2017arXiv171003047L}, Litebird \citep{2023PTEP.2023d2F01L}, and CLASS \citep{2020JLTP..199..289D}, have emerged or are scheduled for launch. 
Due to the solar system's position within the Galactic plane, CMB measurements are inevitably subjected to strong interference from Galactic foreground signals, such as compact sources, synchrotron radiation, free-free emission, molecular emission lines, spinning dust emission, and thermal dust emission. 
Consequently, CMB observations must carefully separate these foreground signals \citep{2016A&A...594A..10P}. 
For instance, \cite{2016MNRAS.458.2032R} discusses the consequences of incorrect modeling of dust emission in detail. 
Among these foreground components, thermal dust emission, originating from Galactic dust particles warmed by interstellar radiation, predominates in the CMB foreground at frequencies exceeding 80 GHz \citep{2013A&A...553A..96D}. 
Improper modeling of thermal dust emission, especially in 100 - 150 GHz bands, may significantly interfere with the search for primordial gravitational waves \citep{2015PhRvL.114j1301B}. 
Thus, modeling the thermal dust component has received particular attention. 

The first full-sky dust emission map at $100\,\mu\mathrm{m}$ (3000 GHz) was created in 1998 \citep{1998ApJ...500..525S} by combining data from the Diffuse Infrared Background Experiment (DIRBE, \citealp{1998ApJ...508...25H}) and the Infrared Astronomical Satellite (IRAS, \citealp{1984ApJ...278L...1N}). 
Subsequently, \cite{1999ApJ...524..867F} developed a thermal dust emission model compatible with the measured maps from 30 to 3000 GHz by merging data sets from DIRBE, the Far-InfraRed Absolute Spectrophotometer (FIRAS, \citealp{1992ApJ...397..420B}), and the Differential Microwave Radiometer (DMR, \citealp{1990ApJ...360..685S}), all aboard COBE, which was launched in 1989. 
This model, featuring two modified blackbody spectra, was interpreted to represent two interstellar dust components speculated by the authors: silicon-based and carbon-based dust grains. 
However, the two-component model still shows biases compared to actual measurements, indicating the potential for more complex dust components along the line of sight than those accounted for by the silicate-carbonaceous model \citep{2018ApJ...853..127H, 2022A&A...659A..70D, 2023MNRAS.519.4370M, 2018A&A...610A..16G, 2015MNRAS.451L..90T, 2017PhRvD..95j3511P}. 
In addition, beyond the focus on physical components, \cite{2017MNRAS.472.1195C, 2023A&A...669A...5V, 2021MNRAS.503.2478R} employ moment expansion method to analyze and separate complicated foreground components containing thermal dust emission while considering spatial averaging effects including averaging along line of sight, over beam of telescopes, and in spherical harmonic degrading process. 

Furthermore, some data-driven methods, such as GNILC (generalized needlet internal linear combination, \citealp{2011MNRAS.418..467R}) and GNILC-moment method \citep{2024JCAP...06..018C}, 
estimate the foreground signal with better model independence, which is very useful. However, these methods only deal with the involved frequency bands, particularly those with high SNR (signal-to-noise ratio). 
Hence, they alone cannot extrapolate the dust emission from $\sim$500 to $\sim$100 GHz, which is useful in detecting primordial gravitational waves. Therefore, one still has to deal with the dust models in an advanced task. 

The \textit{Planck} satellite, launched in 2009, delivered unparalleled full-sky observations of CMB anisotropies with superior accuracy compared to COBE and WMAP, covering wavelengths from submillimeter to centimeter scales.
The \textit{Planck} data products include three Low Frequency Instrument (LFI) maps at 30, 44, and 70 GHz channels \citep{2014A&A...571A...2P} and six High Frequency Instrument (HFI) maps at 100, 143, 217, 353, 545, and 857 GHz channels \citep{2014A&A...571A...6P}, respectively. 
However, the maps available to determine the thermal dust emission are mainly at 353, 545, and 857 GHz, since dust emission is dominant in these bands (\citealp[Fig.~51]{2016A&A...594A..10P}, and \citealp[Fig.~35]{2020A&A...641A...4P}). 
Because the number of bands from 353 to 857 GHz is not enough to support a model with two dust components, the \textit{Planck} team used a single-component thermal dust emission model featuring three parameters to fit the data from 353 to 857 GHz, and then extrapolated the model down to 100-217 GHz to determine their dust emissions \citep{2016A&A...594A..10P}, which is particularly important for the detection of primordial gravitational waves \citep{2015PhRvL.114j1301B}. 

In fact, the \textit{Planck} team has already recognized potential limitations of the single-component dust emission model, cautioning that the single-component model might be suboptimal for thermal dust emission near the Galactic plane, as indicated by high $\chi^2$ values in their statistical tests \citep{2016A&A...596A.109P}. 
Additionally, previous literature investigated the biases of the single-component model at frequencies below 353 GHz, reporting noticeable discrepancies in the 100 - 150 GHz range \citep{2017PhRvD..95j3517L, 2018ApJ...853..127H, 2024ApJ...970...43S, 2015ApJ...798...88M}. 
Wherein \cite{2015ApJ...798...88M} combined \textit{Planck} HFI maps \citep{2014A&A...571A...6P} and DIRBE/IRAS data \citep{1998ApJ...500..525S} to derive a best-fit two-component thermal dust model. 
Furthermore, they compared observed \textit{Planck} data at each pixel with predictions from their two-component model and the \textit{Planck} single-component model, concluding that the latter underestimated thermal dust emission by 7.9\%, 12.6\% and 18.8\% at 217, 143 and 100 GHz respectively. 

Unlike the method for testing the single-component model in \cite{2015ApJ...798...88M}, in this paper, the morphology of thermal dust emission is taken into account to constrain the model-to-data differences in the \textit{Planck} HFI bands, which not only increases the method's sensitivity but also help to exclude several sources of differences, including noise, systematics, color correction uncertainty, band offset, residual free-free emission, cosmic infrared background (CIB) emission, and zodiacal light. 
Compared with other global tests such as \cite{2024ApJ...970...43S}, our method works in local disks with variable sizes to improve the method's robustness. 
In addition, current works usually require preset dust models, either to fit the observational data or to compare the goodness of two different models; whereas our method can use the data itself to test any dust model, which is much better in objectivity. Because of this, our method can also help determine the number of components and their complexities (like the moment orders in \citealp{2017MNRAS.472.1195C}) for multi-component models in future research. 

This article is organized as follows:
Section~\ref{sec:method} provides a detailed explanation of the fundamental concepts and mathematical framework of the statistical methods used in our analysis, and
Section~\ref{sec:data processing} outlines the pre-processing techniques applied to the \textit{Planck} HFI maps, highlighting the steps taken to ensure data accuracy and reliability. Then the discrepancies between the \textit{Planck} single-component model and \textit{Planck} HFI data are examined in Section~\ref{sec:results}, 
and some potential sources that could be responsible for the discrepancies are discussed in 
Section~\ref{sec: discussion of sources}. 
The main conclusions are given in Section~\ref{sec:conclusion}, and the code for this work is publicly available on GitHub.\footnote{\url{https://github.com/Jia-Rui-Li/Thermal-dust-components}}

\section{Mathematics and method}
\label{sec:method}
Our approach starts with the null hypothesis that the single-component modified black body model is a good approximation of the thermal dust emission in real maps, so the total emission intensity is the sum of the model prediction (major) and residual (minor), as shown below: 
\begin{equation}
\label{equ:error of ratio 01}
\begin{aligned}
x'&= x + a \\
y'&= y + b, 
\end{aligned}
\end{equation}
where $x'$, $y'$ represent the total intensity at two neighboring bands, $x$, $y$ are the model predictions, and $a$, $b$ are overall residuals, including noise, systematic errors, residual CMB, other foreground residuals, etc. 
For simplicity, subscript $i$ for pixel index is omitted.
 
Unlike the approach in \cite{2015ApJ...798...88M} that involves a direct comparison of the dust emission's intensity with a zero-level correction, our method takes the dust emission's morphology in local patches into account, and use the similarity of morphology between two neighboring bands to statistically constrain the model-to-data discrepancies. 
The advantage of this idea is numerous: it is unaffected by the band offset (zero-level), almost unaffected by the dipole signal, and helps to exclude most contamination sources in the final discussion. 

The covariance between samples $x$ and $y$ is: 
\begin{equation}
\label{equ:cov}
S_{xy} = \frac{1}{n-1}\sum{(x-\overline{x})(y-\overline{y})},
\end{equation}
where $n$ is the sample size and $\overline{x}$ is the mean of $x$. 
Similarly, we calculate $S_{x'y'}$. 
For the self-covariance, we have 
\begin{eqnarray}
\label{equ:self-cov}
S_{xx} = && \frac{1}{n-1}\sum{(x-\overline{x})^2}=\sigma_x^2,
\end{eqnarray}
where $S_{xx}$ is the sample variance, and $\sigma_x$ is the standard deviation. 
The slope of the linear regression line between samples $x$ and $y$ is known as: 
\begin{equation}
\label{equ:R}
R = \dfrac{\sum(x-\overline{x})(y-\overline{y})}{\sum(x-\overline{x})^2}
= \frac{S_{xy}}{S_{xx}}. 
\end{equation}
We refer to $R$ as the ``linear ratio'' of the model. 
Similarly, the linear ratio between $x'$ and $y'$ is
\begin{equation}
\label{equ:R'}
R'= \frac{S_{x'y'}}{S_{x'x'}}.
\end{equation}
The Pearson cross-correlation between two data samples at neighboring bands is
\begin{equation}
\label{equ:pearson-CC}
C_{x'y'} = \frac{S_{x'y'}}{\sigma_{x'}\sigma_{y'}}. 
\end{equation}
Obviously, if $C_{x'y'}$ (abbreviated as $C'$ hereafter) is close enough to 1, then the residuals $a$ and $b$ should be relatively small, i.e., the residuals-to-dust ratios $k_1$ and $k_2$ below should be small:
\begin{eqnarray}
k_1 = \frac{\sigma_a}{\sigma_x},\quad k_2 = \frac{\sigma_b}{\sigma_y}.
\end{eqnarray}
In this case, $R$ (from model) must be close to $R'$ (from real data), and their difference will be tightly constrained, which is the main idea of this work. Below, we prove this idea analytically: 

Eqs.~(\ref{equ:R}-\ref{equ:pearson-CC}) can be solved using the sample variance / covariance between thermal dust and residual components, yielding 
\begin{eqnarray}
\label{equ:C}
C' &=& \frac{C_{xy}+k_2 C_{xb}+k_1 C_{ya}+k_1 k_2 C_{ab}}{\sqrt{(1+2k_1 C_{xa}+k_1^2)(1+2k_2 C_{yb}+k_2^2)}} \\ \nonumber
&\approx& C_{xy} + k_1 (C_{ya}-C_{xy}C_{xa}) + k_2 (C_{xb}-C_{xy}C_{yb}), 
\end{eqnarray}
and
\begin{eqnarray}
\label{equ:R'/R}
\frac{R'}{R} &=& \frac{C_{xy} + k_2C_{xb} + k_1C_{ya} + k_1 k_2C_{ab}}{C_{xy}(1+2k_1 C_{xa}+k_1^2)} \\ \nonumber
&\approx& 1 - 2 k_1 C_{xa} + (k_1 C_{ya} + k_2 C_{xb})/C_{xy}. 
\end{eqnarray}
The upper rows of Eqs.~(\ref{equ:C}-\ref{equ:R'/R}) are analytic, and the lower rows are asymptotic expansions. One can see that their departures from 1 are at the same level, which is the residual-to-dust ratio times the residual-to-dust cross-correlation. 
Therefore, the value $|1-R'/R|$ is tightly constrained by $|1-C'|$, indicating that, once $|1-C'|$ is sufficiently small, $|1-R'/R|$ must also be small. This connection is the core of this work because $C'$ is model-free and provides a constraint that is completely determined by the real data. 
However, we also note that in some extreme cases, the ratio $R'/R$ can deviate significantly from 1, even when $C'$ is close to 1. 
For instance, if $C_{xy}=1$, $C_{xa} = C_{ya} = 0.5$ and $C_{yb} = C_{xb}=0$, and $k_1=0.5$, then we have $C' = 0.945$ but $R'/R = 0.714$. 
However, this kind of scenario is very unlikely and also violates the null hypothesis. 

For more realistic cases, the above conclusion is tested with a simulation based on samples of $k_1$, $k_2$, $C_{xy}$, $C_{xa}$, $C_{xb}$, $C_{ya}$, $C_{yb}$, and $C_{ab}$, which are assumed to be independent of each other. 
Here, residual and dust emission are expected to be uncorrelated, as well as residuals in different bands; therefore, five of these six cross-correlations should be around zero, except for $C_{xy}$, which should be close to 1. 
The samples of these 
cross-correlations can be drawn from Gaussian random variables via the Fisher transformation~\citep{doi:10.1093...biomet...10.4.507}:
\begin{equation}
\label{equ:fisher-transform}
\begin{dcases}
Z=&\frac{1}{2}\ln\left(\frac{1+C}{1-C}\right), \\
C=&\tanh Z. 
\end{dcases}
\end{equation}
where $Z$ is Gaussianly distributed and is easy to sample. 
For example, assume $C_{xy}\sim0.99$ and the degree of freedom (smaller than the sample size, also the number of sky pixels, due to smoothing and spatial correlation between pixels) is $N = 100$, then the realizations of $Z$ can be drawn from a Gaussian distribution with a mean of $2.6467$ and a standard deviation of $1/\sqrt{N-3} = 0.1015$. 
Then the sample of $C$ is obtained from the sample of $Z$ with Eq.~(\ref{equ:fisher-transform}). 
$C_{xa}$, $C_{xb}$, $C_{ya}$, $C_{yb}$, and $C_{ab}$ are treated in the analogous manner. 

In addition, for $k_1$ and $k_2$, we have
\begin{eqnarray}
\label{equ:k1_expectation_k1}
k_1^2 &=& \frac{\sigma_a^2}{\sigma_x^2} = \frac{\displaystyle\sum_{i=1}^N (a_i-\bar{a})^2}{\displaystyle\sum_{i=1}^N (x_i-\bar{x})^2}
= \frac
    {
    \displaystyle(\sigma_a')^2\sum_{i=1}^N \left(\frac{a_i}{\sigma_a'}-\frac{\overline{a}}{\sigma_a'}\right)^2
    }
    {
    \displaystyle(\sigma_x')^2\sum_{i=1}^N \left(\frac{x_i}{\sigma_x'}-\frac{\overline{x}}{\sigma_x'}\right)^2
    }\nonumber \\ 
&=& (k_1')^2 
\frac
    {
    \displaystyle\sum_{i=1}^N \left(\frac{a_i}{\sigma_a'}-\frac{\overline{a}}{\sigma_a'}\right)^2
    }
    {
    \displaystyle\sum_{i=1}^N \left(\frac{x_i}{\sigma_x'}-\frac{\overline{x}}{\sigma_x'}\right)^2
    }, 
\end{eqnarray}
where $\sigma_a'$, $\sigma_x'$ and $k_1' \equiv \sigma_a' / \sigma_x'$ are normalization factors. It is then evident that
$k_1^2$ and $k_2^2$ follow the F-distribution with degrees of freedom $(N, N)$. 
Therefore, $k_1$ and $k_2$ are drawn by multiplying their
normalization factors $k_1'$ and $k_2'$ (close but not exactly equal to expectation) by the square root of the F-distributed samples.  

\begin{figure}
\centering
\includegraphics[width=0.48\textwidth]{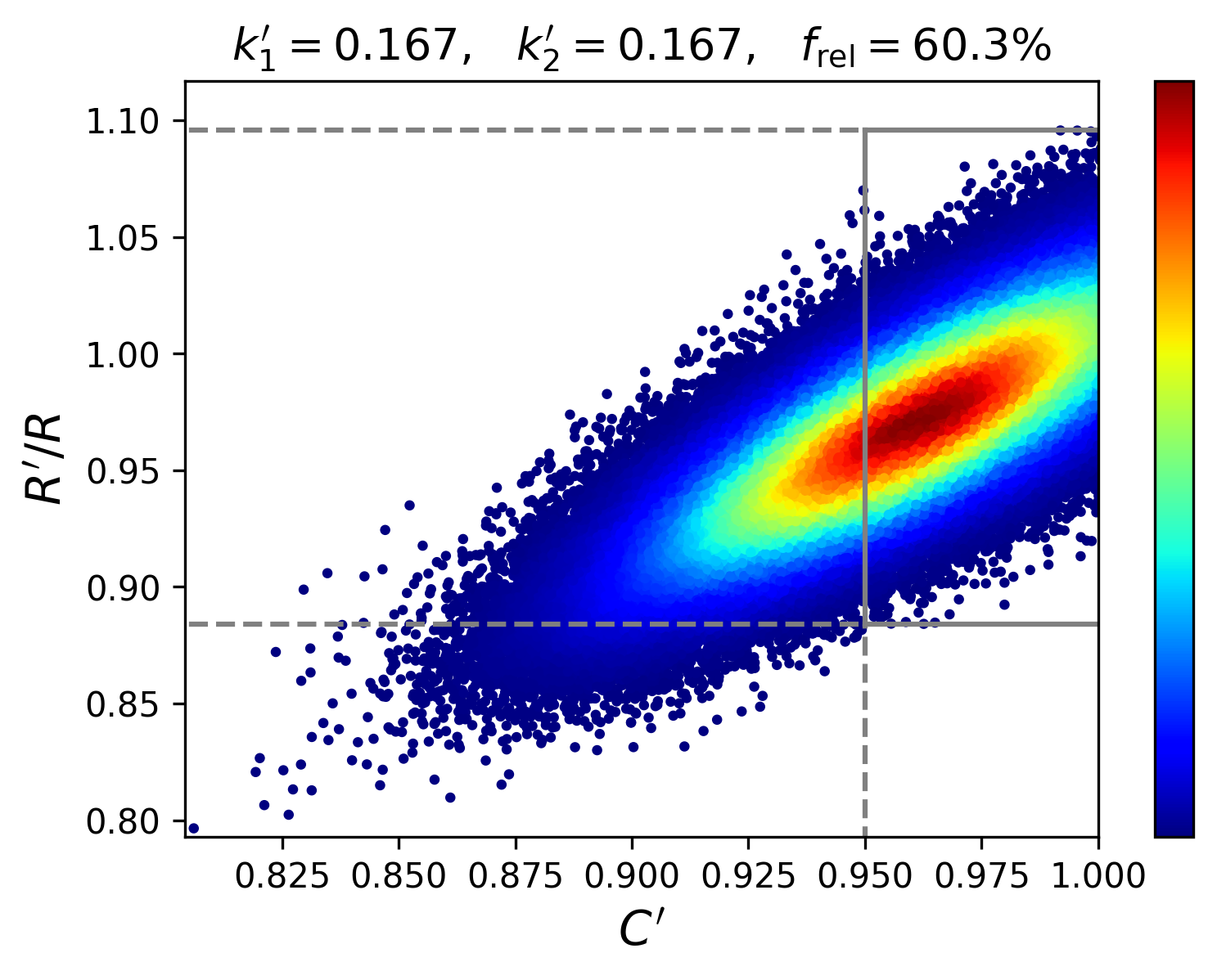}
\caption{Scatter and contour plot of simulated $R'/R$ and $C'$ values, focusing on the departure of $R'/R$ from 1 with the constraint $C' \geq 0.95$ (see Section \ref{sec:method} and Eq.~(\ref{equ:C}-\ref{equ:R'/R})). 
The normalization factors of $k_1$ and $k_2$ are set to $k_1' =  k_2' = 0.167$. 
The color gradient from blue to red indicates an increasing sample density. 
The gray vertical line marks the range of $0.95 \leq C' \leq 1$, and the gray horizontal lines mark the maximum and minimum values of $R'/R$ in this range.
Therefore, the regions of interest is within the rectangular gray frame. 
In this simulation, $60.3\%$ (called $f_{\rm{rel}}$) of the scatter points are within the gray frame, enough for a statistical analysis. }
\label{fig:cc-rr_density}
\end{figure}

One of the simulation results is shown in Fig.~\ref{fig:cc-rr_density}, which is also part of the figure in Subsection~\ref{sub: sim of data to model departure}. 
Because the Fisher transformation is an approximate approach and the random variables $C_{xy}$, $C_{xa}$, $C_{xb}$, $C_{ya}$, $C_{yb}$, and $C_{ab}$ are drawn independently, a tiny fraction of the $C'$ samples can produce values greater than 1, which are excluded. This plus a further constraint of $C' \geq 0.95$ leaves the samples of interest in the gray frame in Fig.~\ref{fig:cc-rr_density}, which will be used later in the statistical analysis. 
From the gray frame in Fig.~\ref{fig:cc-rr_density}, it is evident that when $C' \geq 0.95$, $R'/R$ are approximately in the range $1\pm 0.1$; thus, the difference between $R$ (only from the model) and $R'$ (only from the sky maps) should be less than 10\%, which gives a tight and robust constraint on the model error. 

Interestingly, the above conclusion does not bind to the single-component dust emission model, because it only requires to assume that the model is correct. 
This fact means that the above method can be used to test the matching of any dust model with the observed data.

\section{Data Processing}
\label{sec:data processing}

This section outlines the details in data processing, including the preprocessing of \textit{Planck} HFI data and the division of the dust emission maps into mosaic disks for statistical analysis.

\subsection{Data sets}
\label{subsec:data sets}
The \textit{Planck} HFI sky maps \citep{2020A&A...641A...3P},\footnote{HFI\_SkyMap\_100/143/217/353/545/857\_2048\_R3.01\_full.fits} HFI detector parameters,\footnote{HFI\_RIMO\_R3.00.fits} and CMB map from SMICA method \citep{2020A&A...641A...7P}\footnote{COM\_CMB\_IQU-smica\_2048\_R3.00\_full.fits} used in this study are sourced from the latest \textit{Planck} 2018 release, PR3 \citep{https://doi.org/10.26131/irsa558, https://doi.org/10.26131/irsa559}. 
Commander templates for free-free emission,\footnote{COM\_CompMap\_freefree-commander\_0256\_R2.00.fits} synchrotron radiation,\footnote{COM\_CompMap\_Synchrotron-commander\_0256\_R2.00.fits} carbon monoxide (CO) emission,\footnote{COM\_CompMap\_CO21-commander\_2048\_R2.00.fits} and 94/100 GHz molecular emission line\footnote{COM\_CompMap\_xline-commander\_0256\_R2.00.fits} are sourced from the \textit{Planck} 2015 release (PR2, \citealp{2016A&A...594A..10P}), representing the most recent Galactic foreground results provided by \textit{Planck} team. 
Additional auxiliary data sets, such as point source masks,\footnote{HFI\_Mask\_PointSrc\_2048\_R2.00.fits} Galactic plane masks without apodization,\footnote{HFI\_Mask\_GalPlane-apo0\_2048\_R2.00.fits} and compact source catalogs \citep{2016A&A...594A..26P},\footnote{COM\_PCCS\_030/044/070\_R2.04.fits \citep{https://doi.org/10.26131/irsa443, https://doi.org/10.26131/irsa466, https://doi.org/10.26131/irsa458}} \footnote{COM\_PCCS\_100\,/143\,/\,217\,/\,353/545/857\_R2.01.fits \citep{https://doi.org/10.26131/irsa470, https://doi.org/10.26131/irsa474, https://doi.org/10.26131/irsa456, https://doi.org/10.26131/irsa446, https://doi.org/10.26131/irsa462, https://doi.org/10.26131/irsa464}} \footnote{COM\_PCCS\_100/143/217/353/545/857-excluded\_R2.01.fits \citep{https://doi.org/10.26131/irsa452, https://doi.org/10.26131/irsa475, https://doi.org/10.26131/irsa447, https://doi.org/10.26131/irsa461, https://doi.org/10.26131/irsa477, https://doi.org/10.26131/irsa451}}  are also obtained from PR2. 
Furthermore, three different single-component thermal dust emission models are derived from the \textit{Planck} 2013 release (PR1, \citealp{2014A&A...571A..11P}),\footnote{HFI\_CompMap\_ThermalDustModel\_2048\_R1.20.fits} from PR2 \citep{2016A&A...596A.109P},\footnote{COM\_CompMap\_Dust-GNILC-Model-Opacity\_2048\_R2.01.fits} \footnote{COM\_CompMap\_Dust-GNILC-Model-Spectral-Index\_2048\_R2.01.fits} \footnote{COM\_CompMap\_Dust-GNILC-Model-Temperature\_2048\_R2.01.fits} and from \cite{2019A&A...623A..21I, 2019yCat..36230021I}.\footnote{\url{https://cdsarc.cds.unistra.fr/ftp/J/A+A/623/A21/}}

\subsection{The single-component thermal dust emission model}
\label{subsec:dust model}
The single-component modified blackbody model describes thermal dust emission with three parameters, i.e., optical depth, spectral index, and equilibrium temperature: 
\begin{equation}
\label{equ:single_component_model}
I_\nu = \tau_{\nu_0} \left(\frac{\nu}{\nu_0}\right)^\beta B_\nu(T), 
\end{equation}
where $I_\nu$ represents the thermal dust emission intensity at frequency $\nu$, measured in units of $\mathrm{MJy\,sr^{-1}}$ or $\mathrm{W\,m^{-2}\,Hz^{-1}\,sr^{-1}}$,\footnote{Since $\mathrm{W\,m^{-2}\,sr^{-1}\,Hz^{-1}} = 10^{20}\,\mathrm{MJy\,sr^{-1}}$, both units are referred to as ``SI'' units for reasons of expediency hereafter. } 
the optical depth of interstellar dust at the reference frequency $\nu_0$ is denoted as $\tau_{\nu_0}$, $T$ is the thermal equilibrium temperature of the dust particles, $B_\nu(T)$ is the Planck function and $\beta$ is the spectral index, typically around 1.55 \citep{2020A&A...641A...4P}. 

For cross-checking, the parameters of the single-component dust models are selected from three works:
\textit{Planck} team produced their initial thermal dust emission model in 2013 \citep{2014A&A...571A..11P} by minimum variance fitting (referred to as M13), which did not take the CIB contamination into account.  
In subsequent updates, they used GNILC method to subtract CMB anisotropies, CIB and instrumental noise from the \textit{Planck} 353, 545, and 857 GHz temperature maps \citep{2016A&A...594A...8P}, and produced dust emission maps at these three channels without using a specific dust spectral model. 
Then, along with the $100\,\mu\mathrm{m}$ (3000 GHz) map combining the IRIS map \citep{2005ApJS..157..302M} and the result from \cite{1998ApJ...500..525S}, they obtained the dust model parameters for the single modified blackbody spectrum \citep{2016A&A...596A.109P} by minimum variance fitting (referred to as M15). 
Furthermore, \cite{2019A&A...623A..21I} exploited a sparsity-based and parametric method known as ``parameter recovery exploiting model informed sparse estimates'' (PREMISE) method to obtain another single modified blackbody model (referred to as M19), based on the same data sets used in M15. 

In this paper, we examine the matching of M13, M15, and M19 with the data maps at \textit{Planck} HFI bands. 
We refer to the predicted dust emission map from the single-component model as the ``dust model map.''

\subsection{Re-beaming and in-painting}

The \textit{Planck} HFI maps and foreground templates have different HEALPix\footnote{\url{https://healpix.sourceforge.io}} resolutions and beam width, as indicated in Table \ref{tab:properties input maps}. 
To compare the maps at different frequencies, the maps with high resolutions ($N_\mathrm{side}$ = 2048) are re-beamed to the 100 GHz map's beam width of $9\overset{\prime}{.}66$, and the maps with low resolutions ($N_\mathrm{side}$ = 256) are upgraded to $N_\mathrm{side}=2048$ with a harmonic domain operation. 
To minimize the Gibbs phenomenon during spherical harmonics transform, the point sources in the \textit{Planck} HFI maps are removed by the \textit{Planck} intensity point source masks as shown in Section~\ref{sec: further masking}, 
just as an instance at 857 GHz. 
Then we iterate 2000 times to in-paint the holes after masking \citep{7544859}. 

\begin{table}
\centering
\begin{tabular}{lllr}
\hline
Maps & FWHM & Unit & $N_{\rm side}$\\
\hline
100 GHz map & $9\overset{'}{.}66$ & $\mathrm{K_{CMB}}$ & 2048\\
143 GHz map & $7\overset{'}{.}22$ & $\mathrm{K_{CMB}}$ & 2048\\
217 GHz map & $4\overset{'}{.}90$ & $\mathrm{K_{CMB}}$ & 2048\\
353 GHz map & $4\overset{'}{.}92$ & $\mathrm{K_{CMB}}$ & 2048\\
545 GHz map & $4\overset{'}{.}67$ & $\mathrm{MJy\,sr^{-1}}$ & 2048\\
857 GHz map & $4\overset{'}{.}22$ & $\mathrm{MJy\,sr^{-1}}$ & 2048\\
CMB SMICA map & $5'$ & $\mathrm{K_{CMB}}$ & 2048\\
CO (2-1) high SNR map & $7\overset{'}{.}5$ & $\mathrm{K_{RJ}\,km\,s^{-1}}$ & 2048\\
Synchrotron map & $1^\circ$ & $\mathrm{K_{RJ}}$ & 256\\
Free-free map & $1^\circ$ & $\mathrm{K}$ and $\mathrm{pc\,cm^{-6}}$ & 256\\
94/100 GHz emission line & $1^\circ$ & $\mu\mathrm{K_{CMB}}$ &256\\
\hline
\end{tabular}
\caption{Characteristics of the input HFI maps, CMB, CO emission map, synchrotron, free-free, and 94/100 GHz molecular emission map from \textit{Planck} products, including their FWHM, unit, and HEALPix resolution \citep{2020A&A...641A...1P, 2016A&A...594A..10P, 2020A&A...641A...7P}. }
\label{tab:properties input maps}
\end{table}

\subsection{Unit conversion and color correction}
\label{subsection_Unit_conversion_Colour_correction}
Due to the transmission spectra of \textit{Planck} broadband photometric receivers at each channel and the involvement of multiple unit systems in data processing, this subsection details the necessary unit conversions and color corrections \citep{2014A&A...571A...9P}. 

All photometric detectors determine the flux intensity of astronomical objects by quantifying the electromagnetic energy accumulated on the detectors per unit time, per unit area, per unit solid angle, and per unit frequency interval (i.e., spectral energy distribution): 
\begin{equation}
\int J_\nu \mathcal{T}(\nu) \dd\nu, 
\end{equation}
where $\mathcal{T}(\nu)$ is the transmission spectrum of the detector, and $J_\nu$ is the spectral energy distribution of the emission source, expressed in $\mathrm{W\,m^{-2}\,Hz^{-1}\,sr^{-1}} = 10^{20}\,\mathrm{MJy\,sr^{-1}}$. 
By observing a standard astronomical object with a known spectral energy distribution (denoted as $J_{0,\nu}$), as the calibration source, the accumulated electromagnetic radiation energy (or equivalently power) on the detector can be converted into physical quantities such as thermodynamics brightness differential temperature (in units of $\mathrm{K_{CMB}}$) or brightness temperature (in units of $\mathrm{K_{RJ}}$). 
This process is known as calibration \citep{2014A&A...571A...9P}. 
Since energy can be equivalently converted to other physical quantities, unit conversion naturally occurs in this process. 
In a given unit system X, the detector reading $s_\mathrm{X}$ is: 
\begin{equation}
s_\mathrm{X} = A_\mathrm{X}\int J_\nu \mathcal{T}(\nu) \dd\nu, 
\end{equation}
where $A_\mathrm{X}$ is the calibration coefficient that depends on the chosen unit system. 
When observing a target astronomical object with a calibrated detector, the reading depends on the spectral energy distribution of both the calibration source and the target object. 
Since the transmission spectrum of the detector is not an ideal $\delta$-function and the calibration source and the target source have different spectral energy distribution, 
color correction is necessary. 
Because the calibration coefficient $A_\mathrm{X}$ for unit conversion also depends on the transmission spectrum of the detector and the spectral energy distribution of calibration source, unit conversion and color correction are performed simultaneously. 
The specific cases for unit conversions and color corrections involved in this study are detailed in Appendix~\ref{app:unit_conversion_colour_correction}.

\subsection{Subtracting non-dust components}
\label{subsec:dust data}

\begin{figure}[!htb]
\centering
\includegraphics[width=0.23\textwidth]{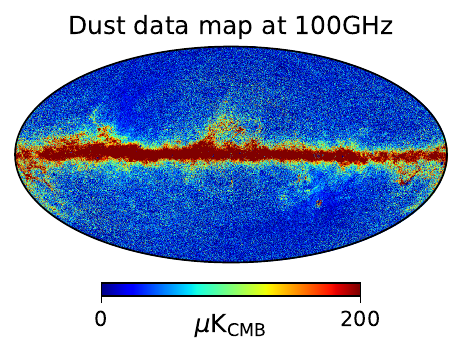}
\includegraphics[width=0.23\textwidth]{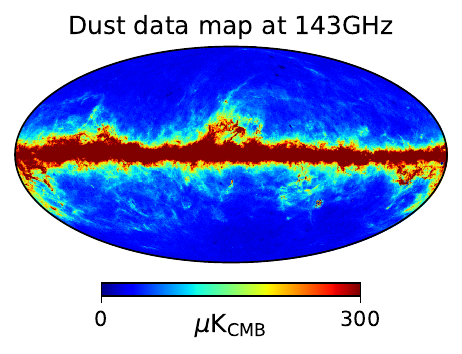}

\includegraphics[width=0.23\textwidth]{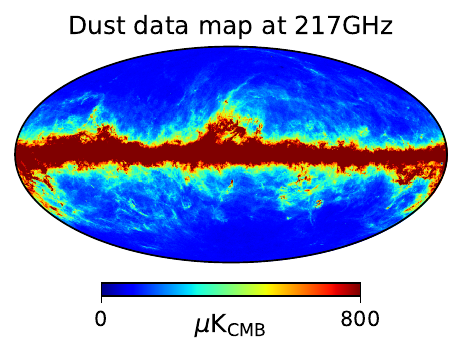}
\includegraphics[width=0.23\textwidth]{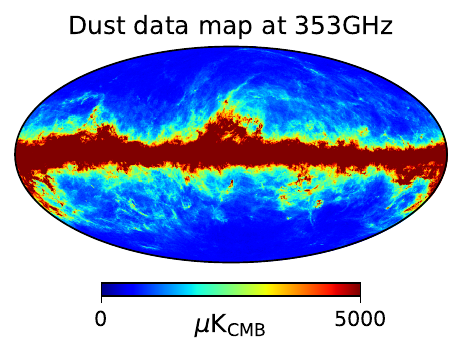}

\includegraphics[width=0.23\textwidth]{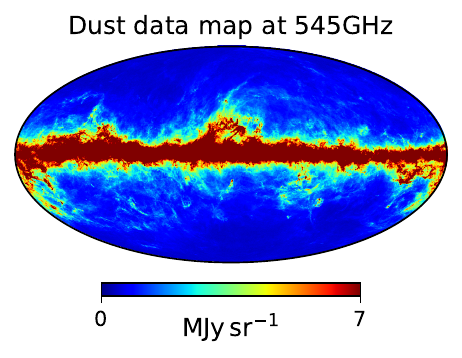}
\includegraphics[width=0.23\textwidth]{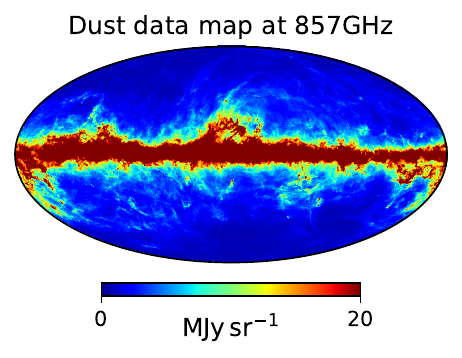}
\caption{
The thermal dust component from \textit{Planck} HFI maps shown in a linear color scale. 
The emission intensities in the left and right panels correspond to $x'$ and $y'$ in Eq. (\ref{equ:error of ratio 01}) respectively. 
}
\label{fig:thermal dust component}
\end{figure}

\begin{figure}[!htb]
\centering
\includegraphics[width=0.49\textwidth]{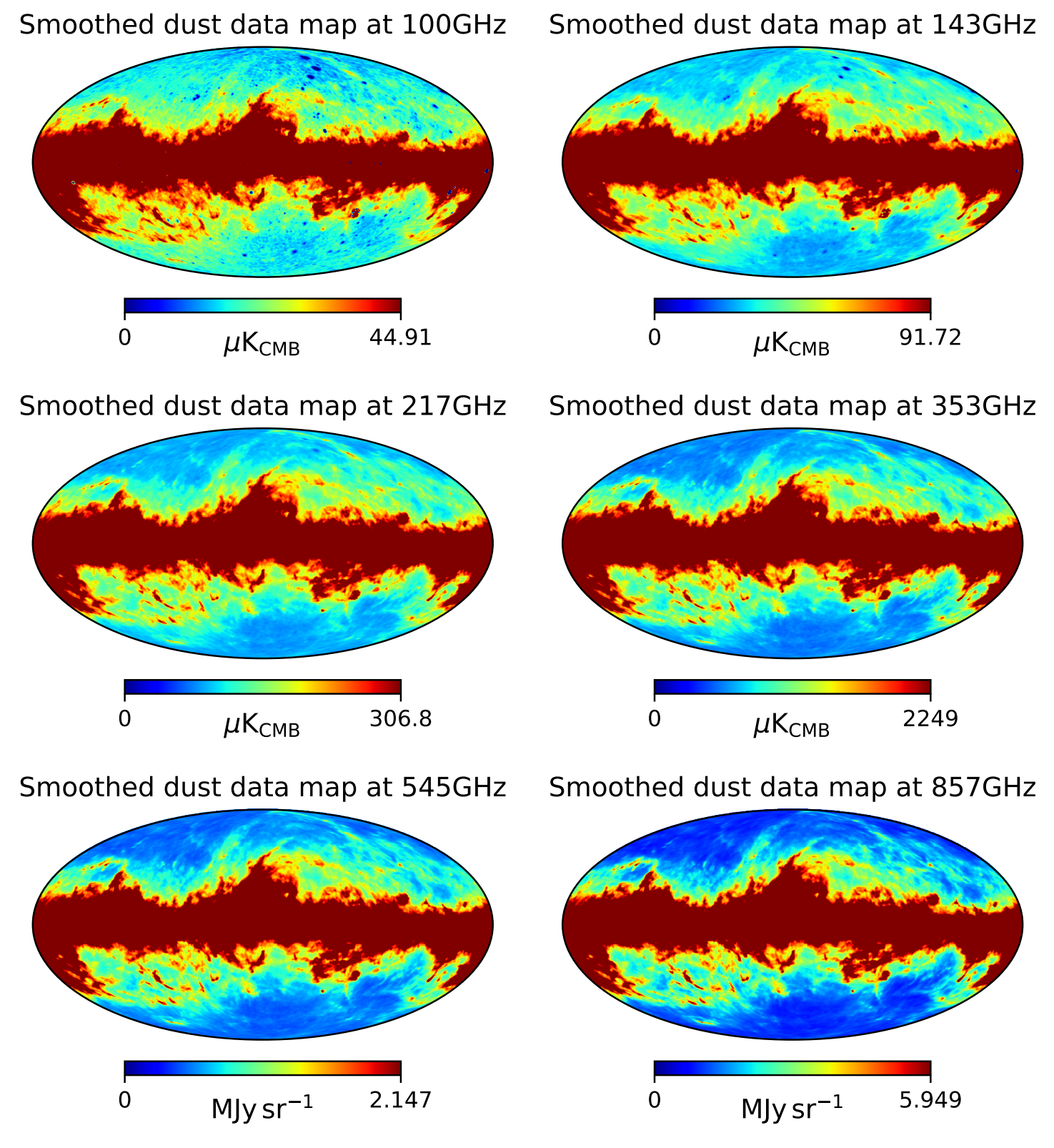}
\caption{
The counterpart of Fig.~\ref{fig:thermal dust component}, but smoothed with $1^\circ$ FWHM, to show the morphology of the dust data maps more clearly. 
The maximum of the color bar is set to make 70\% of the smoothed dust map with intensities smaller than this maximum. 
}
\label{fig:smoothed thermal dust component}
\end{figure}

The non-dust components are subtracted from the PR3 HFI maps, including CMB anisotropies, free-free emission, synchrotron radiation, CO emission lines, and 94/100 GHz lines. 
In addition, the zodiacal light component has been removed from the PR3 maps by \textit{Planck} team \citep{2020A&A...641A...3P}. 
Details of the subtraction are provided in Appendix~\ref{app:removing other components}, and the results of subtraction are illustrated in Fig.~\ref{fig:thermal dust component} and Fig.~\ref{fig:smoothed thermal dust component} (after smoothing), 
from which (especially from Fig.~\ref{fig:smoothed thermal dust component}) one can observe similar morphology at all six HFI channels, confirming that these maps are dominated by the thermal dust emission. 
The sky regions on these maps correspond to $x'$ or $y'$ in Eq. (\ref{equ:error of ratio 01}), and we refer to these sky maps as the ``dust data maps'' in this paper.  

\subsection{Further masking}
\label{sec: further masking}
\subsubsection{Compact source mask}
Some sky regions in Figs.~\ref{fig:thermal dust component}-\ref{fig:smoothed thermal dust component} have unwanted cold spots, because the non-dust foreground templates (especially free-free emission) might be affected by the compact point sources in relevant pixels \citep{2016A&A...594A..25P}. 
Therefore, in addition to the initial point source mask, 
further masking is adopted using the \textit{Planck} Catalogues of Compact Sources \citep{2016A&A...594A..26P} at both LFI and HFI channels (as the free-free template is affected by compact sources in LFI channels). 
\begin{figure}[!htb]
\includegraphics[width=0.23\textwidth]{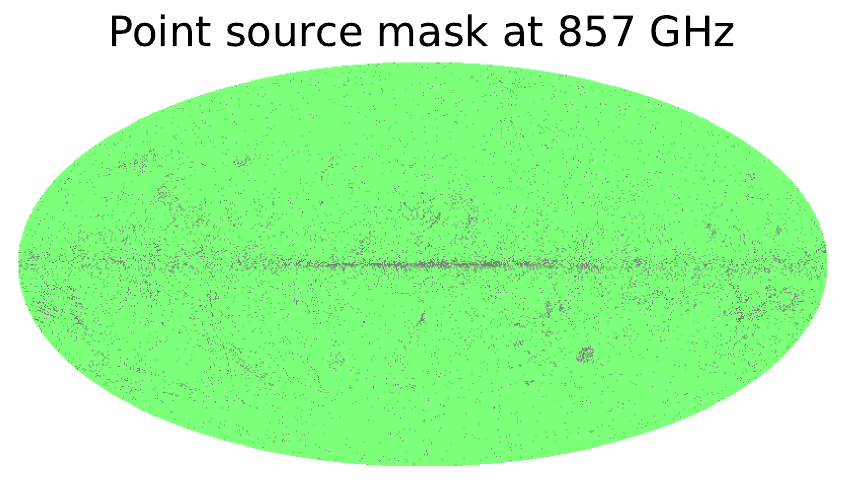}
\includegraphics[width=0.23\textwidth]{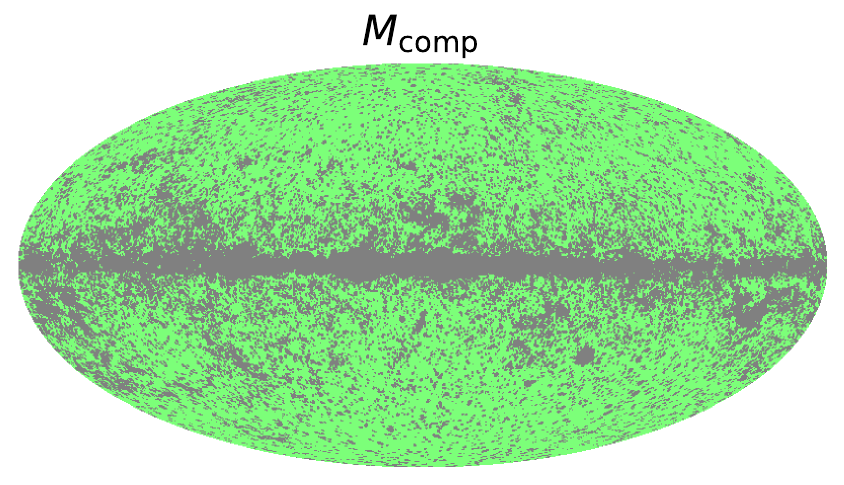}

\includegraphics[width=0.23\textwidth]{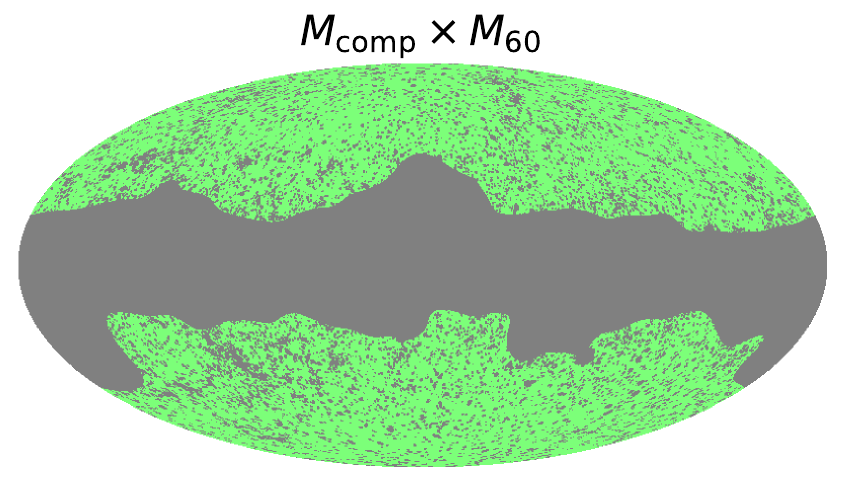}
\includegraphics[width=0.23\textwidth]{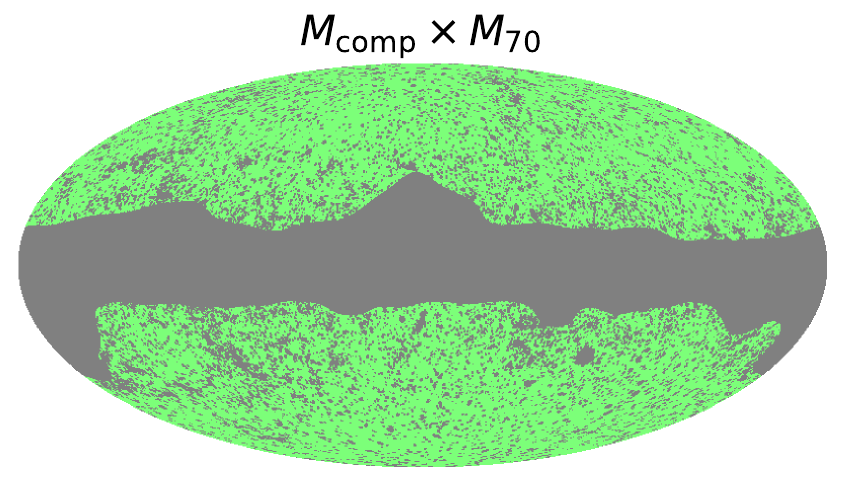}

\includegraphics[width=0.23\textwidth]{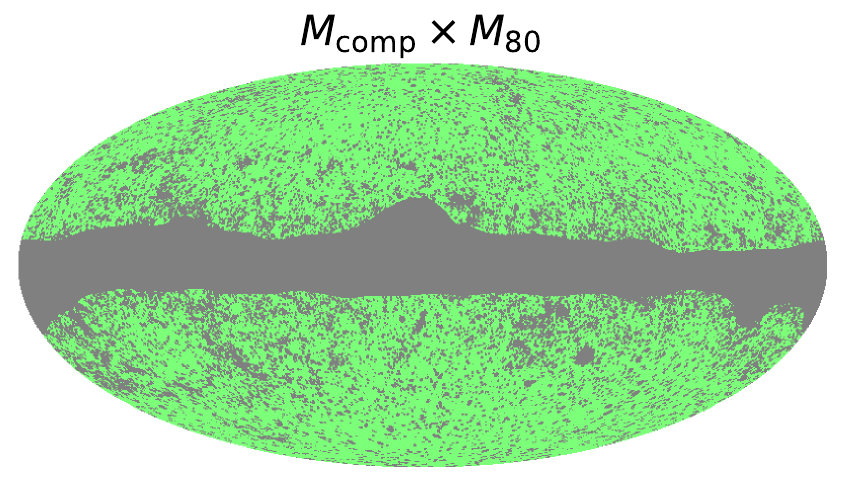}
\includegraphics[width=0.23\textwidth]{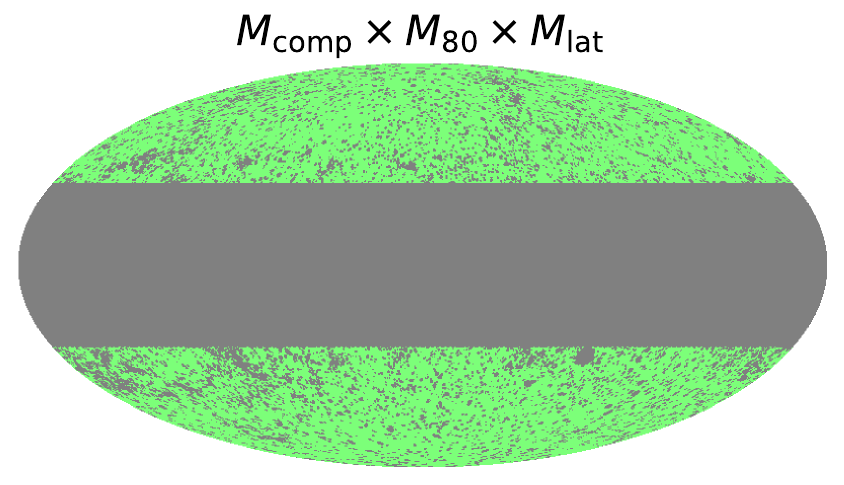}
\caption{ Different masking schemes employed in this paper, for preprocessing or cross-check in the statistics.  
\textit{Top left}: Point source mask at 857 GHz, from HFI\_Mask\_PointSrc\_2048\_R2.00.fits. 
\textit{Top right}: Compact source mask $M_\mathrm{comp}$, incorporating \textit{Planck} Catalogue of Compact Sources at each LFI and HFI channel. 
\textit{Middle left}: Compact source mask $M_\mathrm{comp}$ combined with Galactic plane mask $M_{60}$, where 60\% of the sky is available in $M_{60}$. 
\textit{Middle right}: Compact source mask $M_\mathrm{comp}$ combined with Galactic plane mask $M_{70}$, where 70\% of the sky is available in $M_{70}$. 
\textit{Bottom left}: Compact source mask $M_\mathrm{comp}$ combined with Galactic plane mask $M_{80}$, where 80\% of the sky is available in $M_{80}$. 
\textit{Bottom right}: Compact source mask $M_\mathrm{comp}$ combined with Galactic plane mask $M_{80}$, and regions with low Galactic latitude ($|b| < 30^\circ$) have been masked. 
}
\label{fig:masks}
\end{figure}

In order to generate a sky mask from the catalogs while preserving as much sky area as possible, the following strategy is adopted: at each frequency channel, for the top 10\% of the strongest compact sources, the mask aperture is set to four times the FWHM of the compact sources. 
For the next 10\% to 30\%, it is three times the FWHM, 
and for the remaining compact sources, the mask aperture is twice the FWHM.\footnote{For example, the compact source ``PCCS2E 143 G353.07+16.91'' in 143 GHz channel, located at Galactic longitude $353\overset{\circ}{.}072$, Galactic latitude $16\overset{\circ}{.}918$, has a flux density, as determined by the detection method, of 7639.504 MJy and a Gaussian fit effective FWHM of $26\overset{\prime}{.}942$. } 
Finally, the masks from all channels are combined to create the overall mask map, denoted $M_\mathrm{comp}$, as shown in the top right panel of Fig.~\ref{fig:masks}. 

\subsubsection{Galactic plane mask}
In the Galactic plane and regions close to it, the complexity of Galactic foreground components introduces significant uncertainties in the signal. 
To mitigate this, we use the Galactic plane masks provided by the \textit{Planck} team, which allow 60\%, 70\%, and 80\% of the sky to be observed (referred to as $M_{60}$, $M_{70}$, and $M_{80}$, respectively, and collectively as $M_\mathrm{plane}$), to exclude the regions near the Galactic plane. 
The middle left panel of Fig.~\ref{fig:masks} displays the combined mask with $M_\mathrm{comp}$ and $M_{60}$. 
Similarly, the middle right and bottom left panels of Fig.~\ref{fig:masks} are for $M_{70}$ and $M_{80}$. 

\subsubsection{Low Galactic latitude mask}
Due to the relatively clean foreground in the mid- to high-Galactic latitude regions, these areas are of specific interest for the CMB community. 
To assess the robustness of our conclusions in these regions, we will examine the impact of excluding the low latitude areas (with Galactic latitude $|b|<30^\circ$). 
For such a case, we superimpose a low-Galactic mask, as shown in the bottom right panel of Fig.~\ref{fig:masks}, where the mask against the low latitude regions is denoted as $M_\mathrm{lat}$. 

In summary, we employ 
\begin{equation}
\label{equ:mask}
M_\mathrm{tot} = M_\mathrm{comp} \times M_\mathrm{plane} \left( \times M_\mathrm{lat} \right), 
\end{equation}
in the statistics, where $M_\mathrm{tot}$ is the total mask employed, $M_\mathrm{comp}$ is the compact source mask, $M_\mathrm{plane}$ is the Galactic plane mask. 
When mid-high-latitude regions are checked, the low-Galactic mask $M_\mathrm{lat}$ is superimposed. 

\subsection{Statistics in mosaic disks}
\label{subsec:mosaic ratio}
The linear ratios $R$, $R'$ and the cross-correlation $C'$ between neighboring dust data maps are computed in mosaic disks for three pairs of neighboring channels: 100-143, 217-353, and 545-857 GHz pairs. 
The center of mosaic disks are placed on the sky directions defined by the HEALPix resolution $N_\mathrm{side}=64$, and their angular radius is $6^\circ$; 
thus, the centers of two adjacent mosaic disks are separated by approximately $1.1^\circ$.
Meanwhile, to reduce the effects of CIB anisotropies and noise, we smooth the dust model and data maps with a smoothing angle of $2^\circ$ FWHM.

Note that the computations of $R$, $R'$, and $C'$ are all free from constant offsets, which is a significant advantage. 
Moreover, although the absolute values of $R$ and $R'$ depend on the units of two channels in one pair, the ratio $R'/R$ is free from units, because the units of $R$ and $R'$ always cancel each other, and $R'/R$  is always expected to be close to 1.
As mentioned before, $C'$ serves as an indicator: a mosaic disk (sky region) is considered reliable for thermal dust emission only if it exhibits not less than $0.95$ cross-correlation between two bands, 
which ensures that the noise and systematic effects are subdominant. 
The results for $R$, $R'$, and $C'$ are shown in Fig.~\ref{fig:dust_ratio}. 
For real data, if the unmasked area in a mosaic disk is larger than 30\% of the mosaic disk area, then this mosaic disk is ``valid.'' 
Furthermore, if $C'$ of this mosaic disk is not less than 0.95, then this mosaic disk is ``reliable.'' 
For simulations, if $C'$ of this simulation point is not larger than 1, then it is ``valid''; 
if $0.95 \leq C' \leq 1$ of this simulation point, then it is ``reliable.'' 

In summary, the data preprocessing and processing steps described in this section are as follows: 
\begin{enumerate}
\item Resolution unification: The map resolution is unified to $N_{\mathrm{side}}=2048$ and the HFI maps' beam width is unified to a FWHM of $9\overset{\prime}{.}66$. 
During this process, \textit{Planck} point source masks are used for masking, followed by in-painting to alleviate the Gibbs phenomenon. 

\item Subtraction of none-dust components: We subtract CMB anisotropies and other none-dust foreground components to obtain the dust data maps (noted with prime, like $x'$, $y'$).

\item Model prediction: The thermal dust emission in the \textit{Planck} HFI frequency bands is computed using a single-component dust emission model. 
Color correction is included in steps 2 and 3, and the output maps' units are consistent with those of the corresponding \textit{Planck} HFI maps.

\item Mosaic disk analysis: Small mosaic disks are defined on the sky maps, within which $R$, $R'$, and $C'$ are calculated. 
\end{enumerate}

\section{Results}
\label{sec:results}

\subsection{The model-to-data departure}
\label{sub: data to model departure}

\begin{figure*}[!htb]
\centering
\includegraphics[width=0.32\textwidth]{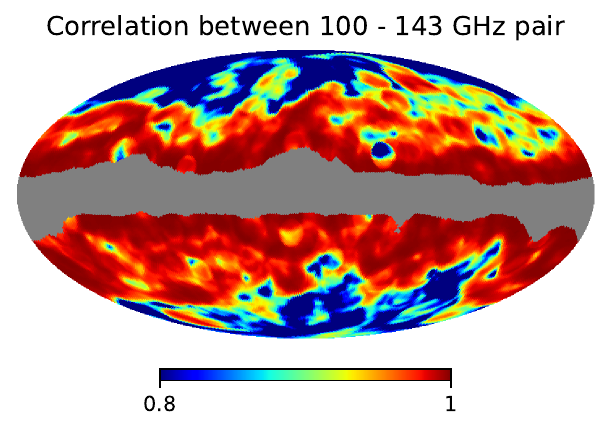}
\includegraphics[width=0.32\textwidth]{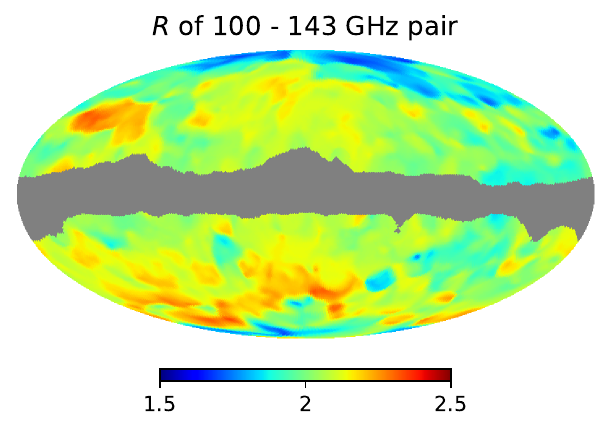}
\includegraphics[width=0.32\textwidth]{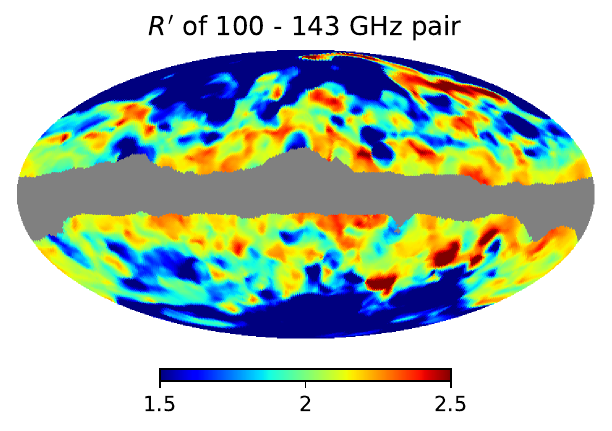}

\includegraphics[width=0.32\textwidth]{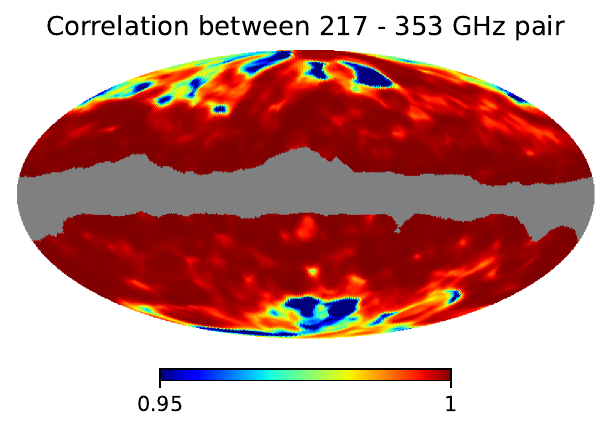}
\includegraphics[width=0.32\textwidth]{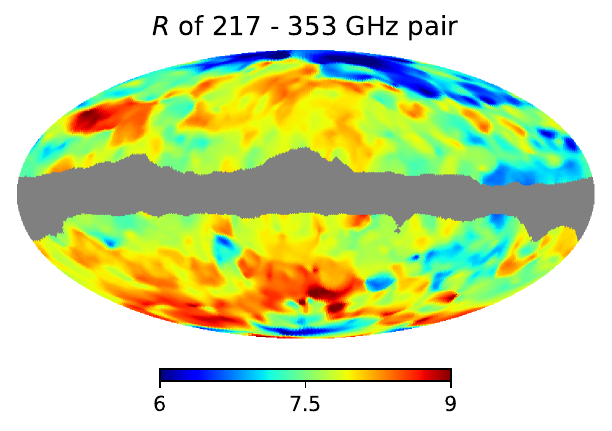}
\includegraphics[width=0.32\textwidth]{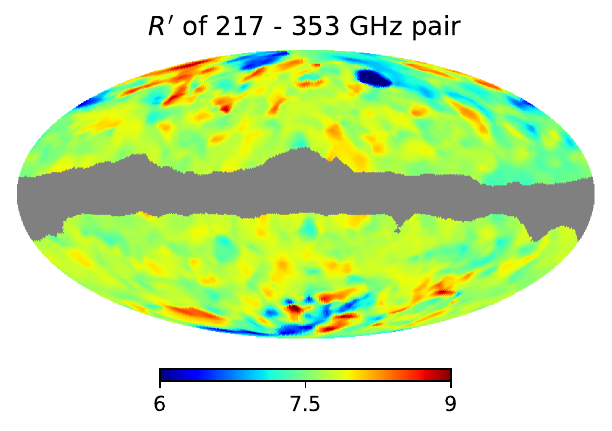}

\includegraphics[width=0.32\textwidth]{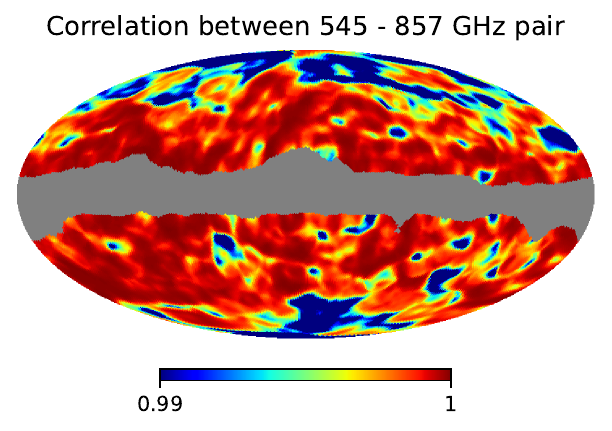}
\includegraphics[width=0.32\textwidth]{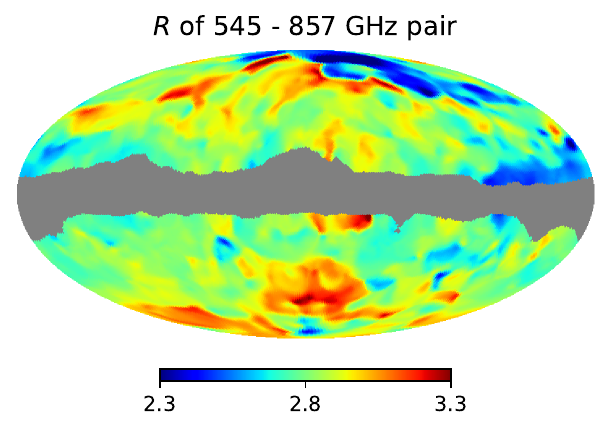}
\includegraphics[width=0.32\textwidth]{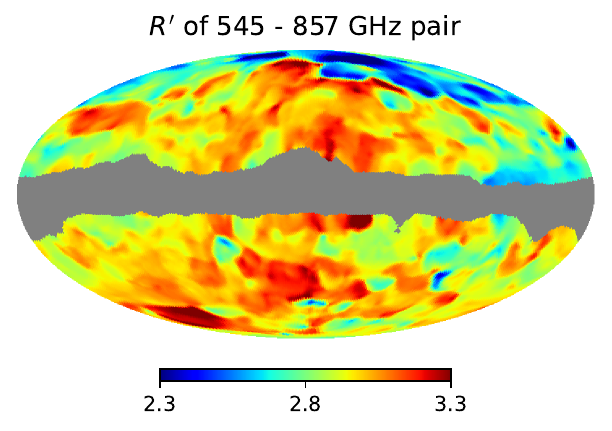}
\caption{Maps for $C'$ (\textit{left}), $R$ (\textit{middle}), and $R'$ (\textit{right}) between two neighboring bands. 
\textit{From top to bottom}: the 100-143, 217-353, and 545-857 GHz pairs. 
The input maps are smoothed by $2^\circ$ and the angular radius of the mosaic disks is $6^\circ$. 
$M_\mathrm{tot} = M_\mathrm{comp} \times M_{80}$. 
}
\label{fig:dust_ratio}
\end{figure*}

As mentioned in Section~\ref{sec:method}, if the single-component dust emission model is a good approximation at 100-857 GHz, then $R$ and $R'$ should be close to each other for the regions with high mosaic correlation $C'$. 
Fig.~\ref{fig:dust_ratio} shows the linear ratio of the dust model maps ($R$) and dust data maps ($R'$), and mosaic corrections between adjacent dust data maps ($C'$). 
As mentioned above, the linear ratios are the slopes of the linear regression lines between mosaic samples at adjacent HFI channels. 
The linear ratio can be approximately regarded as the intensity ratio of dust maps between adjacent channels. 
This is visually represented in Figs.~\ref{fig:smoothed thermal dust component} and \ref{fig:dust_ratio}: Since the maximum values of the color bars in Fig.~\ref{fig:smoothed thermal dust component} correspond to the 70-th percentile of the maps, they approximately represent the dust emission intensity at the respective frequency channels. 
Accordingly, typical values of $I_{143}/I_{100}$, $I_{353}/I_{217}$, and $I_{857}/I_{545}$ are 2.04, 7.33, and 2.77,\footnote{It should be noted that $R$ and $R'$ depend on unit. 
For 100, 143, 217, and 353 GHz, we employ $\mu\mathrm{K_{CMB}}$ and for 545, 857 GHz, we employ $\mathrm{MJy\,sr^{-1}}$, to be consistent with \textit{Planck} HFI maps. } respectively, which are broadly consistent with the median values shown in Fig.~\ref{fig:dust_ratio}. 
As shown in Fig.~\ref{fig:dust_ratio}, there is no significant departure between $R$ and $R'$ for the 545 - 857 GHz pair; however, for the 217 - 353 GHz pair, the departure is visually higher, and the 100 - 143 GHz pair displays a significant departure.

\begin{figure*}[!htb]
\centering
\includegraphics[width=0.32\textwidth]{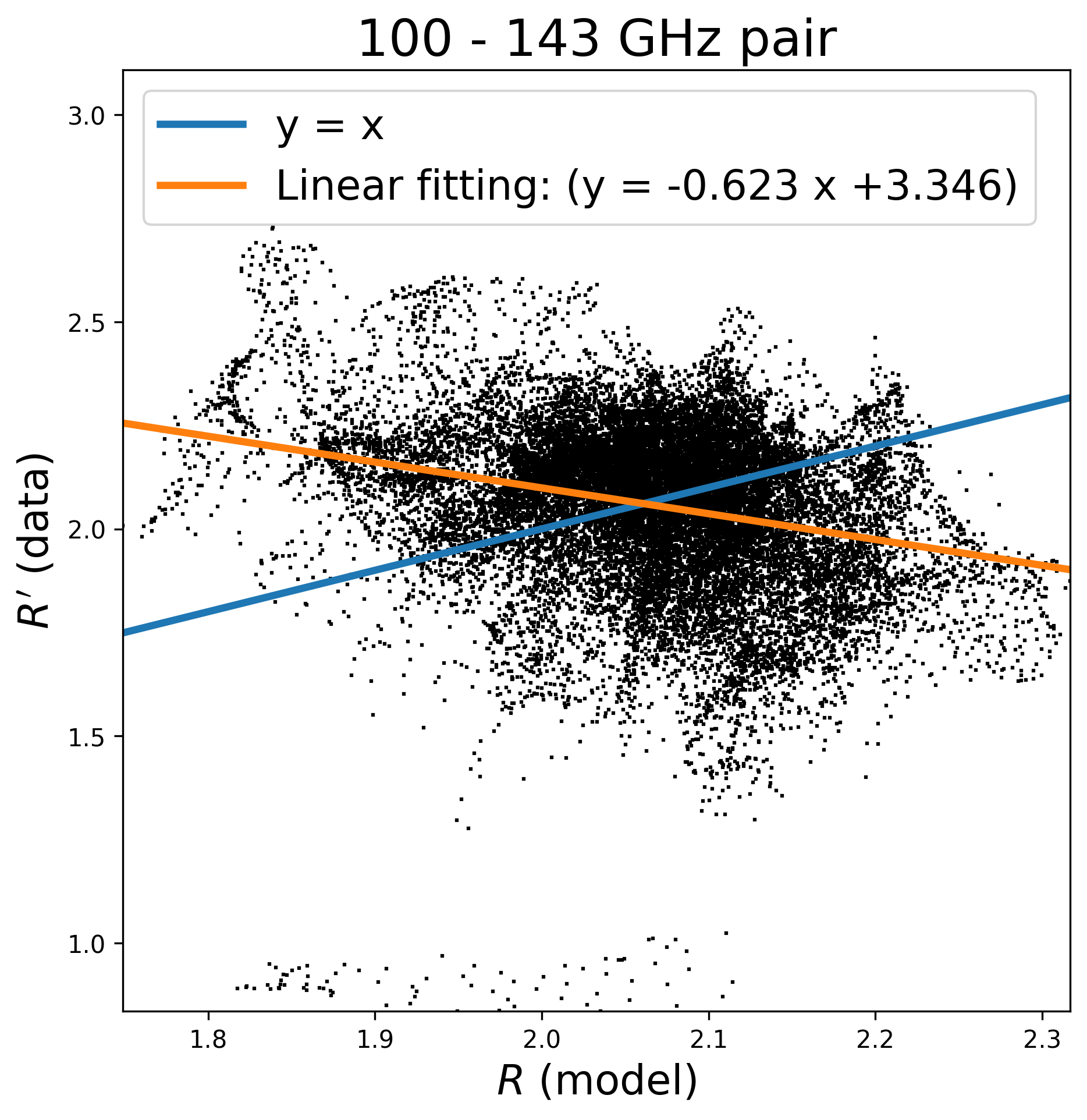}
\includegraphics[width=0.32\textwidth]{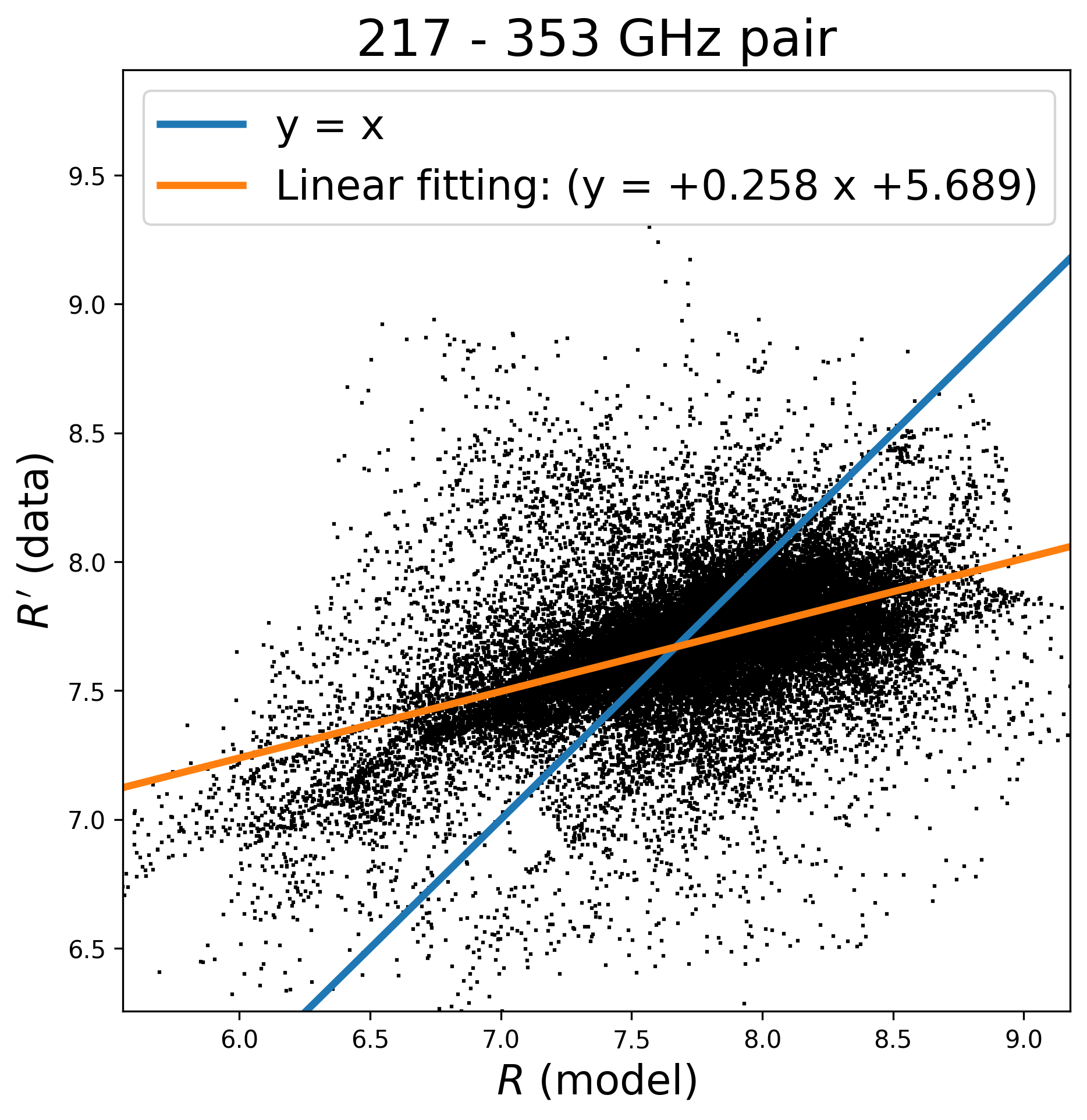}
\includegraphics[width=0.32\textwidth]{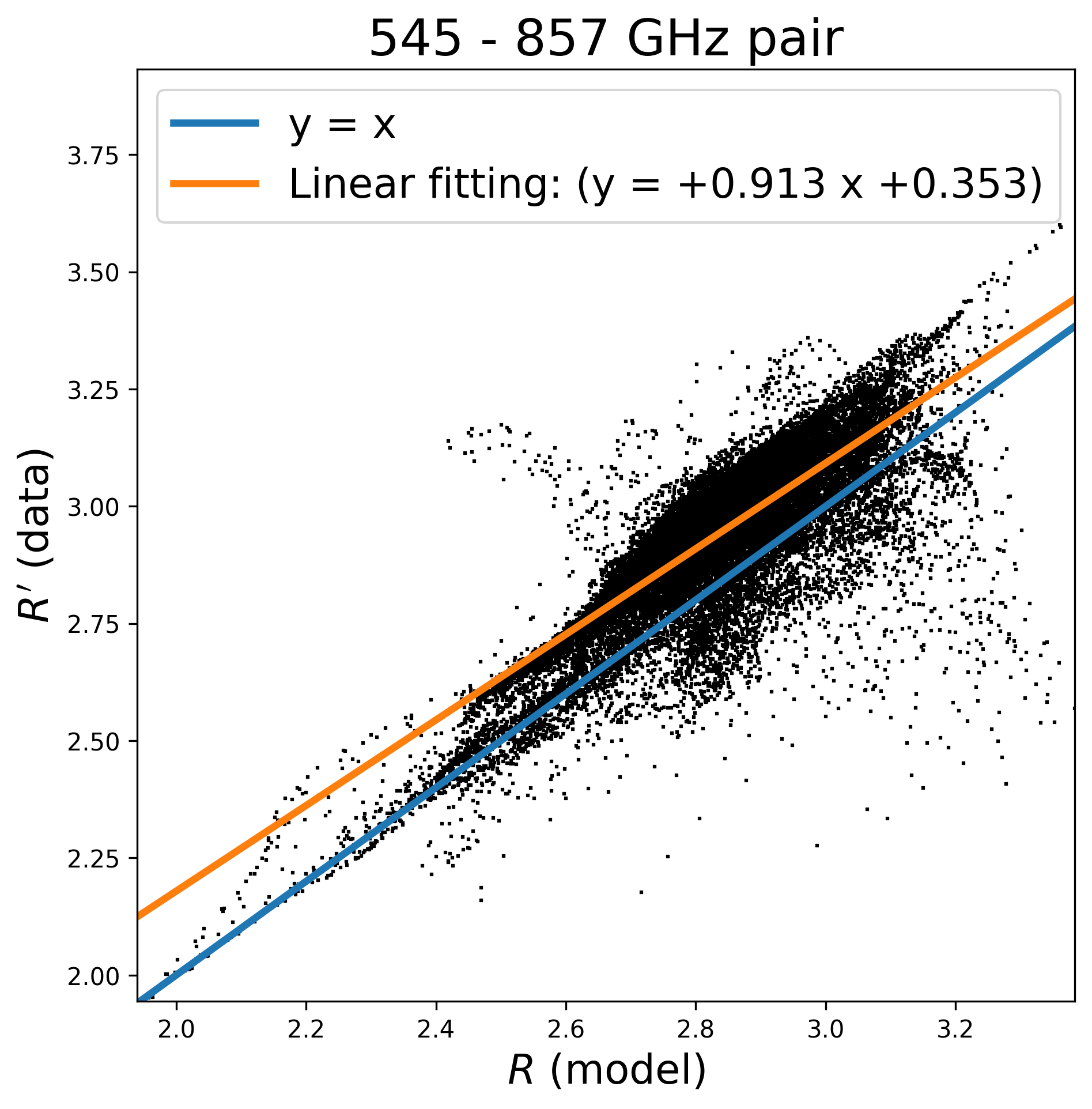}

\includegraphics[width=0.32\textwidth]{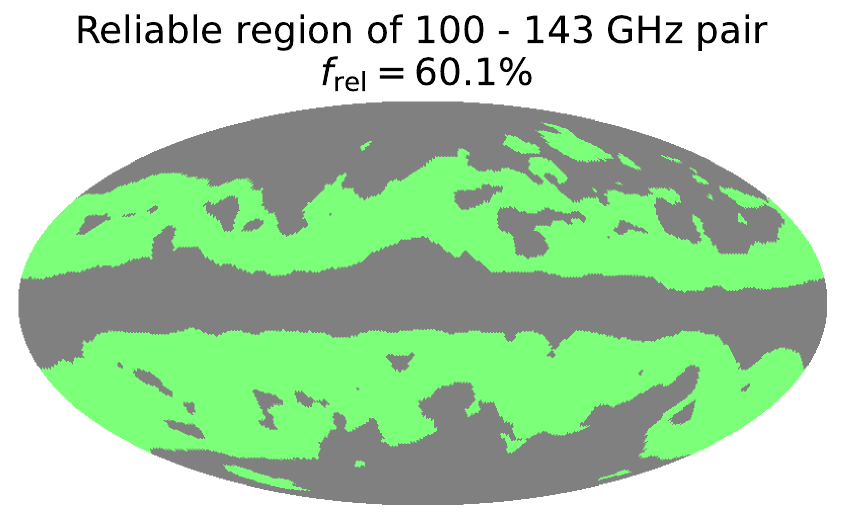}
\includegraphics[width=0.32\textwidth]{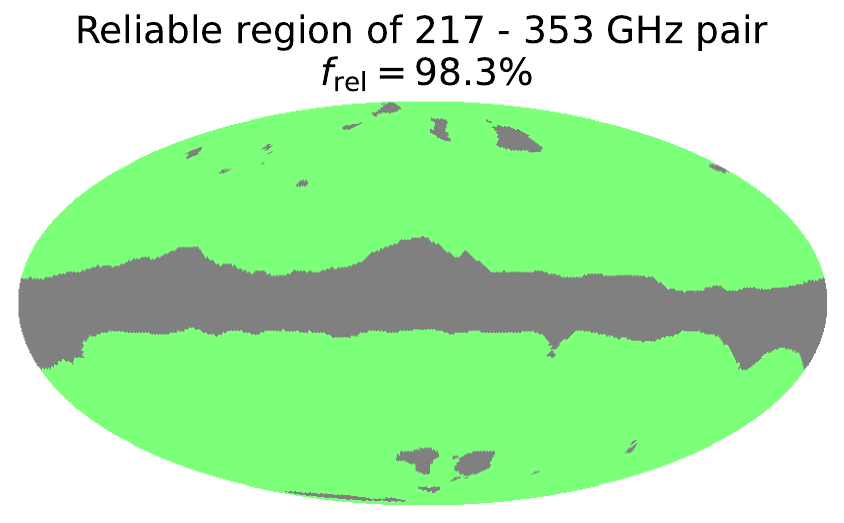}
\includegraphics[width=0.32\textwidth]{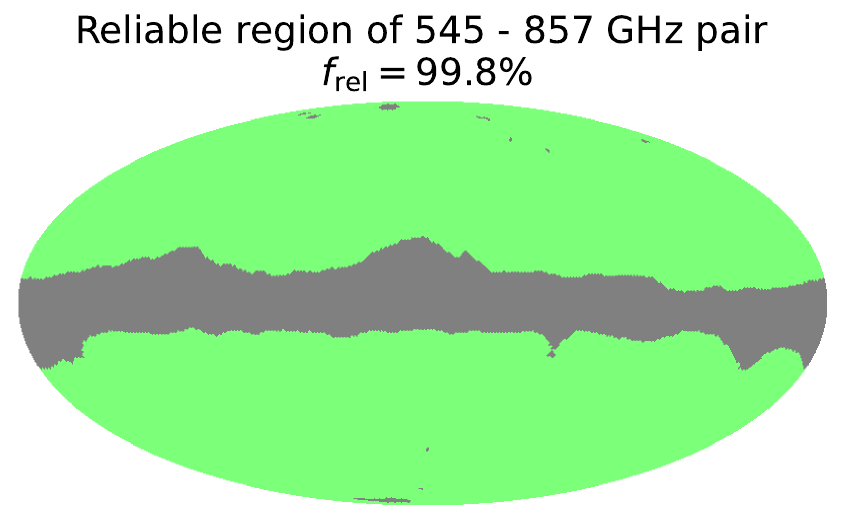}
\caption{Scatter plots of $R'$ versus $R$ in Fig.~\ref{fig:dust_ratio} (\textit{upper}) for the reliable regions with $C' \geq 0.95$ in green (\textit{lower}). 
\textit{From left to right}: the 100-143, 217-353, and 545-857 GHz pairs. 
The blue lines are the $R = R'$ expectation, and the orange lines are the least squares regression lines of the scatter points. 
$M_\mathrm{tot} = M_\mathrm{comp} \times M_{80}$. 
The intercept of the orange line for the 545-857 GHz pair is slightly different from the blue line, which is likely due to the color correction coefficients' variation along with the uncertainty of the dust emission's spectral shape. 
}
\label{fig:dust_ratio_compare}
\end{figure*}

\begin{figure*}[!htb]
\centering
\includegraphics[width=0.32\textwidth]{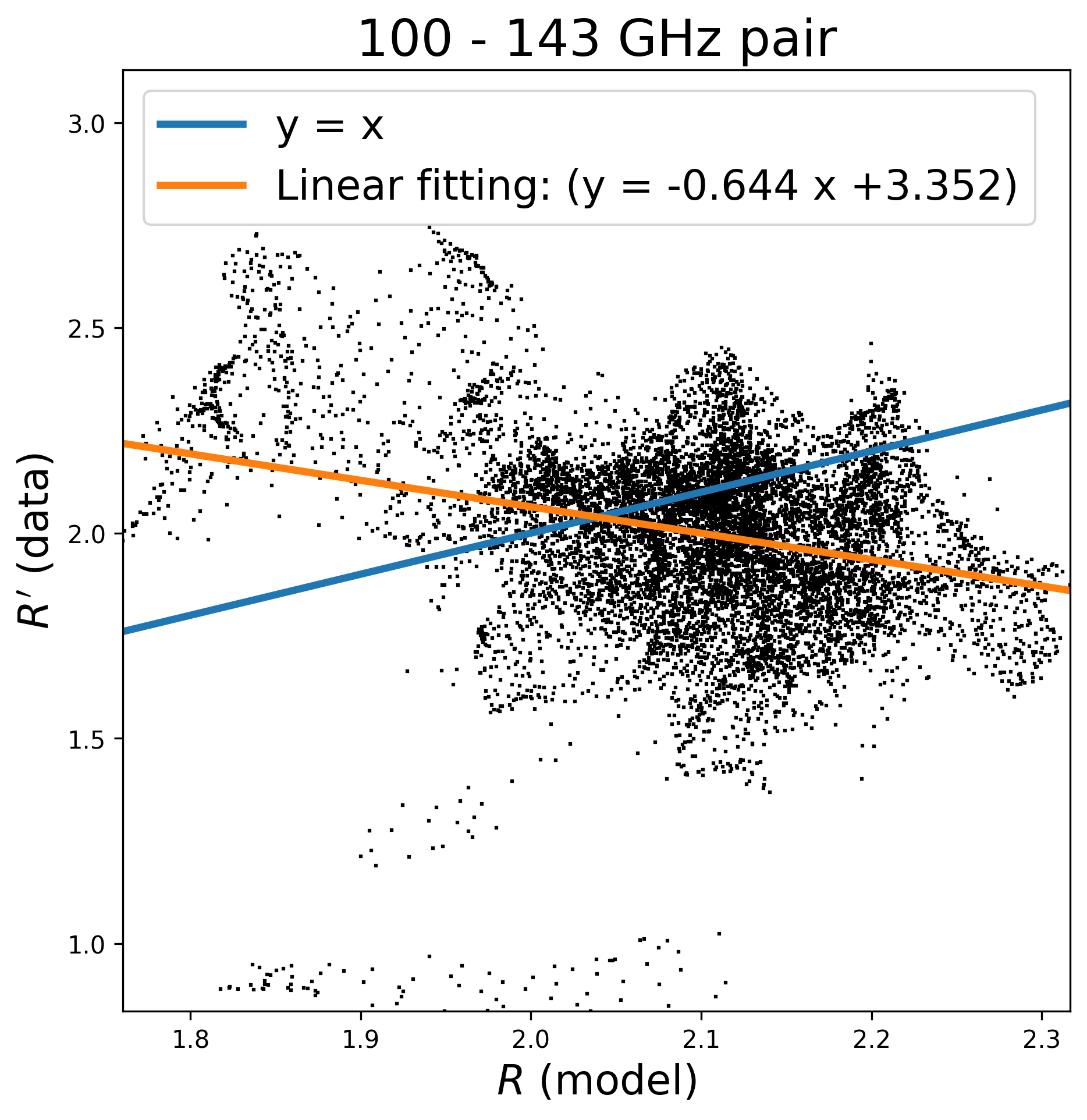}
\includegraphics[width=0.32\textwidth]{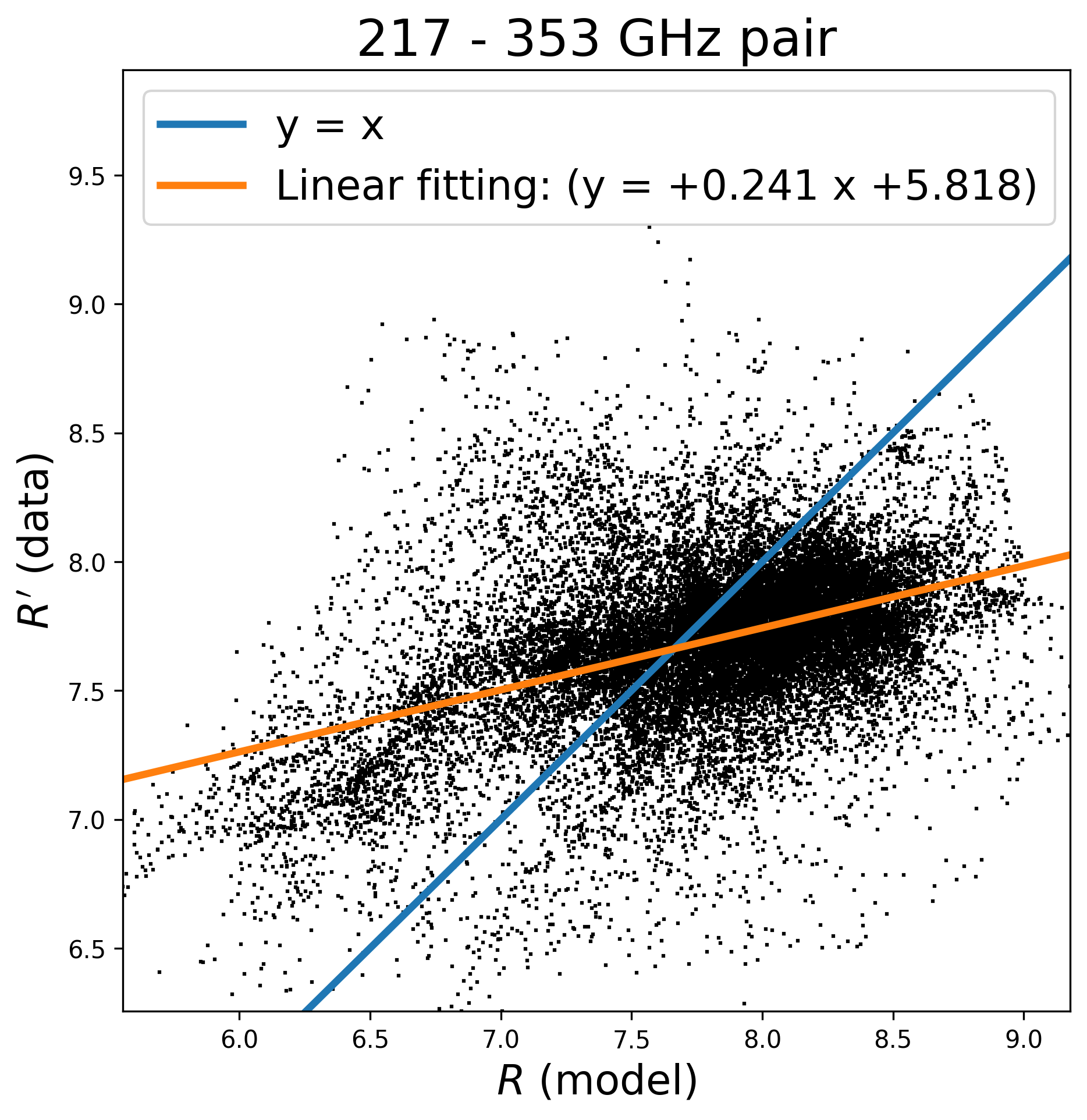}
\includegraphics[width=0.32\textwidth]{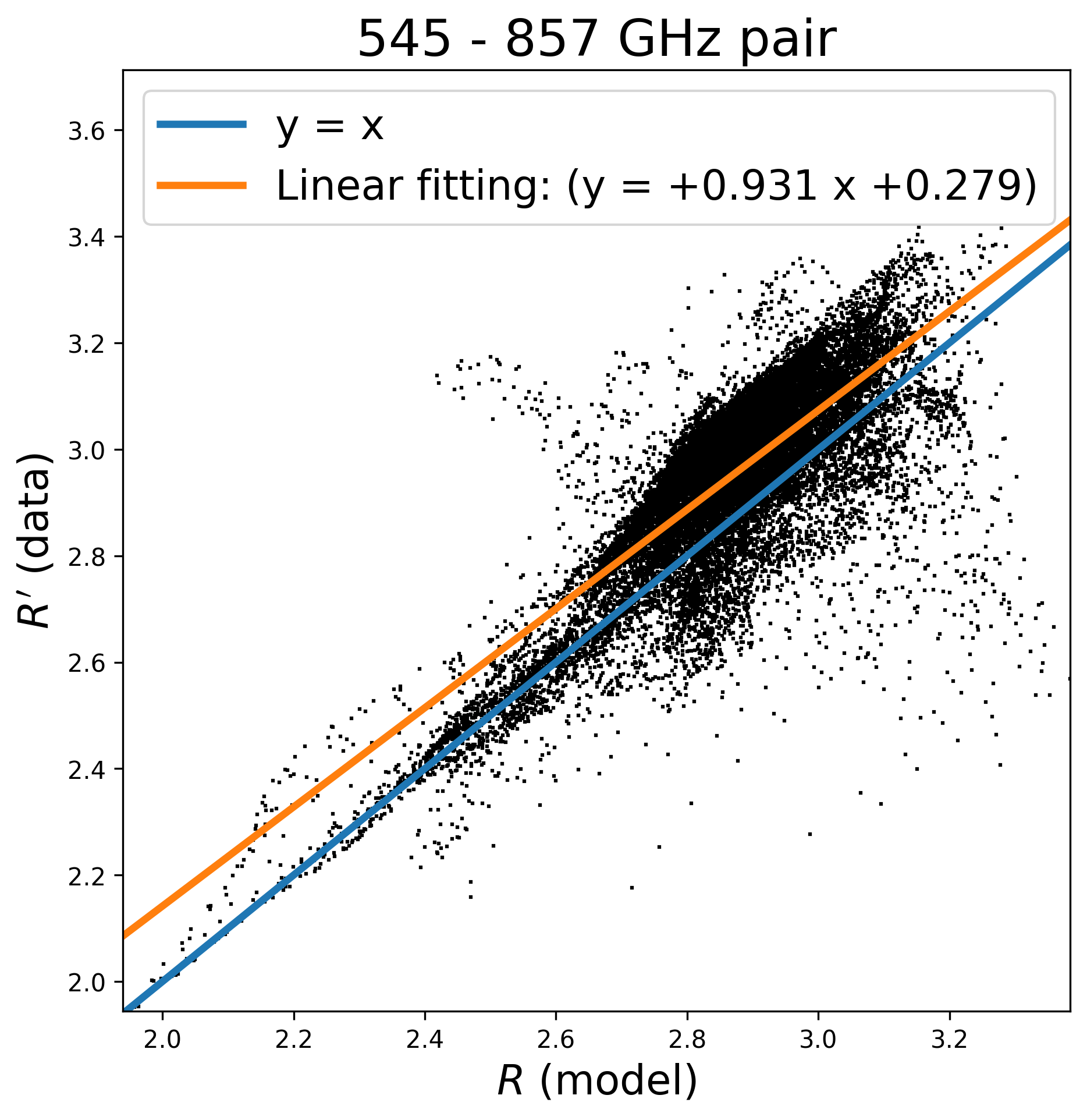}

\includegraphics[width=0.32\textwidth]{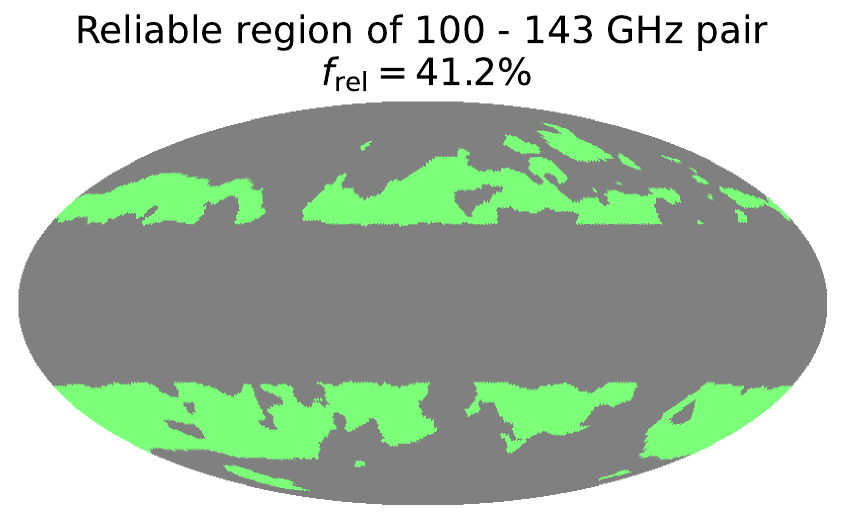}
\includegraphics[width=0.32\textwidth]{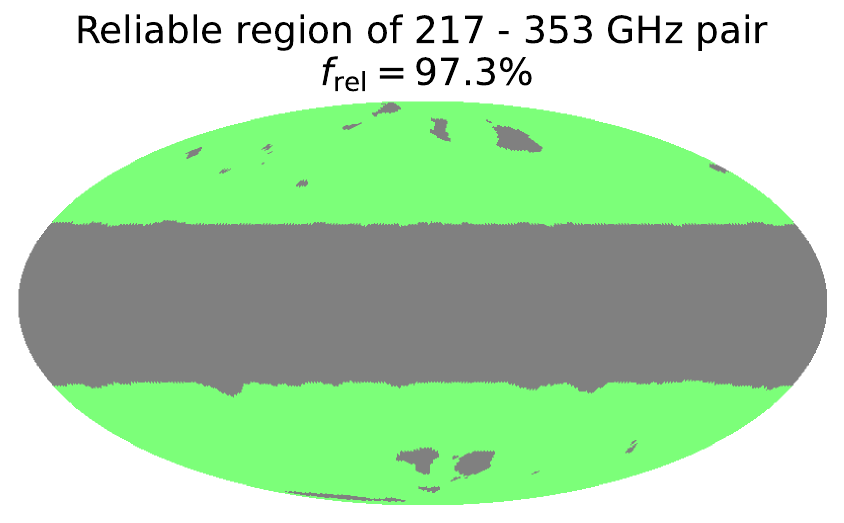}
\includegraphics[width=0.32\textwidth]{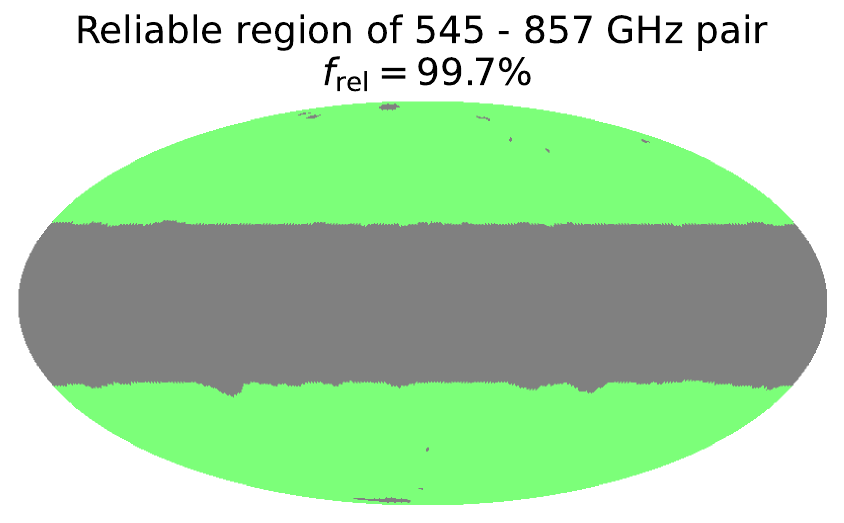}
\caption{Similar to Fig.~\ref{fig:dust_ratio_compare} but excludes the low Galactic latitude regions with $|b|<30^\circ$ 
($M_\mathrm{tot} = M_\mathrm{comp} \times M_{80} \times M_\mathrm{lat}$). }
\label{fig:dust_ratio_compare2}
\end{figure*}

The $R$-to-$R'$ departure is more evident in the scatter plots in Fig.~\ref{fig:dust_ratio_compare}, with $R$ and $R'$ being the horizontal and vertical axes, respectively.
Note that the mosaic disks with $C' < 0.95$ are already excluded in Fig.~\ref{fig:dust_ratio_compare}; therefore, the dots in the scatter plots are expected to align with the $R=R'$ line, which is significantly violated by the pairs 100-143 and 217-353 GHz. A double-check of the Galactic signal (expected to be more complicated) is done by excluding the $|b|<30\degree$ regions (Fig.~\ref{fig:dust_ratio_compare2}), which brings no significant changes. Therefore, the low Galactic latitude signal is not a major cause of the departure.

\subsection{Simulation of the model-to-data departure}
\label{sub: sim of data to model departure}

\begin{figure*}[htb]
\centering
\includegraphics[width=0.32\textwidth]{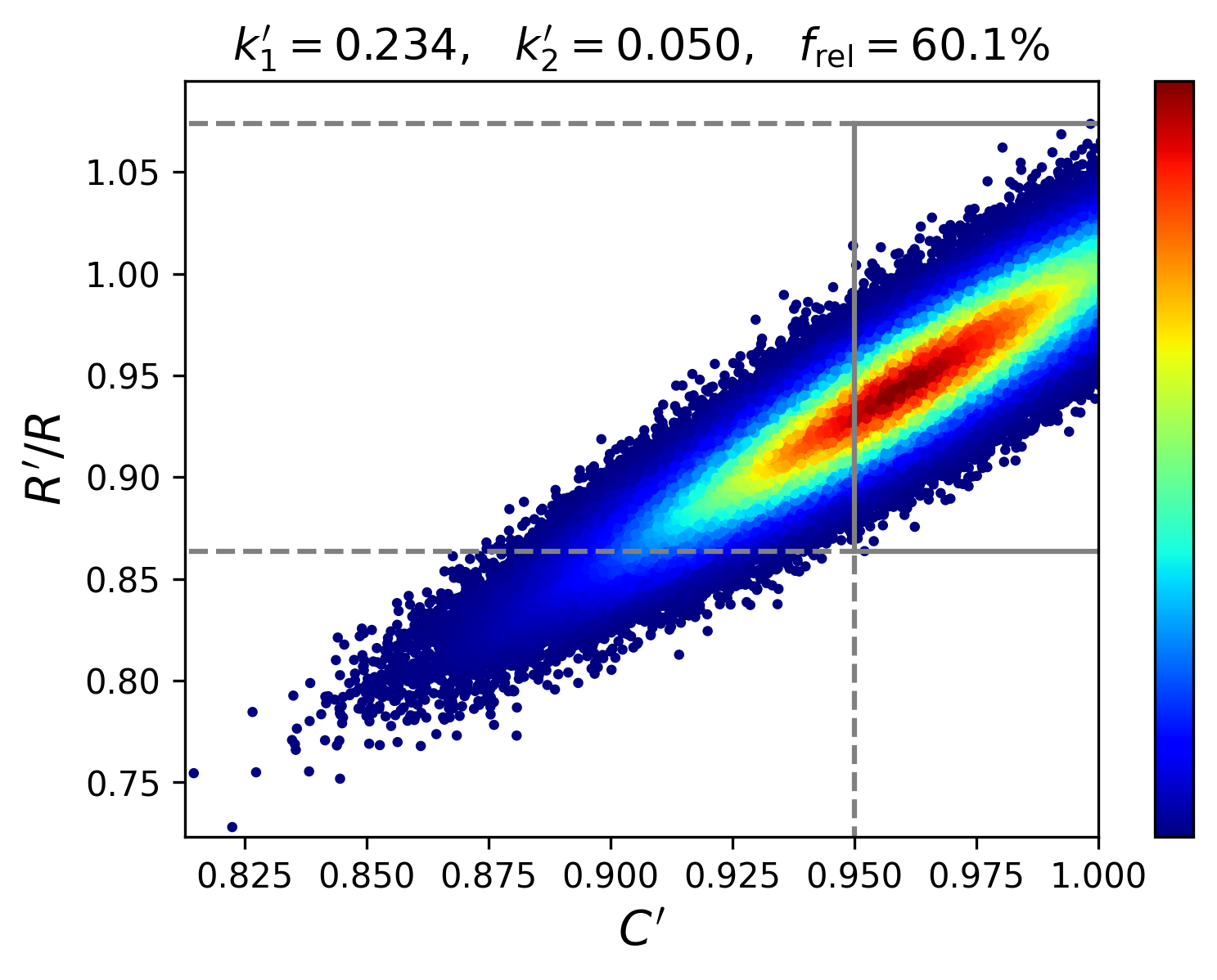}
\includegraphics[width=0.32\textwidth]{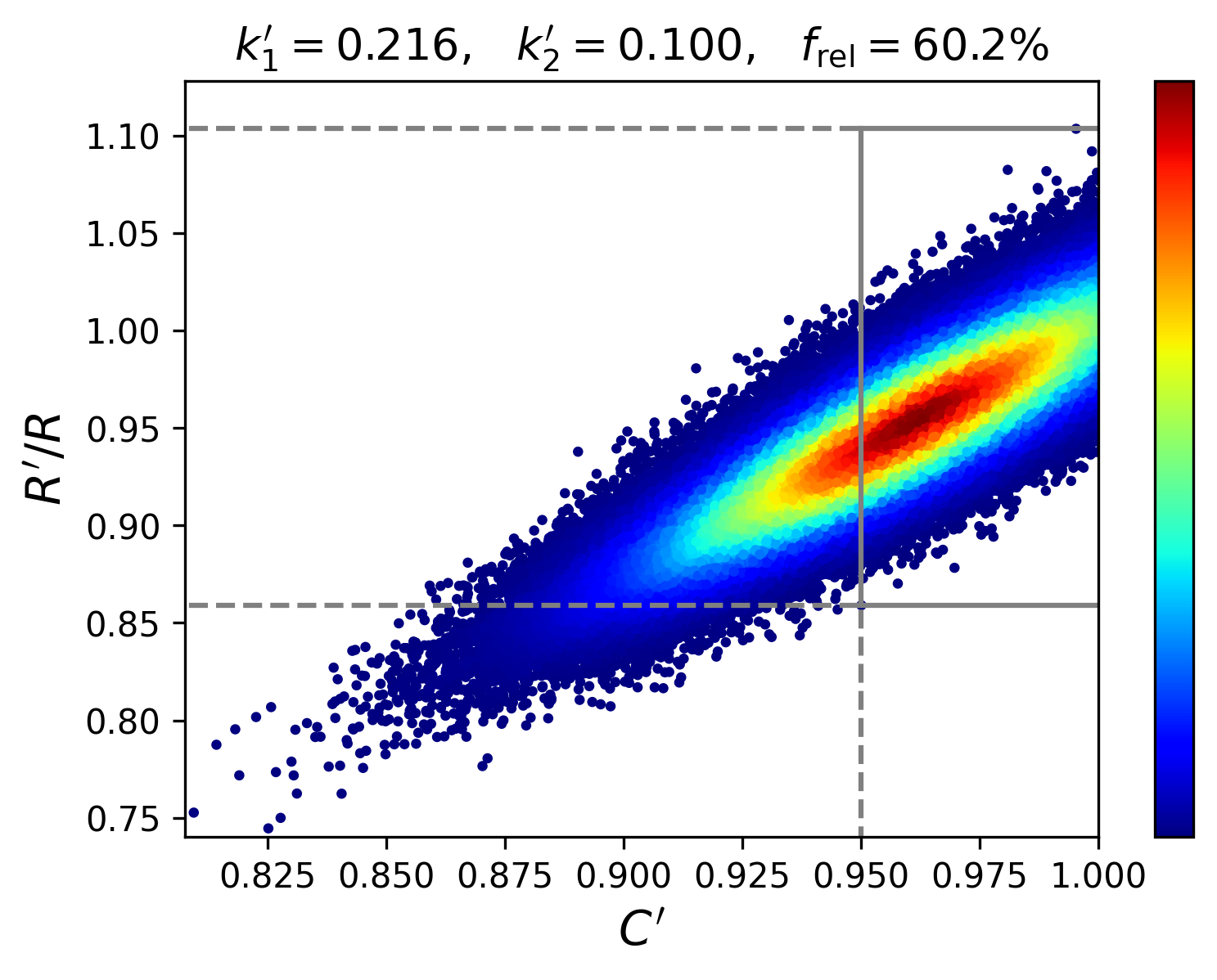}
\includegraphics[width=0.32\textwidth]{cc_rr_143_100_freedom_100_low_galac_mask_no_simulation_3.png}
\caption{Contours of simulated scatter points with different $k_1'$ and $k_2'$. 
The horizontal axis represents $C'$, and the vertical axis represents $R'/R$. 
The values of $k_1'$ and $k_2'$ are chosen such that the simulated $f_\mathrm{rel}$ is approximately 60.1\%, matching the values shown in the bottom left panel of Fig.~\ref{fig:dust_ratio_compare}. 
}
\label{fig:simulation R'-R with low galactic}
\end{figure*}

\begin{figure*}[htb]
\centering
\includegraphics[width=0.32\textwidth]{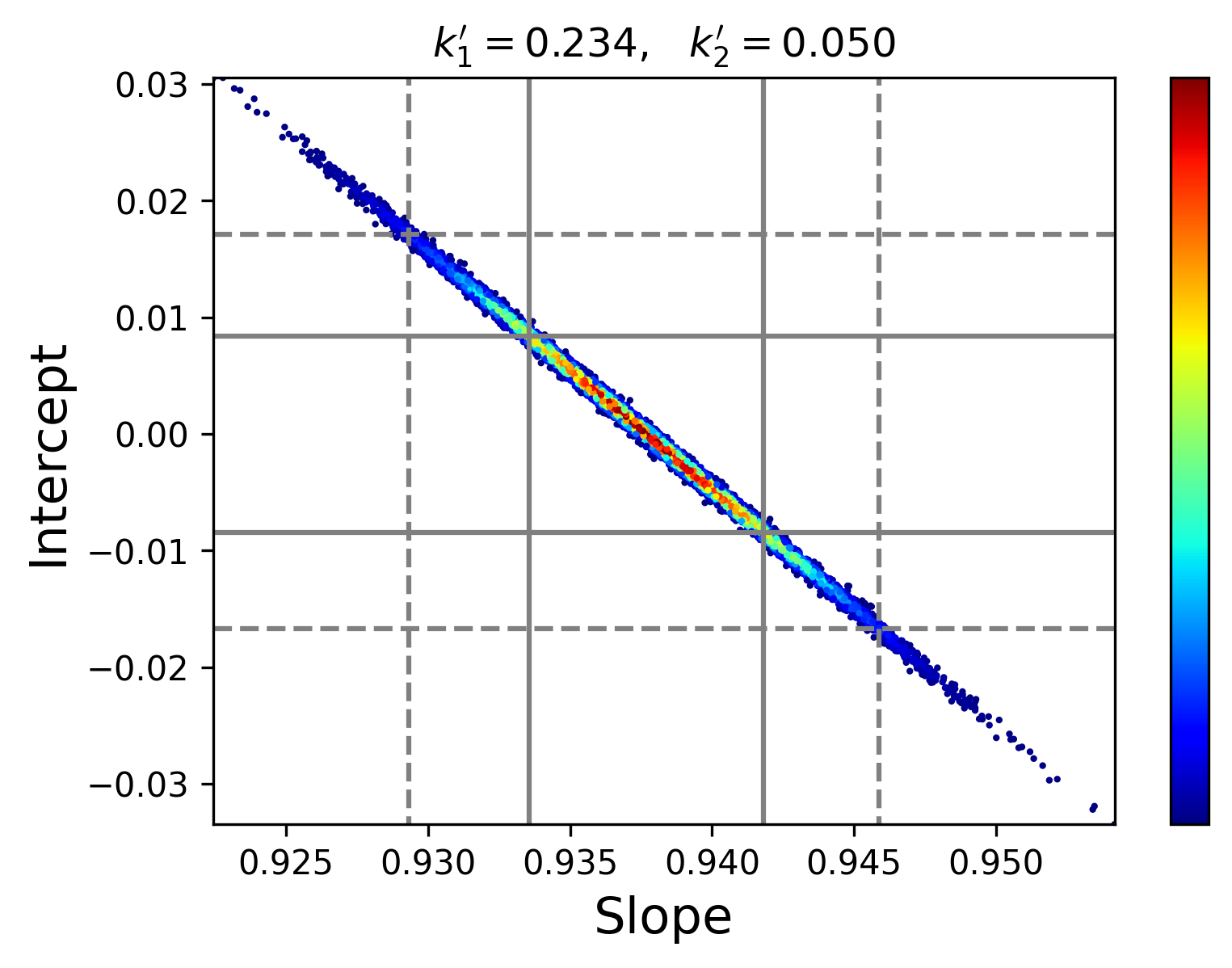}
\includegraphics[width=0.32\textwidth]{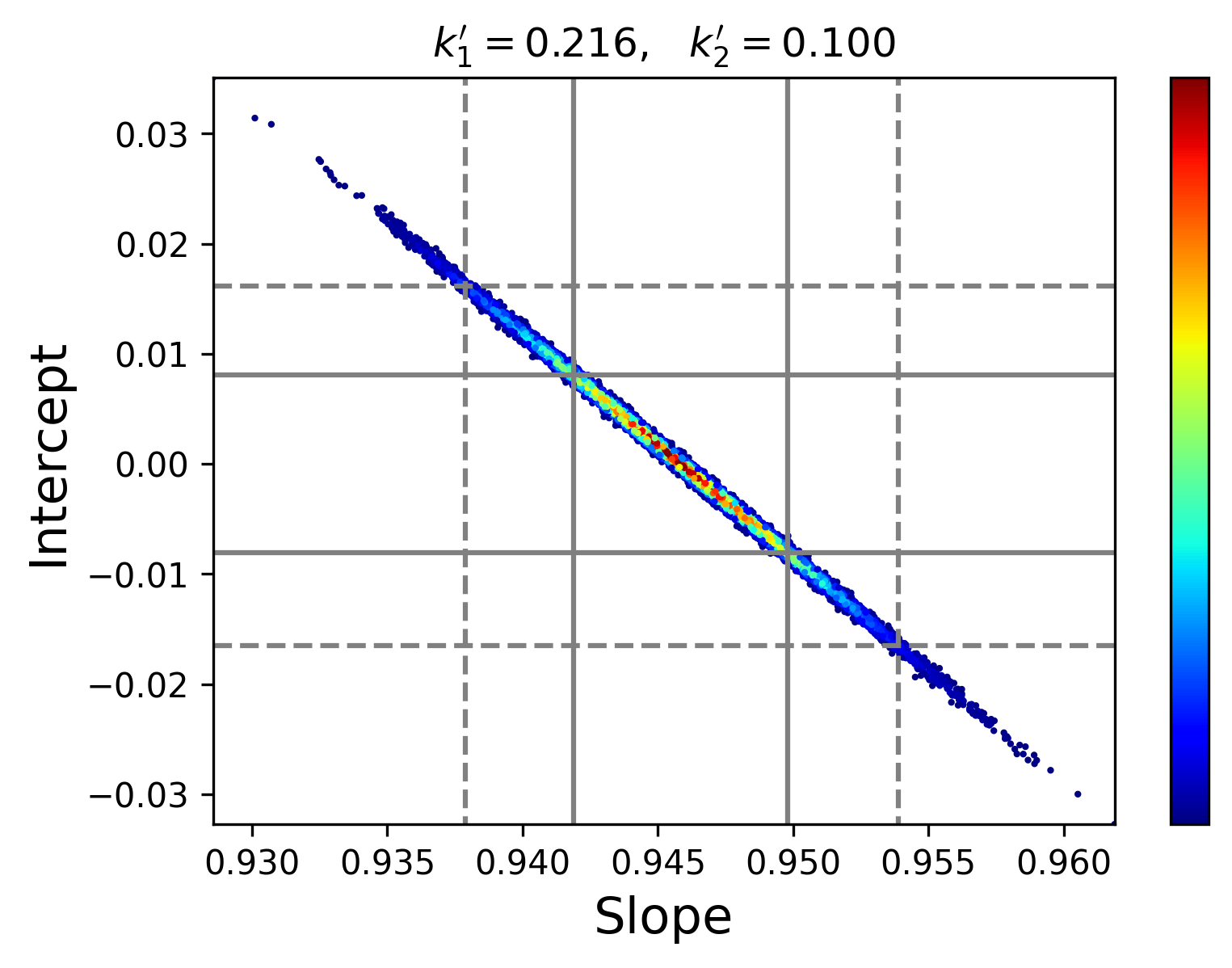}
\includegraphics[width=0.32\textwidth]{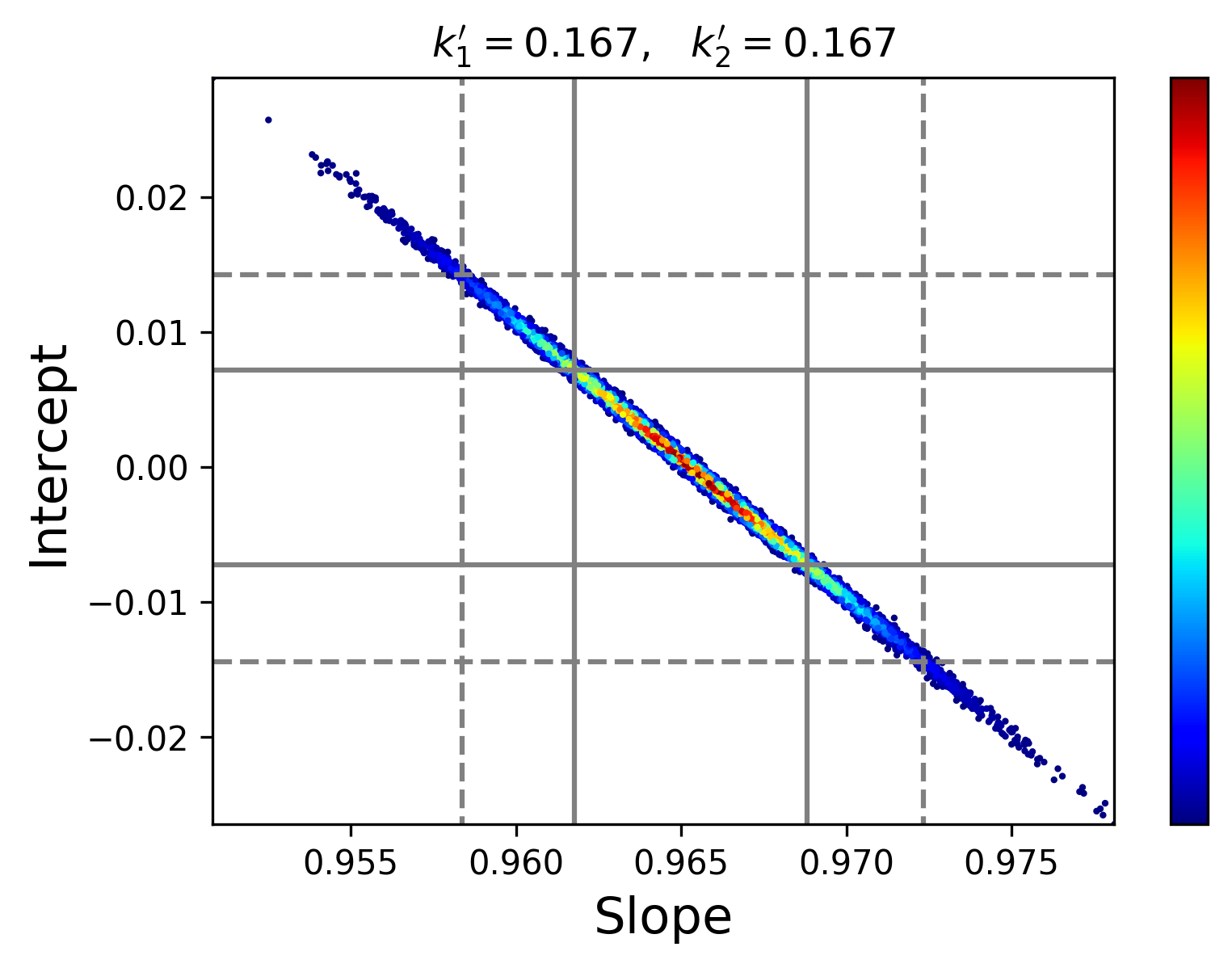}
\caption{Contours of simulated $R'$-$R$ linear regression coefficients with different $k_1'$ and $k_2'$ determined in Fig.~\ref{fig:simulation R'-R with low galactic}. 
The horizontal axis represents slope, and the vertical axis represents intercept. 
Note that each panel in Fig.~\ref{fig:simulation R'-R with low galactic} corresponds to only one point in this figure. 
Simulations for linear regression coefficients are realized 10000 times, with degrees of freedom set to 100. 
The solid gray lines and dashed gray lines indicate the upper and lower bounds corresponding to 1 $\sigma$ and 2 $\sigma$, respectively. }
\label{fig:simulation slope intercept with low galactic}
\end{figure*}

\begin{figure*}[htb]
\centering
\includegraphics[width=0.32\textwidth]{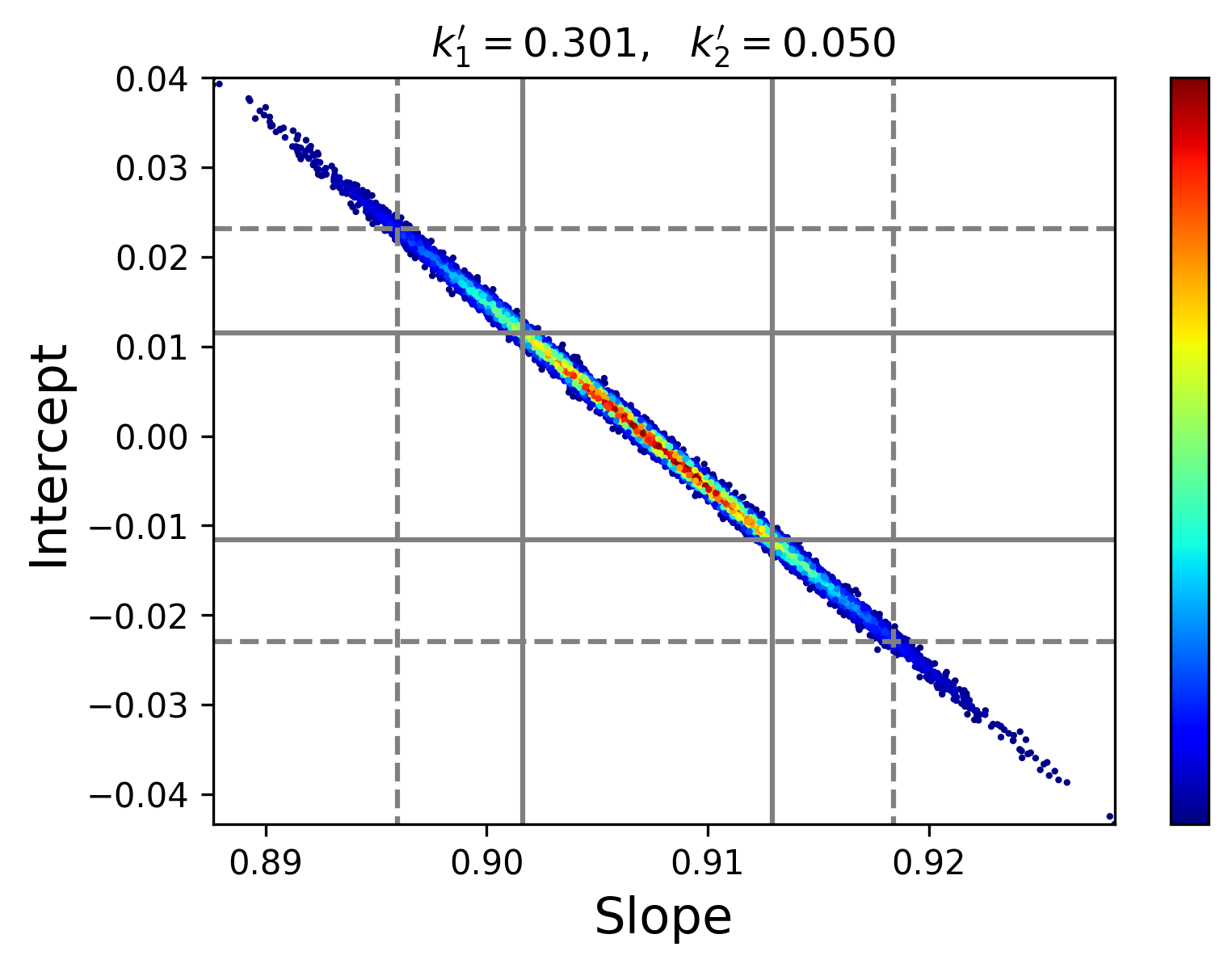}
\includegraphics[width=0.32\textwidth]{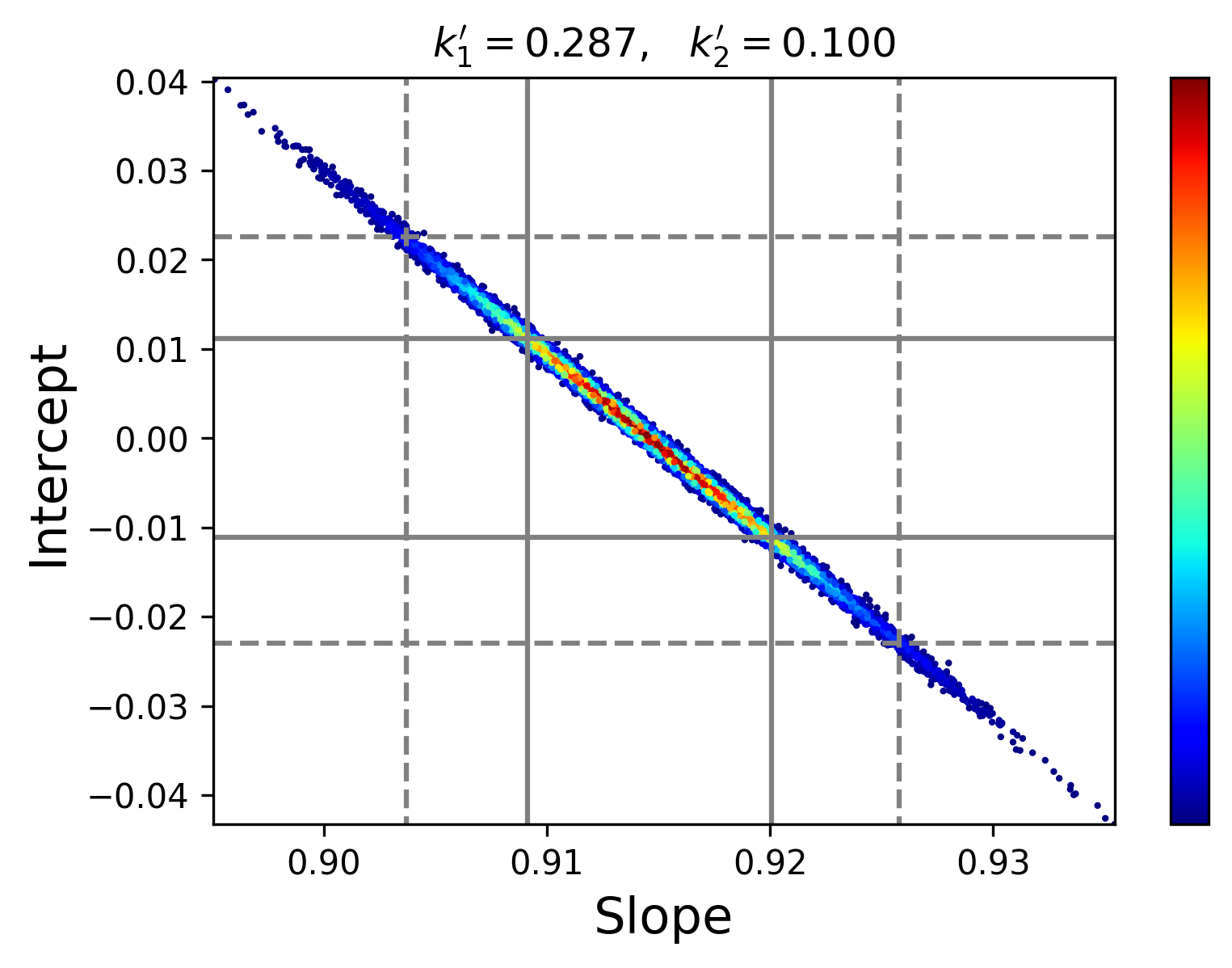}
\includegraphics[width=0.32\textwidth]{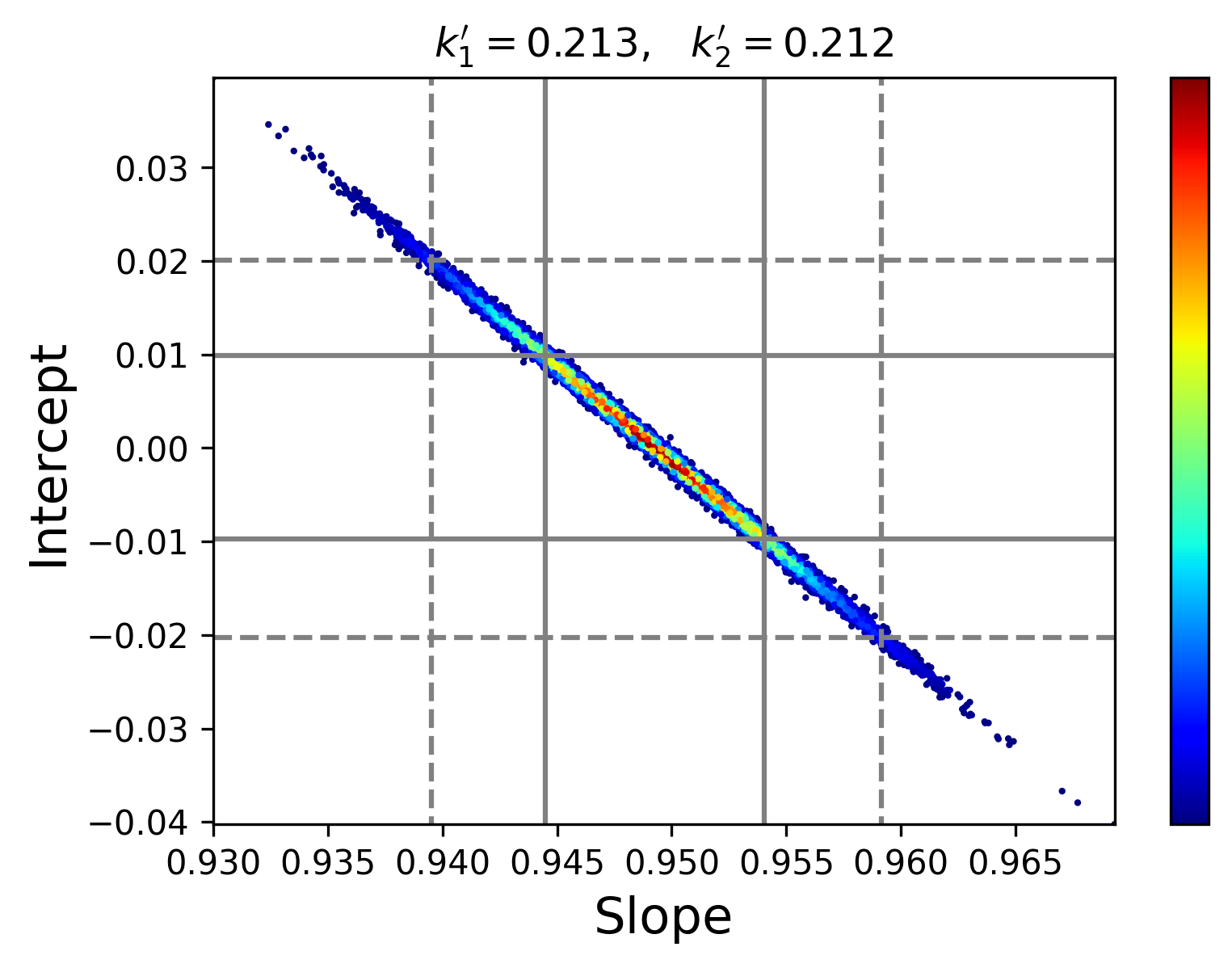}
\caption{Similar to Fig.~\ref{fig:simulation slope intercept with low galactic}, but the values of $k_1'$ and $k_2'$ are chosen such that the simulated $f_\mathrm{rel}$ is approximately 41.2\%, matching the values shown in the bottom left panel of Fig.~\ref{fig:dust_ratio_compare2}. }
\label{fig:simulation slope intercept no low galactic}
\end{figure*}

As mentioned above, if the single-component model is a good approximation, then the expectation is $R \approx R'$ when $C' \geq 0.95$. 
Fig.~\ref{fig:dust_ratio_compare} already departs significantly from this expectation, which cannot be attributed to the low Galactic latitude signals. 
We also confirm that this departure does not change much with conditions such as the smoothing angle or the radius of the mosaic disks (see Appendix~\ref{app:supplement figures} for details). 
To quantitatively evaluate the $R$-to-$R'$ departure in Fig.~\ref{fig:dust_ratio_compare}, simulations are performed to study the distribution of the $R'$-$R$ regression lines (represented by the slopes and intercepts of the orange lines in Fig.~\ref{fig:dust_ratio_compare}). 
This kind of simulation requires specific priors of $k_1'$ and $k_2'$ in Eq.~(\ref{equ:k1_expectation_k1}), as described below:
\begin{enumerate}
\item Because $k_1'$ and $k_2'$ represent the residual-to-dust ratio, they should be much smaller than 1. 

\item Since the dust emission increases as the frequency becomes higher, we expect $k_1' \ge k_2'$. 

\item The comparison is more reasonable if $k_1'$ and $k_2'$ are adjusted to make the simulated $f_\mathrm{rel}$ roughly matches the actual $f_\mathrm{rel}$ for the 100 - 143 GHz pair (see the bottom left panel of Fig.~\ref{fig:dust_ratio_compare}). 
\end{enumerate}
Fig.~\ref{fig:simulation R'-R with low galactic} illustrates the determination of $k_1'$ and $k_2'$ based on the above criteria:
The scatter points in this figure represent the simulation results with the limit $C'\leq 1$, with the gray frames highlighting the subset where $0.95 \leq C' \leq 1$. 
The percentage of these points is $f_\mathrm{rel}$, which is approximate to 60.1\% for the bottom left panel of Fig.~\ref{fig:dust_ratio_compare}; 
hence, three sets of $(k_1', k_2')$ that can make $f_\mathrm{rel} \approx 60.1\%$ are chosen: $(0.234, 0.050)$, $(0.216, 0.100)$, and $(0.167, 0.167)$. 
The sample's degrees of freedom is set to $N=100$, and we also confirm that the results are insensitive to $N$ (see Appendix \ref{app:supplement figures}, 
where we take an extreme case: the degrees of freedom are set to $N = 10$). 
Eventually, the scatter points are simulated based on Eqs.~(\ref{equ:R'/R})-(\ref{equ:k1_expectation_k1}) to obtain simulated slopes and intercepts of $R'$-$R$ regression lines, as shown in Figs.~\ref{fig:simulation slope intercept with low galactic}-\ref{fig:simulation slope intercept no low galactic}.

Fig.~\ref{fig:simulation slope intercept with low galactic} displays the scatter plots of 10,000 simulated $R'$-$R$ slopes and intercepts. 
Similarly, Fig.~\ref{fig:simulation slope intercept no low galactic} shows the case without the $|b|<30\degree$ regions. 
We observe that, 
the $R'$-$R$ slopes never fall below 0.85 in 10,000 simulations, which means the slope of -0.623 for the 100-143 GHz pair (top left panel of Fig.~\ref{fig:dust_ratio_compare}) is far beyond expectation. 
Therefore, the null hypothesis that ``the single-component dust emission model is a good approximation'' is rejected completely.

\begin{figure*}[!htb]
\centering
\includegraphics[width=0.32\textwidth]{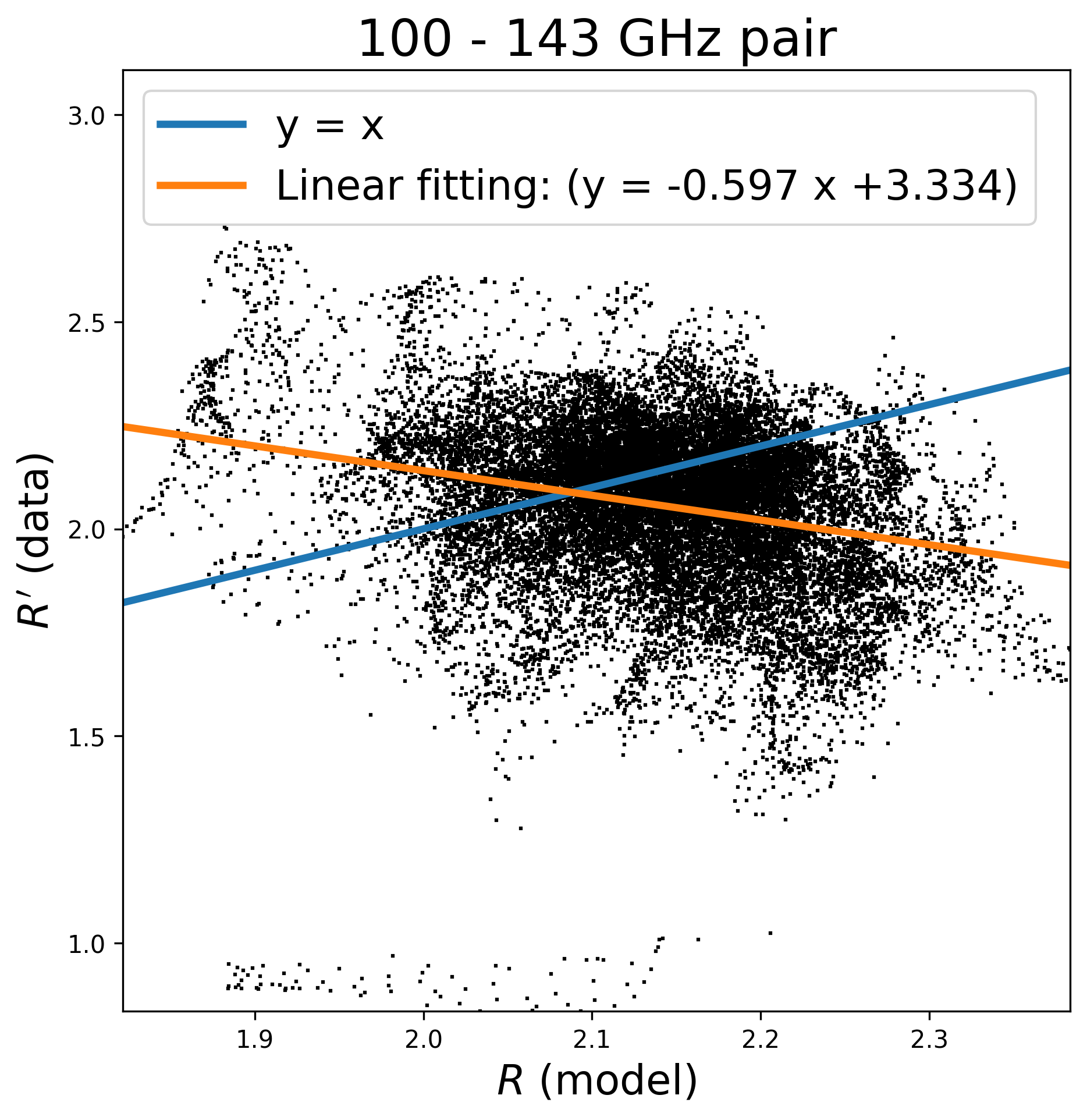}
\includegraphics[width=0.32\textwidth]{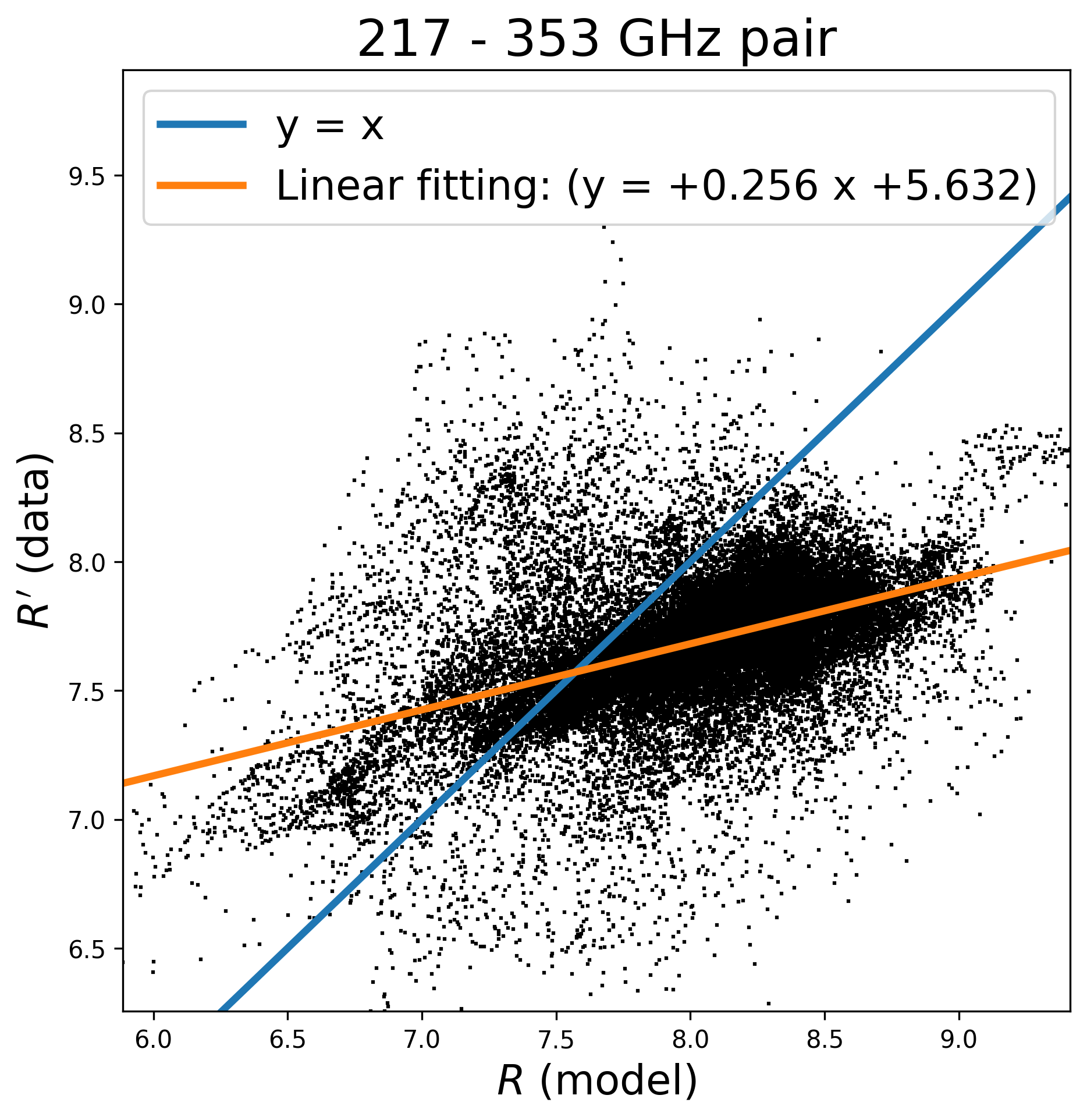}
\includegraphics[width=0.32\textwidth]{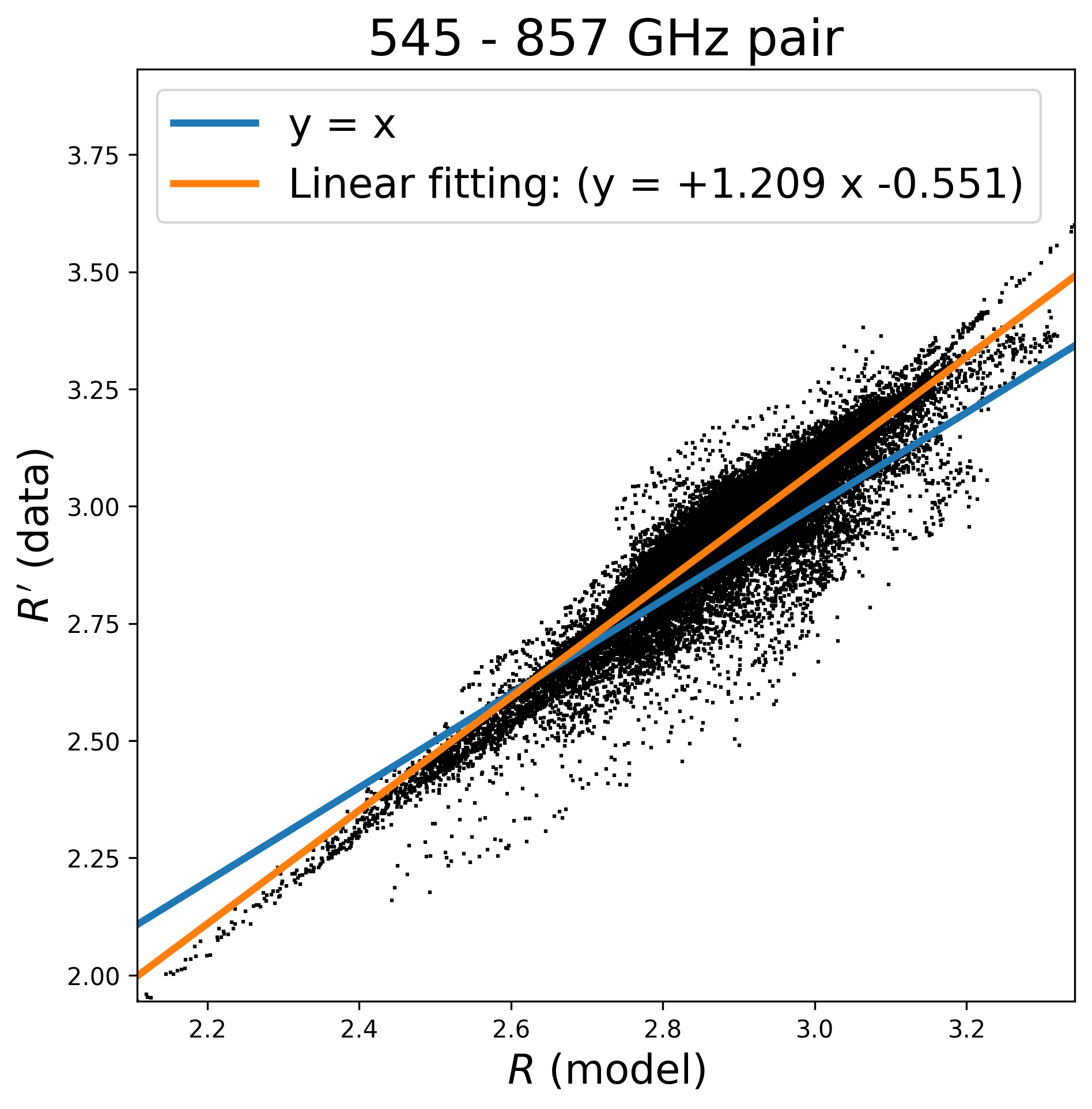}\\
\includegraphics[width=0.32\textwidth]{Scatter_model_2015_color_correction_yes_1_galactic_mask_80_smooth_degree_2_disk_degree_6_low_galac_mask_no_zodiacal_mask_no.png}
\includegraphics[width=0.32\textwidth]{Scatter_model_2015_color_correction_yes_2_galactic_mask_80_smooth_degree_2_disk_degree_6_low_galac_mask_no_zodiacal_mask_no.png}
\includegraphics[width=0.32\textwidth]{Scatter_model_2015_color_correction_yes_3_galactic_mask_80_smooth_degree_2_disk_degree_6_low_galac_mask_no_zodiacal_mask_no.png}\\
\includegraphics[width=0.32\textwidth]{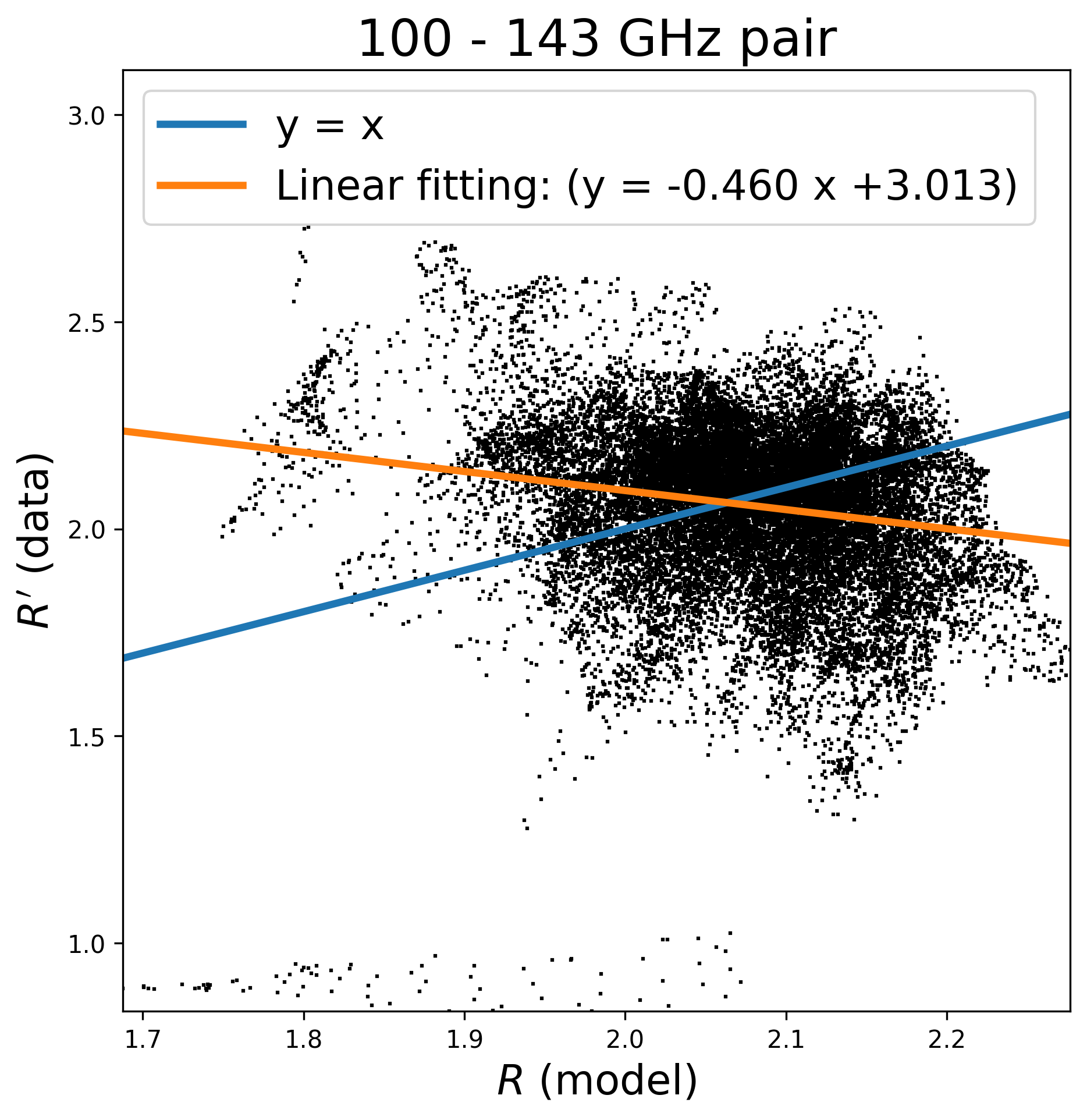}
\includegraphics[width=0.32\textwidth]{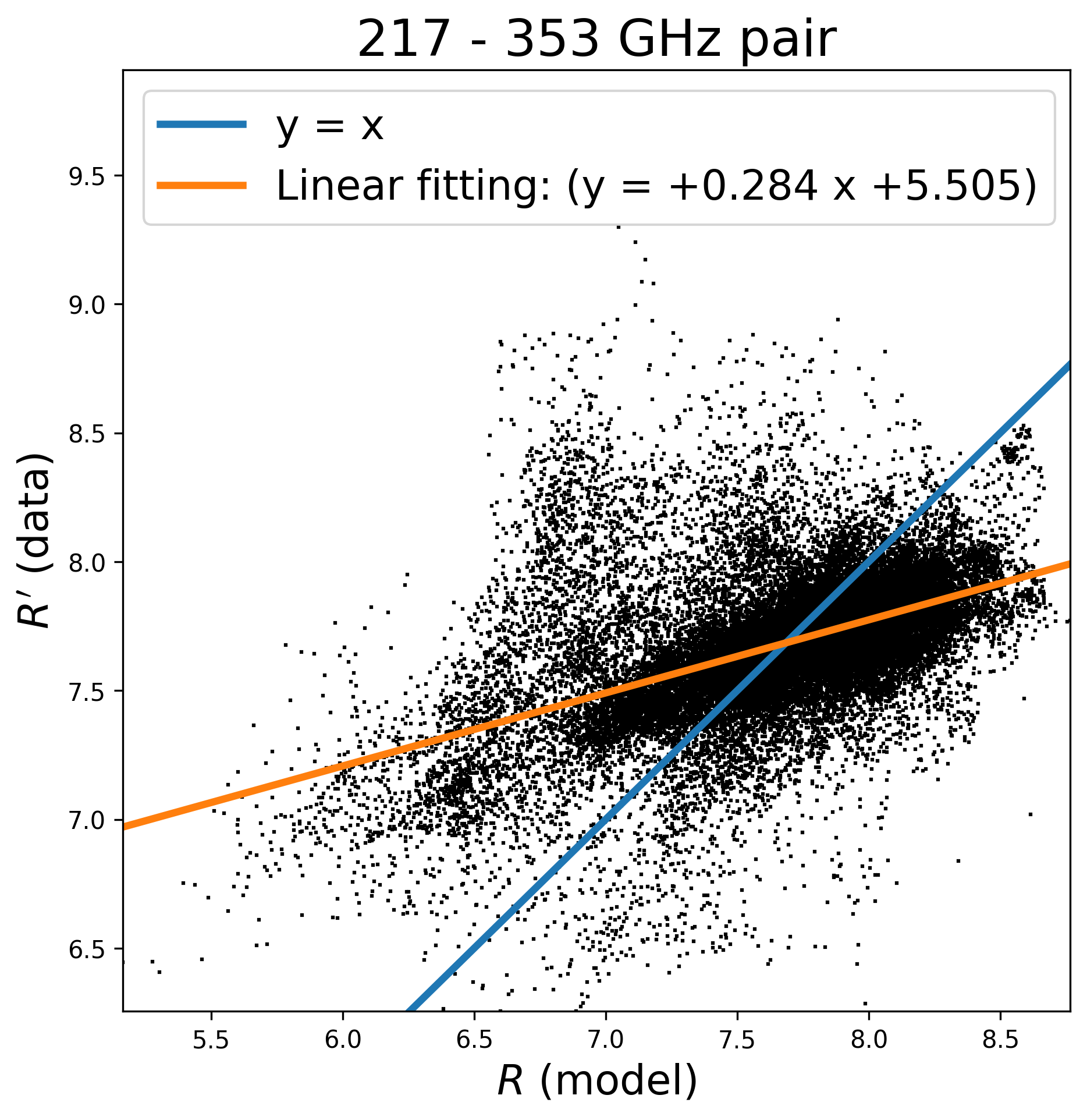}
\includegraphics[width=0.32\textwidth]{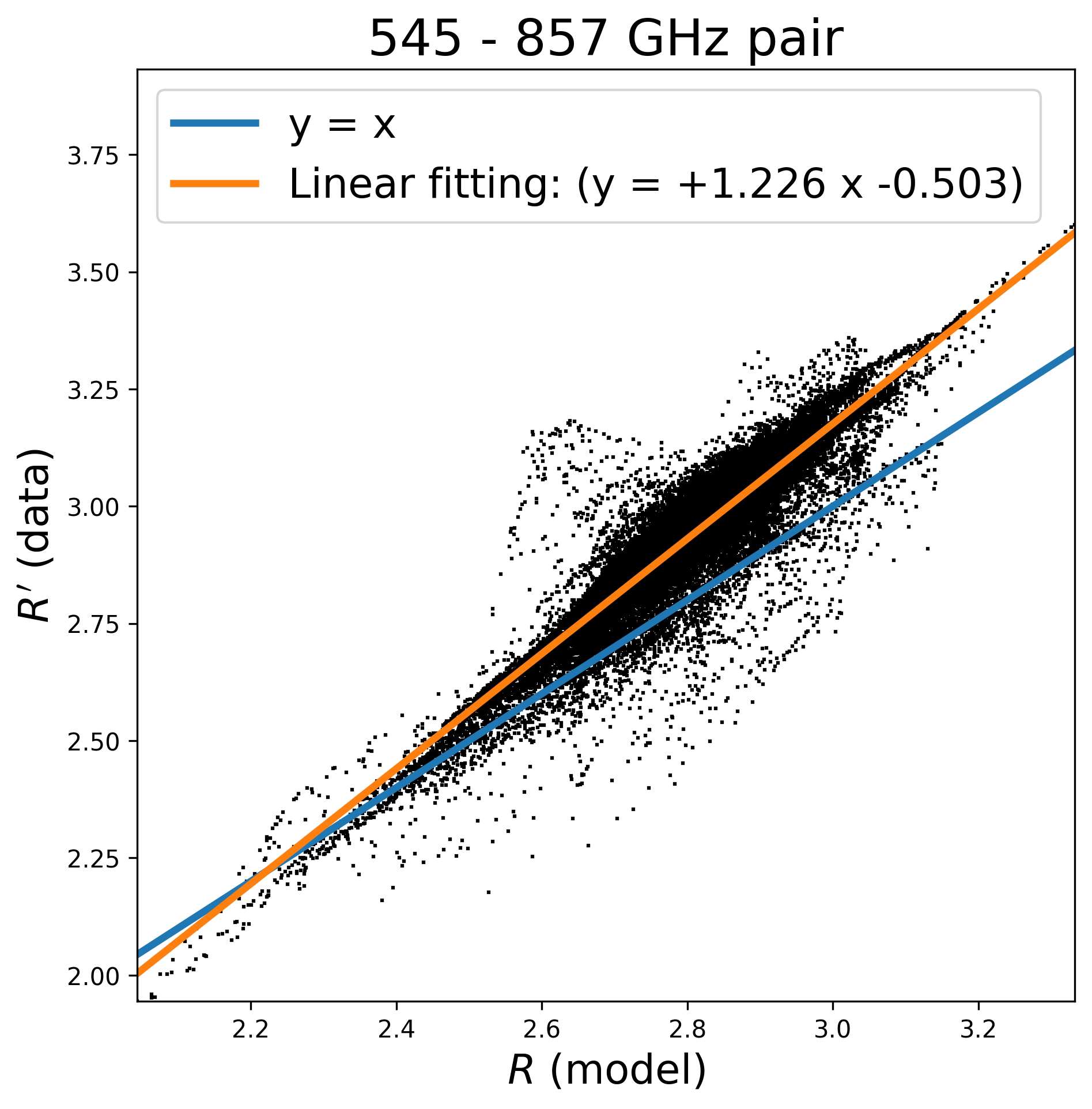}
\caption{Scatter plots of  $R'$ versus $R$ with color correction, with smoothing angle of $2^\circ$ and mosaic disk radius of $6^\circ$. 
$M_\mathrm{tot} = M_\mathrm{comp} \times M_{80}$. 
\textit{From top to bottom}: M13, M15, and M19. 
\textit{From left to right}: 100-143, 217-353, and 545-857 pairs. }
\label{fig:dust_ratio_compare with different models with color correction}
\end{figure*}

\begin{figure*}[!htb]
\centering
\includegraphics[width=0.32\textwidth]{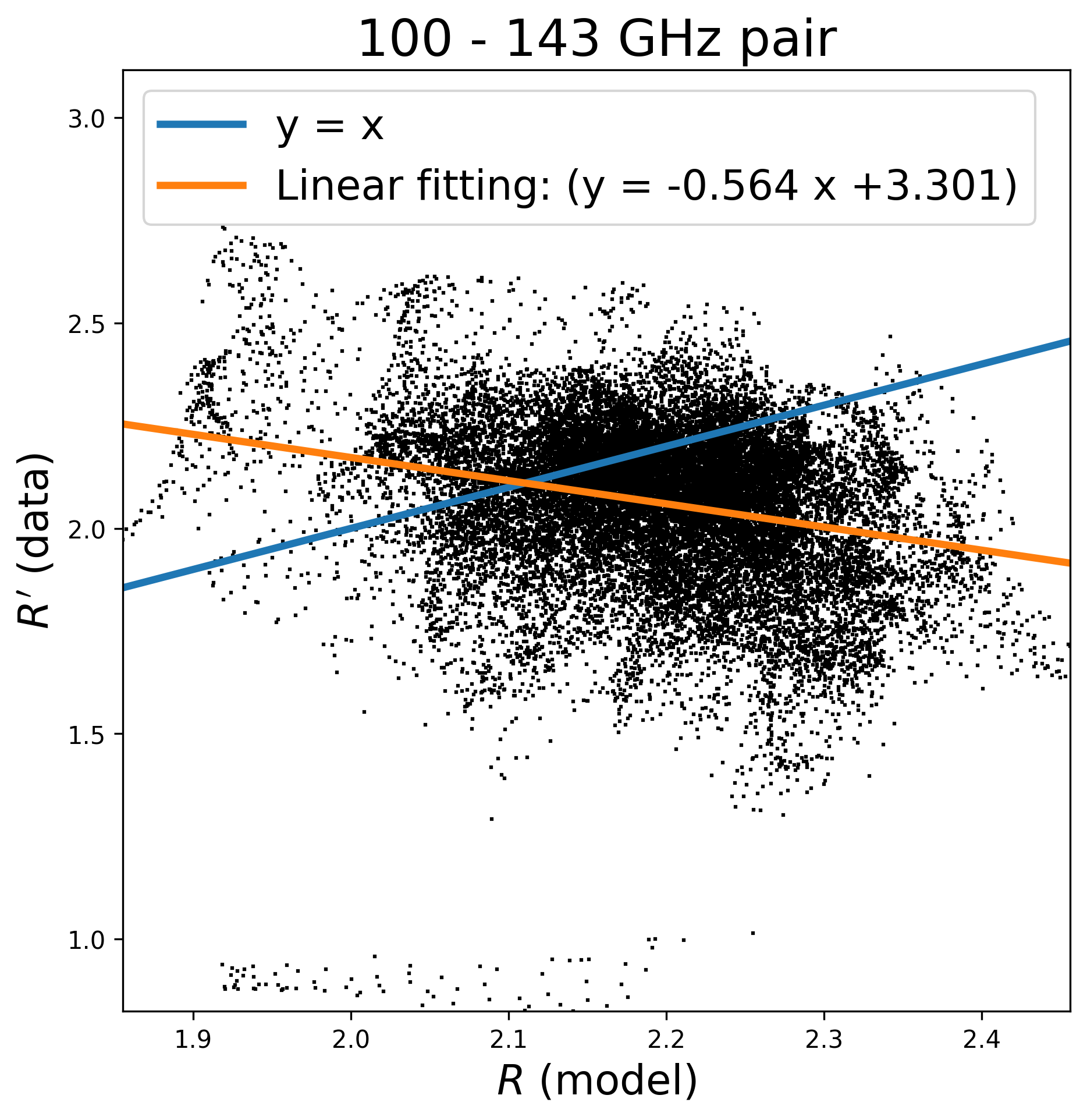}
\includegraphics[width=0.32\textwidth]{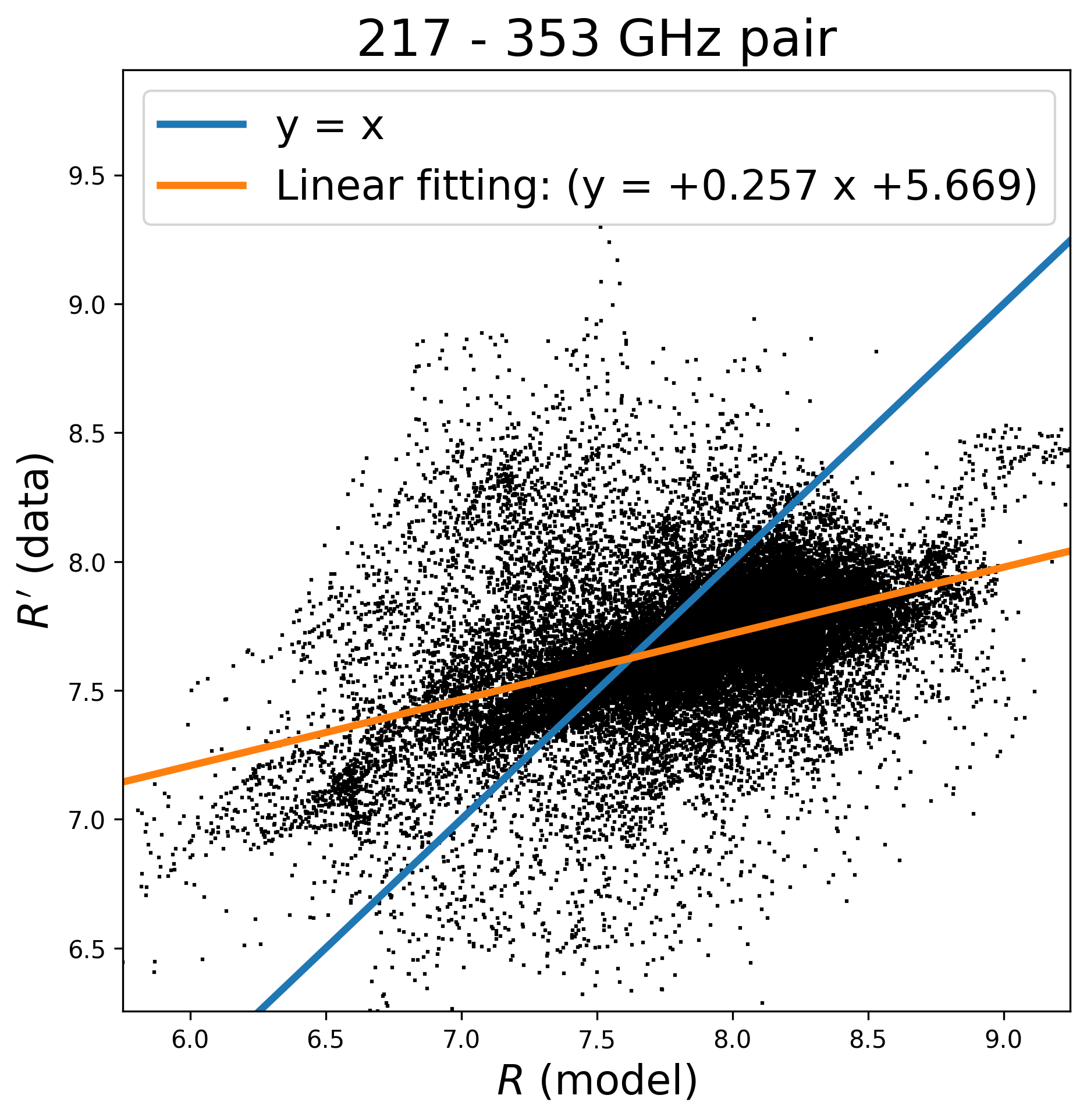}
\includegraphics[width=0.32\textwidth]{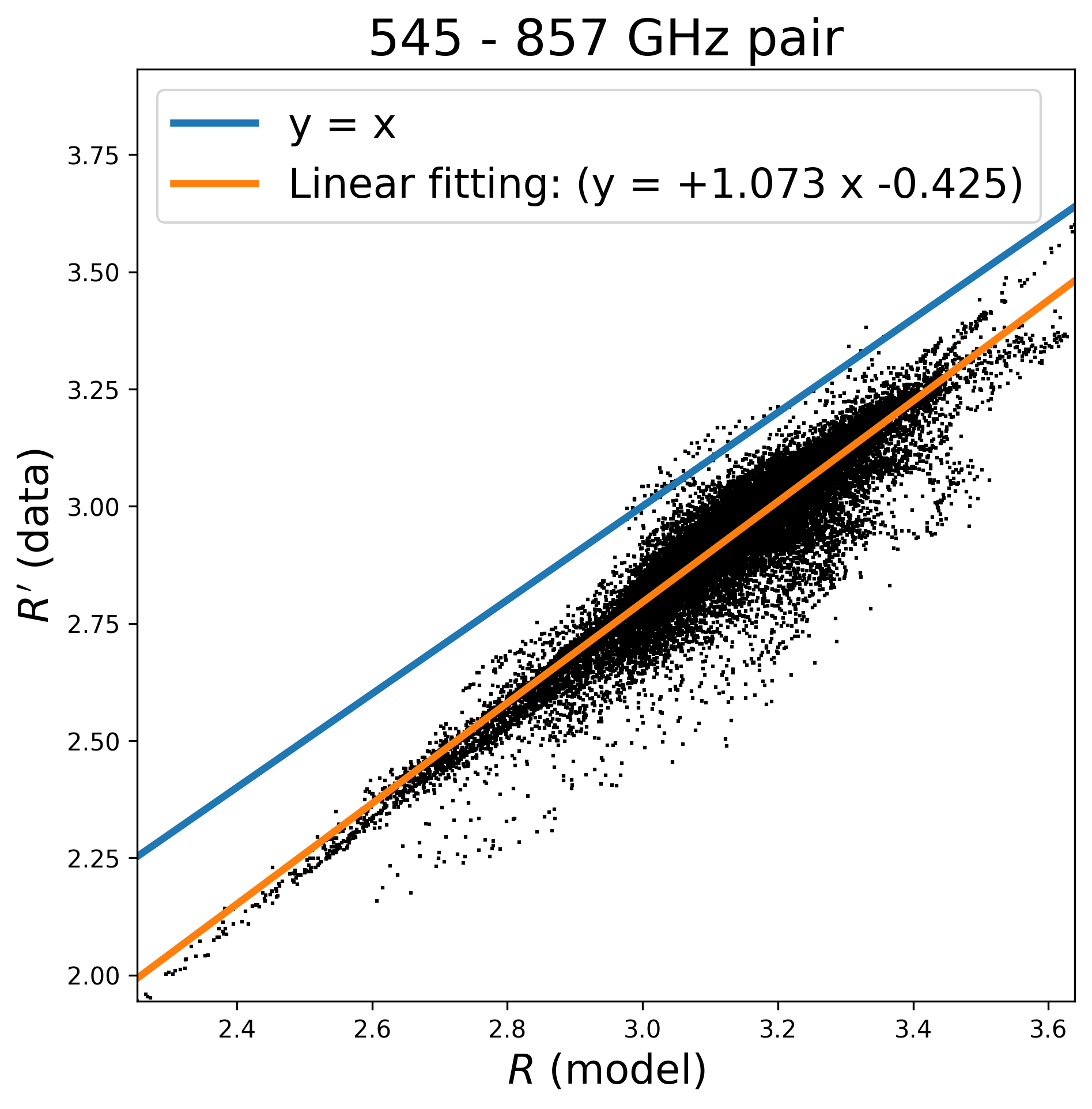}\\
\includegraphics[width=0.32\textwidth]{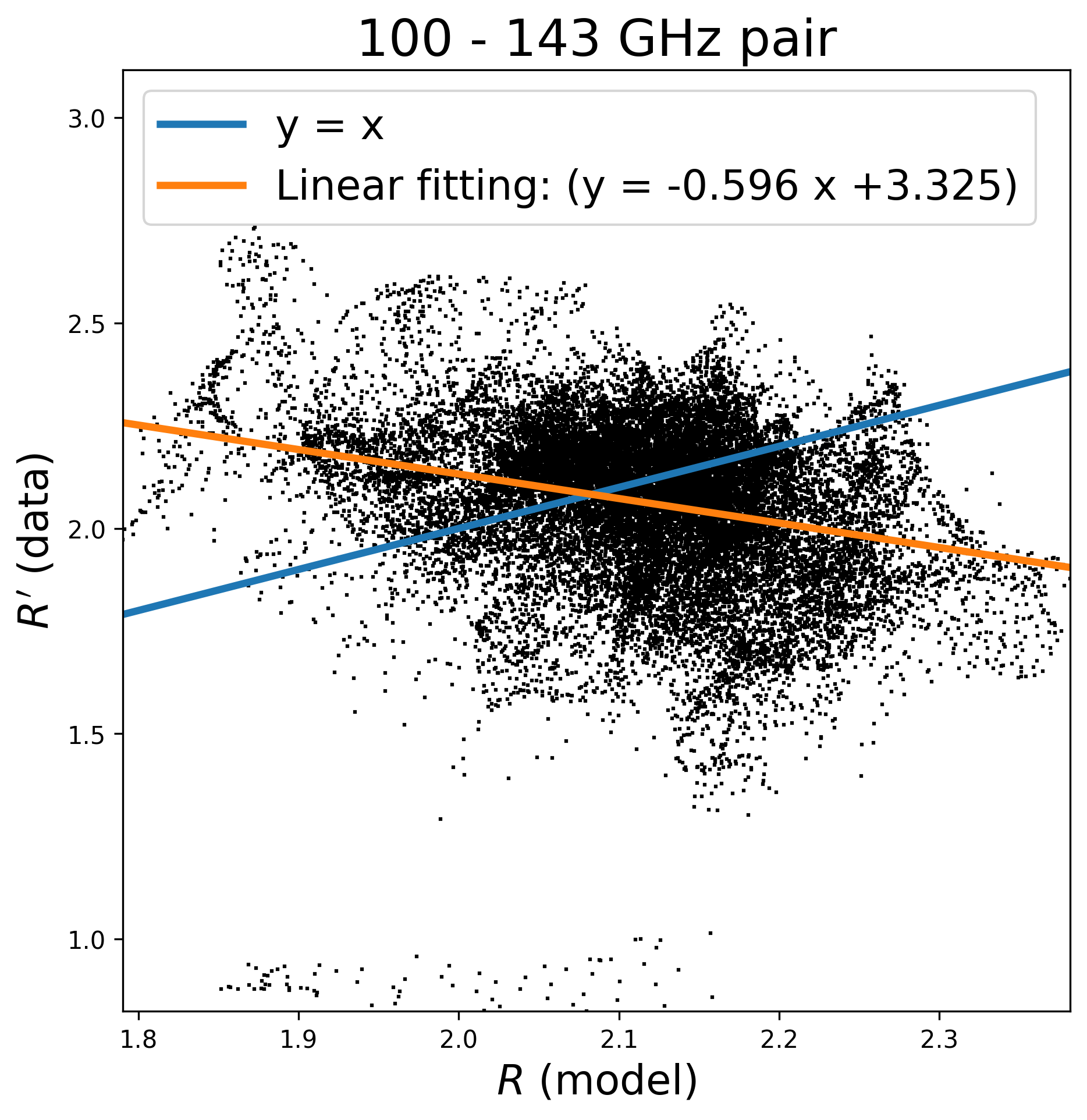}
\includegraphics[width=0.32\textwidth]{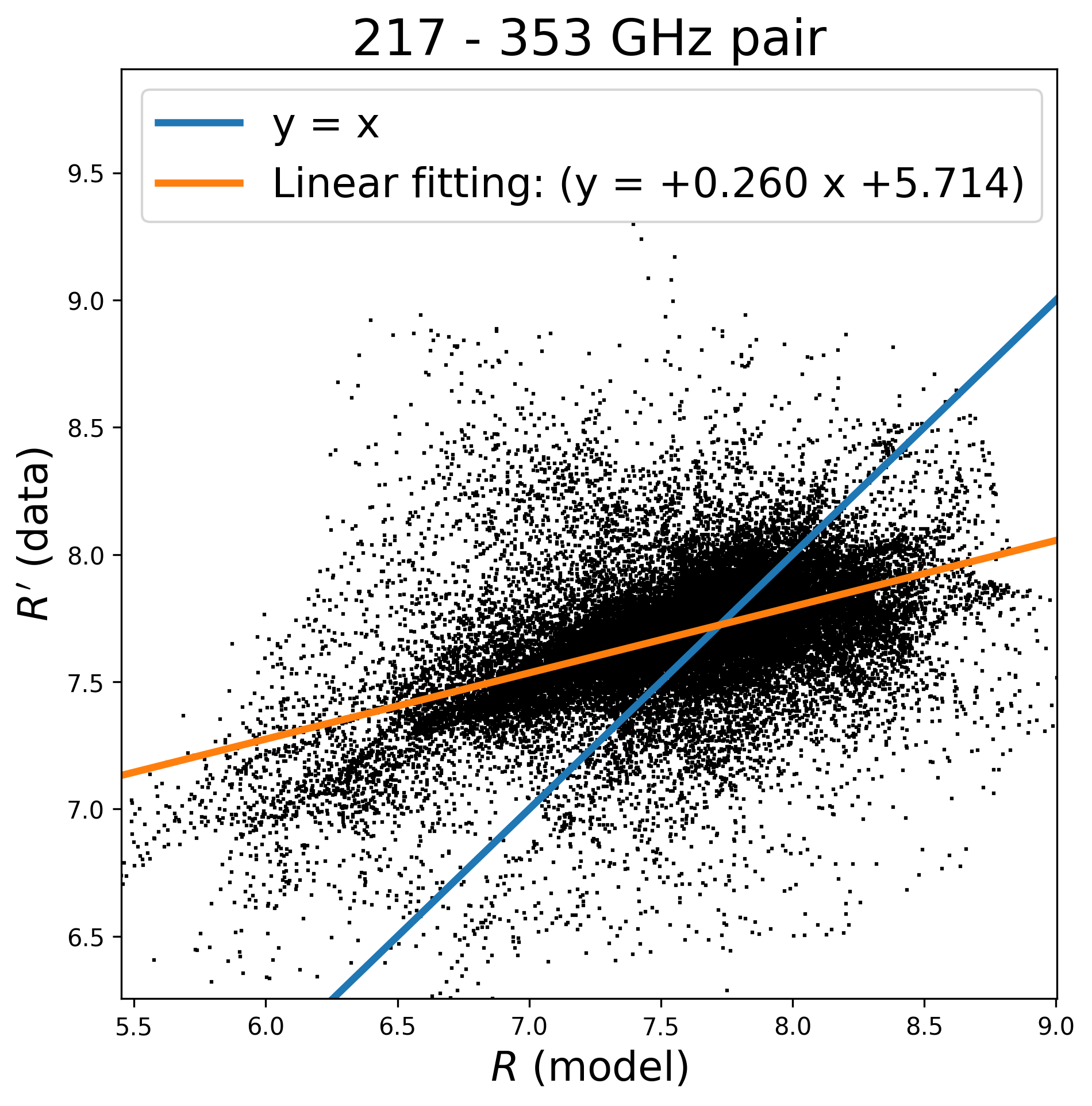}
\includegraphics[width=0.32\textwidth]{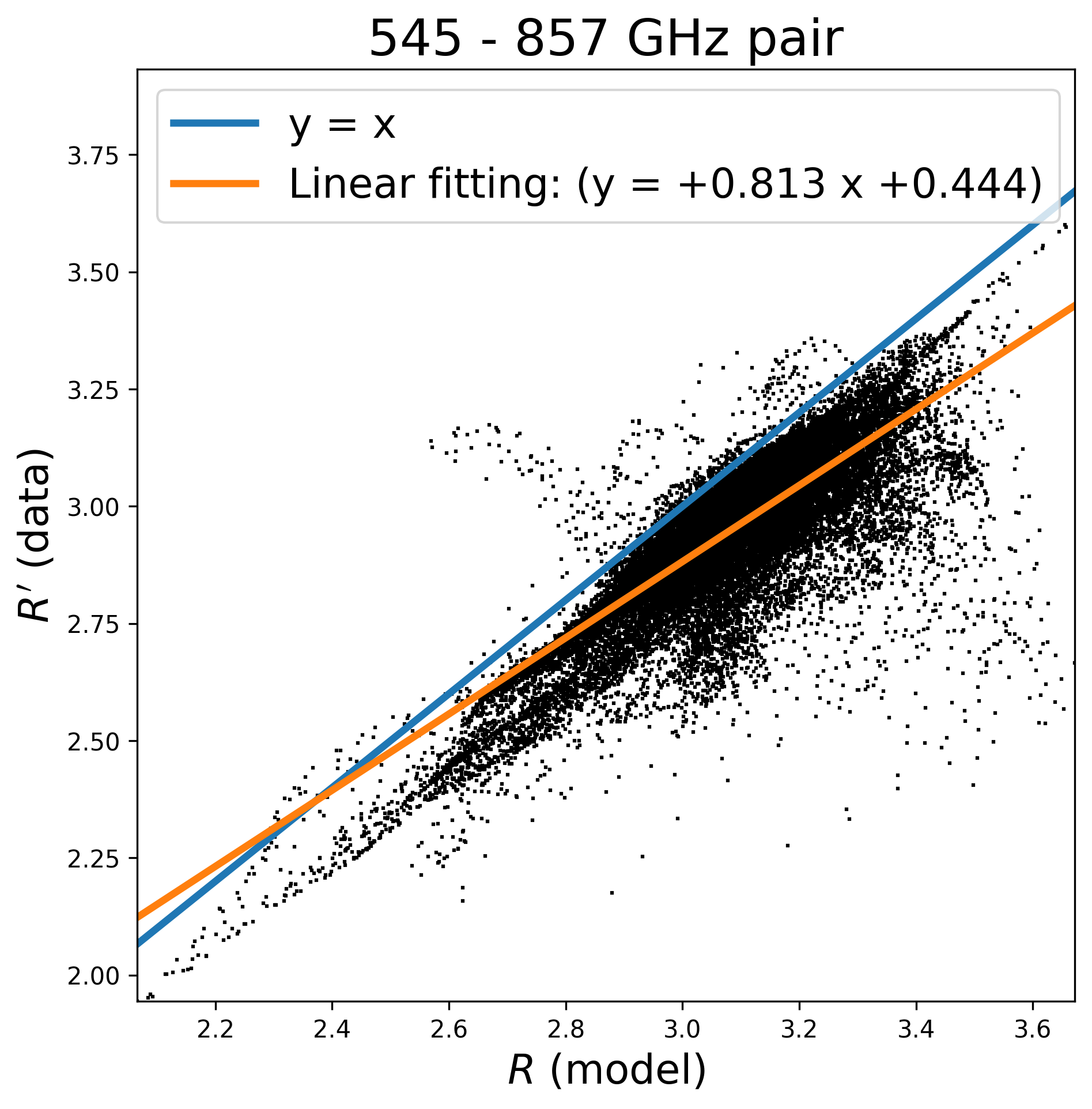}\\
\includegraphics[width=0.32\textwidth]{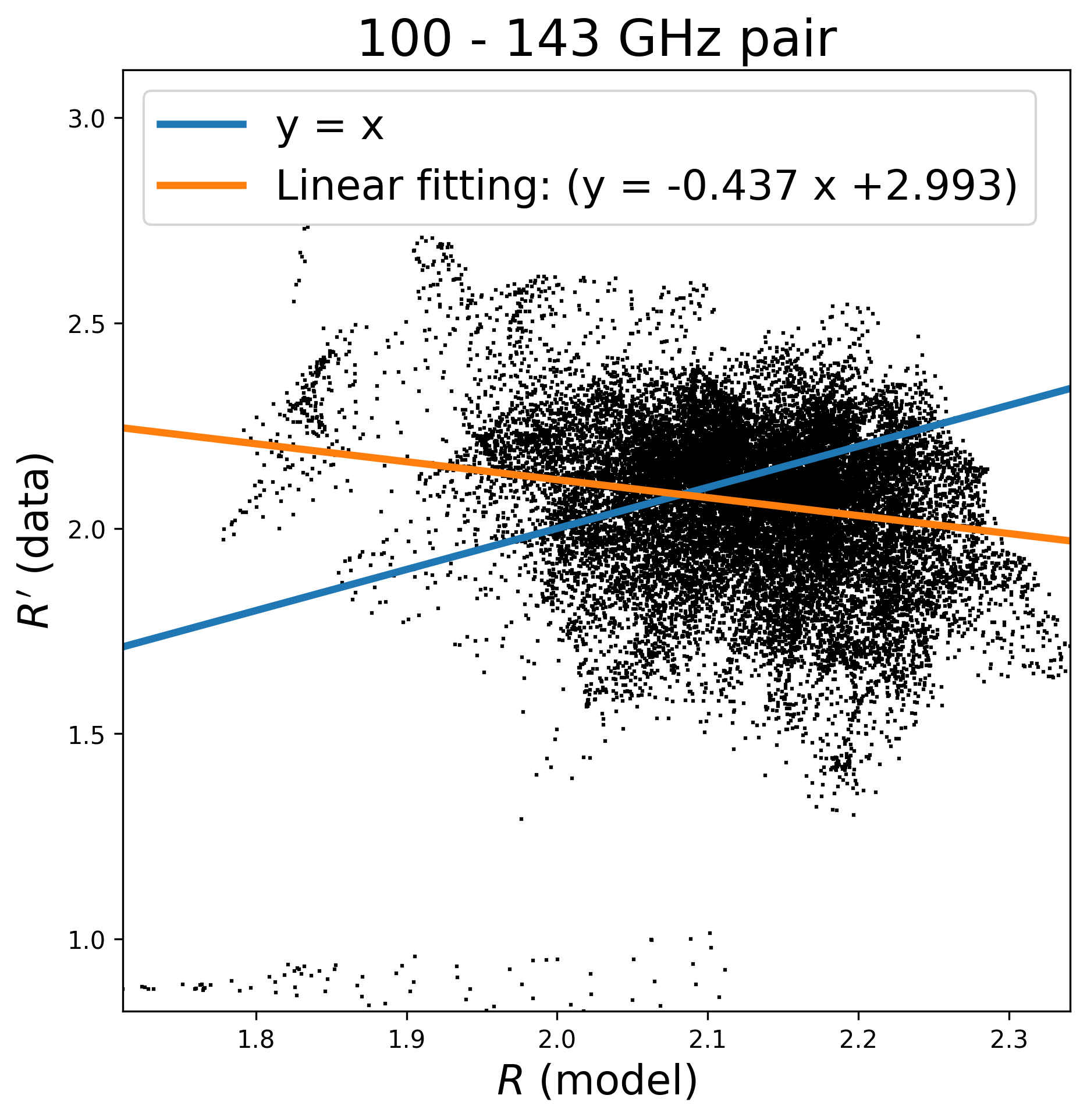}
\includegraphics[width=0.32\textwidth]{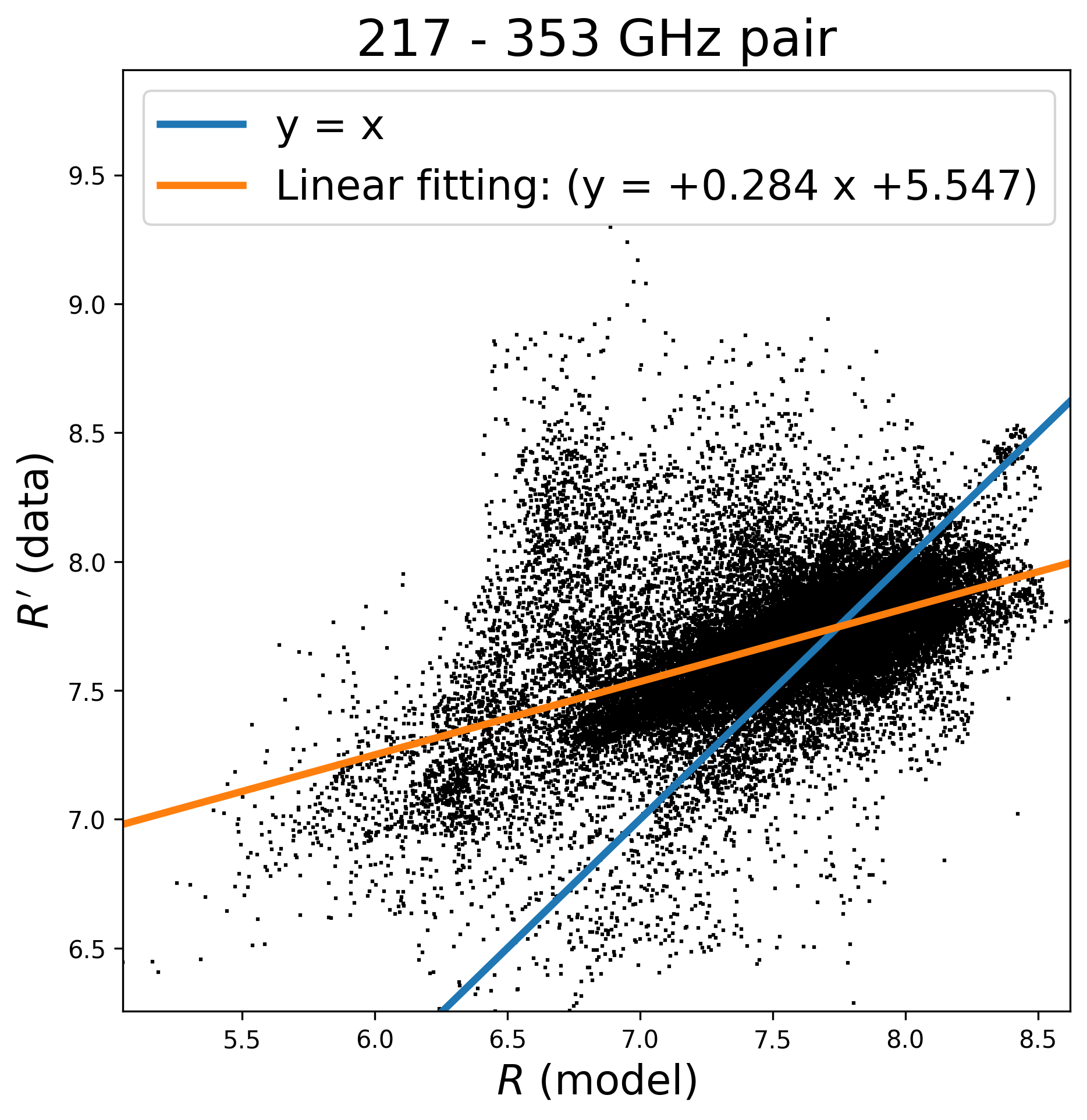}
\includegraphics[width=0.32\textwidth]{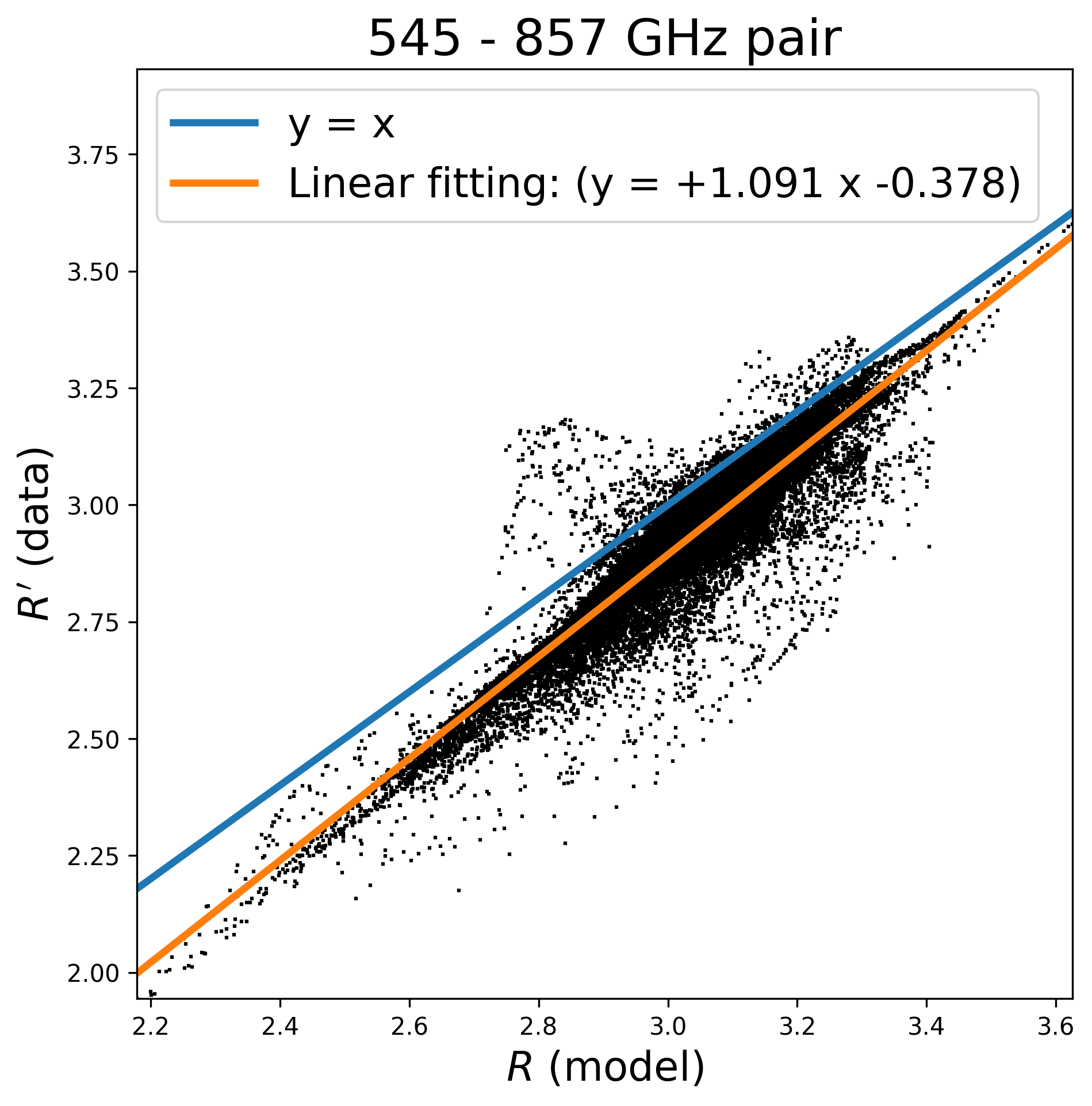}
\caption{The counterpart of Fig.~\ref{fig:dust_ratio_compare with different models with color correction}, but without color correction (except the processing for subtracting CO lines). }
\label{fig:dust_ratio_compare with different models without color correction}
\end{figure*}

\subsection{Test with different single-component dust models}
\label{sub: test with different dust models}

To investigate the robustness of the above conclusion for different single-component dust emission models, in Fig.~\ref{fig:dust_ratio_compare with different models with color correction}, we compare the results obtained from M13, M15, and M19. 
The data preprocessing with all of these models contain color correction. 
The output of three models are similar: the $R'$-$R$ regression line's departure from $R = R'$ is increasing significantly from 545-857 GHz to 100-143 GHz pair, and the slopes are negative for all three models at 100-143 GHz pair.
Similar behaviors of the $R'$-$R$ regression lines in these three models indicate that the model-to-data departure indeed cannot be solved by different options of the single-component models. 

However, certain details warrant further attention. 
As indicated by Eq.~(\ref{equ:R'/R}) and Figs.~\ref{fig:simulation slope intercept with low galactic}-\ref{fig:simulation slope intercept no low galactic}, the presence of residuals suggests that the slopes of the $R'$-$R$ regression lines should be slightly less than 1, at least for frequencies where the single-component model remains valid, such as the 545 - 857 GHz pair. 
However, one can observe that the slopes corresponding to the M13 and M19 models exceed 1.2, which may imply some deficiencies in M13 and M19 compared to M15, even from the perspective of the single-component model.

\section{Discussion about possible sources of the departure}

Statistics in mosaic disks guarantee that both linear ratios and cross-correlation are unaffected by constant offset, and a high $C'$ ensures that the noise and systematics are subdominant. Therefore, the band offset, noise and systematics are all rejected as major contributors to the model-to-data departure observed in Fig.~\ref{fig:dust_ratio_compare}. Below we focus on some other possible sources, and show how they are rejected as well. 
\label{sec: discussion of sources}

\subsection{Color correction}
Color correction has already been taken into account in this paper, as stated in Subsection \ref{subsection_Unit_conversion_Colour_correction}. However,
for comparison and cross-check, we also test the results without color correction. 
This is equivalent to assuming a $\delta$-function-like transmission at the 
reference frequency for all detectors.\footnote{Since the CO emission line spectrum is almost a $\delta$-function itself, color correction is still considered in the subtraction of CO emission lines.}
As illustrated in Fig.~\ref{fig:dust_ratio_compare with different models without color correction}, the slopes of the $R'$-$R$ regression lines change only slightly without the color correction, from -0.623, 0.258, 0.913 to -0.596, 0.260, 0.813, for M15. 
This confirms that color correction is not a major source of the model-to-data departure.

Meanwhile, behavior of the regression lines at 545 - 857 GHz pair for the M13 and M19 models is also interesting:
When the color correction is ignored, the slopes of the regression lines for this pair decrease approximately from 1.2 to 1, demonstrating an improved match. 
This suggests that M13 and M19 might be presented in a ``raw'' mode, ignoring proper color correction.

\subsection{CIB}
The CIB was first detected in 1996 by the FIRAS instrument aboard the COBE satellite \citep{1996A&A...308L...5P}. 
The authors attributed the CIB to high-redshift sources. 
Subsequent studies further identified the physical sources of the CIB as: (1) starlight from extragalactic sources redshifted into the infrared band; (2) starlight absorbed by extragalactic dust and re-emitted at lower frequencies; and (3) electromagnetic radiation originating from active galactic nuclei \citep{2000A&A...360....1G, 2005ARA&A..43..727L, 2014A&A...571A..11P}. 
The CIB does interferes with the modeling of Galactic thermal dust emission, especially in high Galactic latitude regions, where the anisotropies of the CIB significantly amplify the uncertainties in the modeling of thermal dust spectral index and temperature \citep{2014A&A...571A..11P}.
Although the spectral index of thermal dust emission and the CIB in the \textit{Planck} frequency bands are somewhat similar, their morphologies differ significantly. 
Specifically, the angular power spectrum of thermal dust emission is roughly proportional to $\ell^{-2.7}$ \citep{2014A&A...571A..30P}, whereas that of the CIB is about $\ell^{-1}$ \citep{2014A&A...571A..18P}, which is relatively much stronger at smaller angular scales.

For three reasons, the CIB is not a favored explanation of the model-to-data departure in this paper: 
\begin{enumerate}
\item As one can see from Fig.~\ref{fig:dust_ratio_compare}, for the 100 - 143 GHz pair, the comparison is done mainly in the regions with mid-Galactic latitudes, where the CIB is supposed to be subdominant. 
\item The CIB exhibits partial correlations across different frequencies due to the fact that extragalactic sources at varying redshifts emit infrared radiation at different frequencies \citep{2014A&A...571A..11P}. 
Thus when we use only highly correlated patches, the impact of CIB is small. 
\item Because the CIB anisotropies are more pronounced at smaller scales, the CIB tends to behave as a constant offset as the smoothing angle increases \citep{2019A&A...623A..21I}. 
\end{enumerate}
However, as shown in Figs.~\ref{fig:scatter plots with different smoothing and disk angle 143-100}-\ref{fig:scatter plots with different smoothing and disk angle 857-545}, 
we observe almost identical results regardless of the smoothing angle, suggesting that the influence of the CIB is negligible.

\subsection{Free-free and CO emissions}
We have three reasons to dismiss free-free emission as an explanation: 
\begin{enumerate}
    \item Free-free emission is subdominant above 100 GHz \citep{2020A&A...641A...1P}.
    \item We have already subtracted the estimated free-free emission from each band.
    \item Free-free emission is very weak in the Galactic regions with $|b|>30\degree$, but Fig.~\ref{fig:dust_ratio_compare2} shows no essential difference from Fig.~\ref{fig:dust_ratio_compare}, indicating that free-free emission is not a favored explanation, even after considering possible uncertainties of free-free emissions.
\end{enumerate}
The CO line emission can be excluded in a similar way: it mainly affects the low Galactic latitudes and has already been subtracted.

\subsection{Zodiacal light}
The zodiacal light encompasses planetary light, radiation from small celestial bodies and interplanetary dust radiation, dominating the radiation throughout the full sky in the 5000 - 30000 GHz range \citep{1998ApJ...508...44K}. 
The current mainstream model of zodiacal emission is based on the research of \cite{1998ApJ...508...44K}, which divides the zodiacal emission into several components: a diffuse cloud, three dust bands, a circumsolar dust ring, and an Earth-trailing feature \citep{2022A&A...666A.107S}.

We have three compelling reasons to exclude the possibility of zodiacal emission as a candidate for explaining the model-to-data departure in this paper: 
\begin{enumerate}
\item The \textit{Planck} releases since 2015 have already removed the zodiacal emission in their HFI products \citep{2016A&A...594A...8P}, with no zodiacal residuals in \textit{Planck} 2018 HFI maps except the 857 GHz map with tiny zodiacal residuals $\sim 10^{-2}\,\mathrm{MJy\,sr^{-1}}$ \citep{2020A&A...641A...3P}. 
    
\item The zodiacal emission is stronger at higher frequencies, but the departure is stronger at lower frequencies.
    
\item The \textit{Planck} release in 2013 provides HFI maps with the zodiacal emission\footnote{HFI\_SkyMap\_217\_2048\_R1.10\_nominal.fits} and without.\footnote{HFI\_SkyMap\_217\_2048\_R1.10\_nominal\_ZodiCorrected.fits}
We identify regions in the 217 GHz map with differences greater than $5\,\mu\mathrm{K_{CMB}}$ to create a zodiacal mask. 
As shown in Fig.~\ref{fig:dust_ratio_compare_no zodiacal region}, we confirm that incorporating the zodiacal mask does not change the results significantly. 
This rules out zodiacal light as a possible source of the departure. 
\end{enumerate}

\section{Conclusion}
\label{sec:conclusion}
This study investigates the applicability of the single-component thermal dust emission model at the CMB observation frequencies.
Using the latest data release from the \textit{Planck} mission, we derive the thermal dust data maps across the HFI bands, and compute the thermal dust model maps for the same bands based on three different single-component models (M13, M15, and M19). 
The \textit{Planck} HFI frequency channels are divided into three pairs: 100 - 143, 217 - 353, and 545 - 857 GHz, and mosaic disks are used to calculate the correlation coefficients $C'$ (data) and linear ratios $R$ (model) and $R'$ (data), between adjacent frequency bands. 
In each frequency band pair, disks with not less than 0.95 correlation coefficients of the adjacent dust data maps are chosen for comparing the linear ratios $R'$ (data) and $R$ (model), and the accuracy of the single-component model is tested for these three frequency pairs. 

Both analytic equations and simulations show that, when the cross-correlation between two mosaic disks is not less than 0.95, the slopes of the $R'$-$R$ regression lines 
should be not less than 0.85 (from 10,000 simulations). 
However, all M13, M15, and M19 models give the slopes
that are below -0.43 in the 100-143 GHz pair. 
Such a big departure confirms that the single-component thermal dust model is \emph{not} a valid approximation of the thermal dust emission in the 100-857 GHz bands. 

Our analysis is novel in a number of aspects: 
We use regions with strong correlation coefficients to cast a model-free constraint on the model-to-data departure, and the method based on mosaic disks ensures that the results are unaffected by band offsets. 
Meanwhile, with our method, the impact of various sources can be easily excluded, like noise, systematics, color correction, free-free, CIB, zodiacal light, etc, which makes the conclusion in this work strong and robust. 
Meanwhile, the criterion used in this work to test a dust model is completely determined by the observational data (without model dependence), which makes it useful in determining the content and complexity of future multi-component dust emission estimations.

The minimal model dependency of our method allows it to be employed in the next step to test a variety of combined dust models, like a multi-component thermal dust model. 
It should also be noted that the possibility of anomalous microwave emission (AME) is not excluded in this work, although the \textit{Planck} team suggests an upper frequency limit of about 70 GHz for its effect \citep{2016A&A...594A..10P}. The main reason is that the AME belongs to the family of dust emissions, which could be regarded as a potential additional dust component.

Due to the fact that \textit{Planck} team only provides polarization sky maps at 100, 143, 217, and 353 GHz in the HFI bands, and considering the complexities such as depolarization along the line of sight, this study focuses on intensity analysis. 
Future work may extend this method to polarization to further verify the robustness of the single-component model across different observational parameters. 

\begin{acknowledgments}
This work is supported in part by National Key R\&D Program of China (2021YFC2203100, 2021YFC2203104), by NSFC (12433002, 12261131497), by CAS young interdisciplinary innovation team (JCTD-2022-20), by 111 Project (B23042), by Fundamental Research Funds for Central Universities, by CSC Innovation Talent Funds, by USTC Fellowship for International Cooperation, by USTC Research Funds of the Double First-Class Initiative, by the Anhui Provincial Natural Science Foundation 2308085MA30, and by the Anhui project Z010118169. 
\end{acknowledgments}

\appendix


\begin{figure*}[!htb]
\centering
\includegraphics[width=0.32\textwidth]{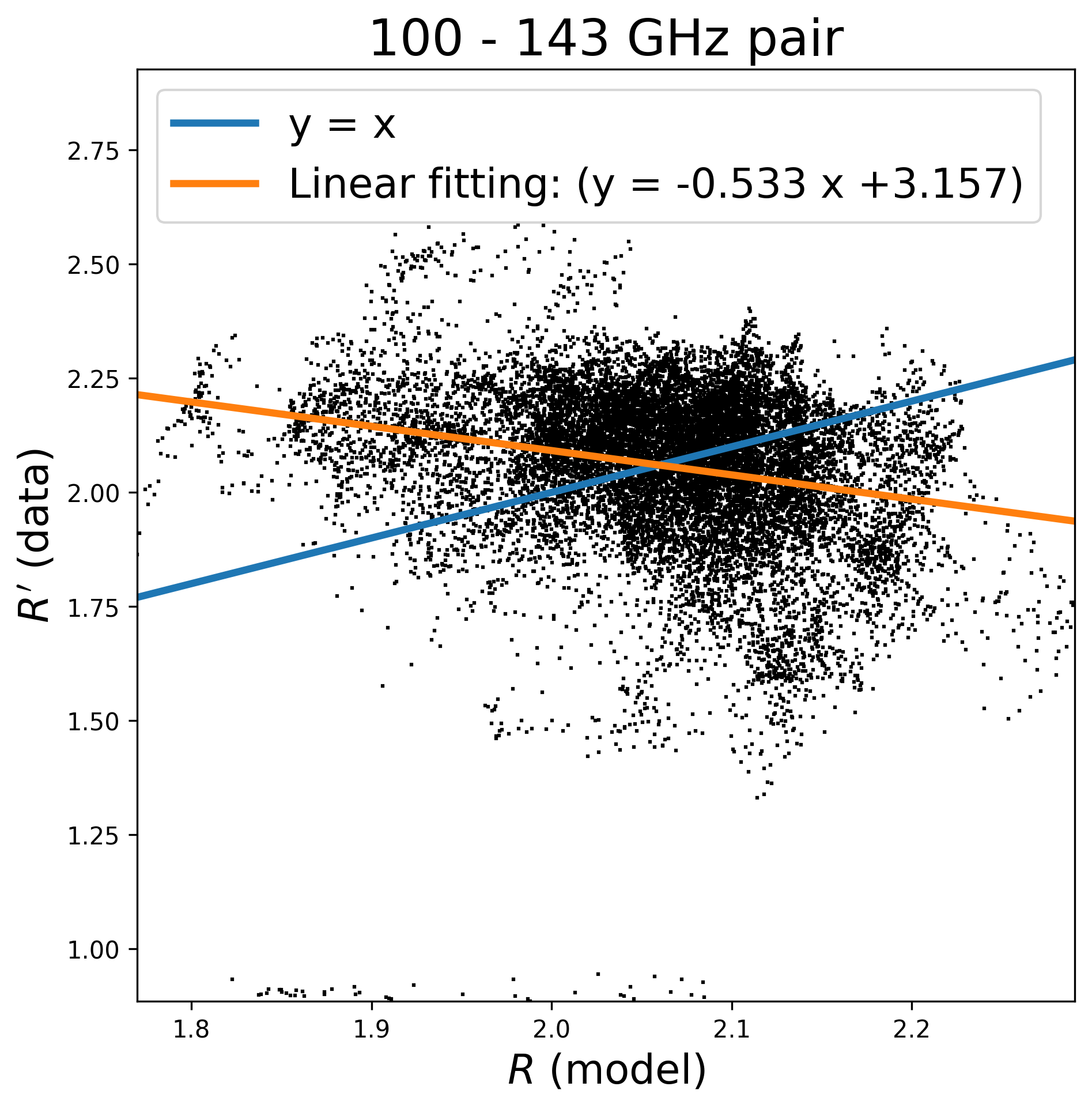}
\includegraphics[width=0.32\textwidth]{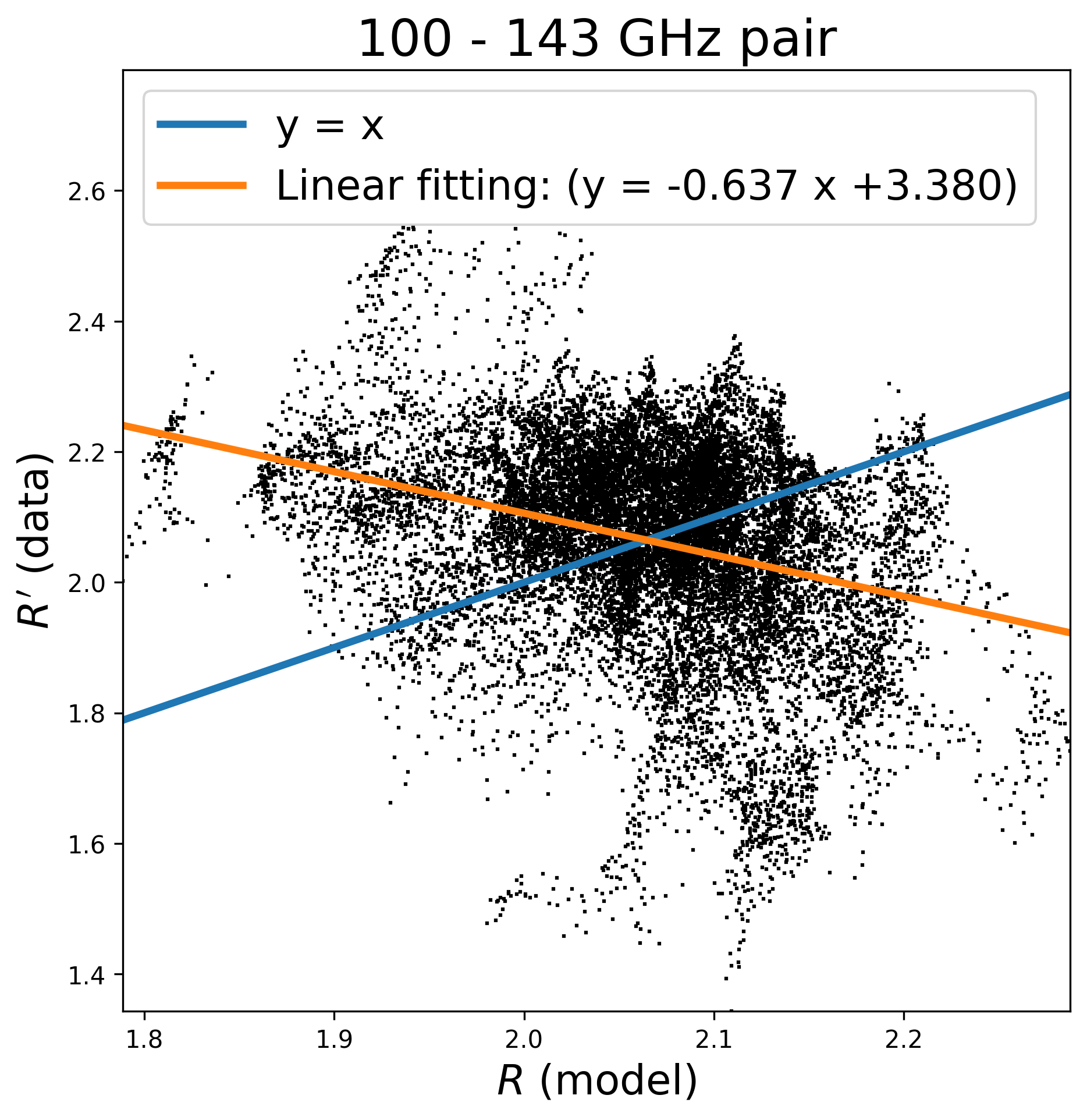}
\includegraphics[width=0.32\textwidth]{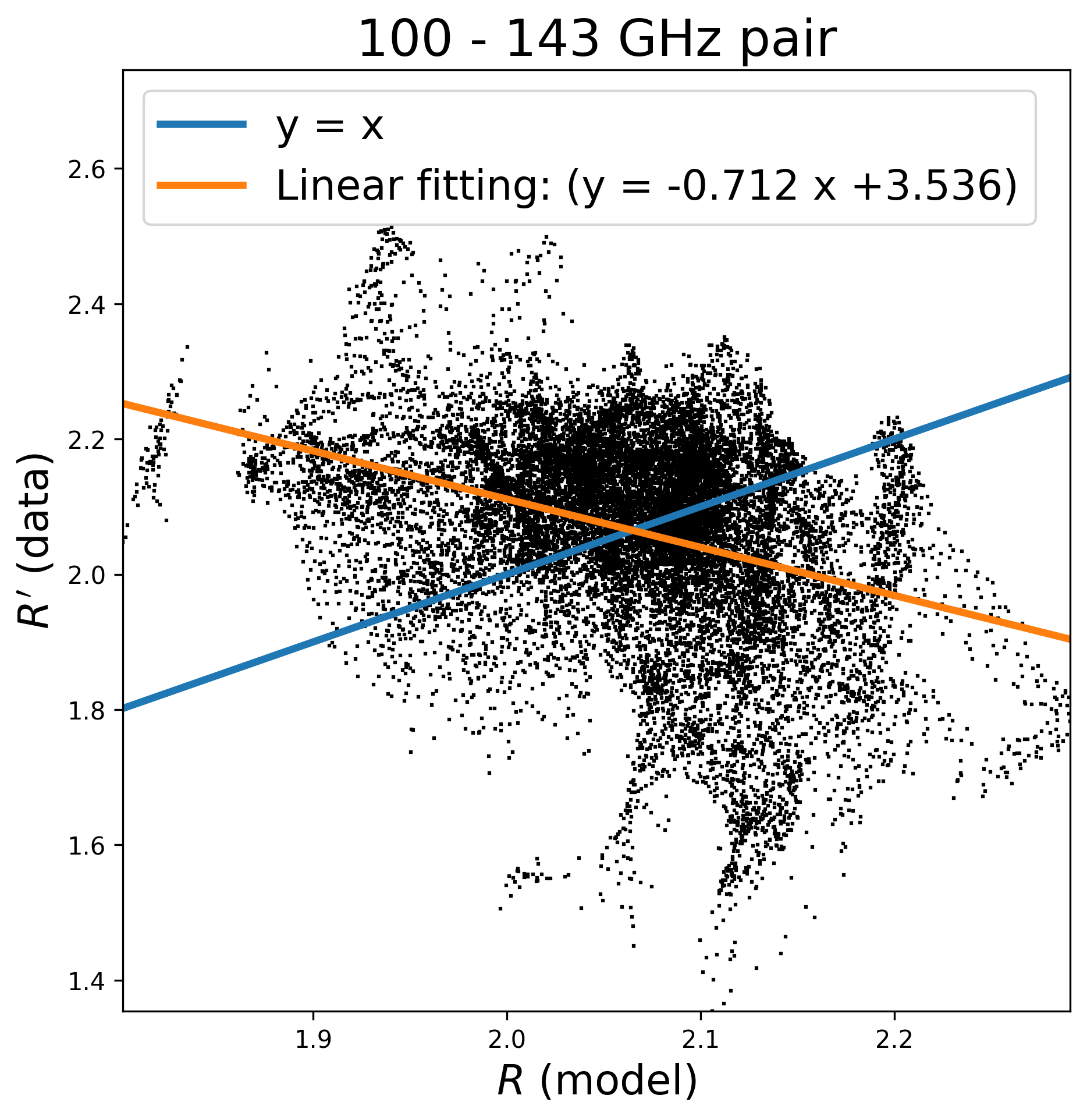}\\
\includegraphics[width=0.32\textwidth]{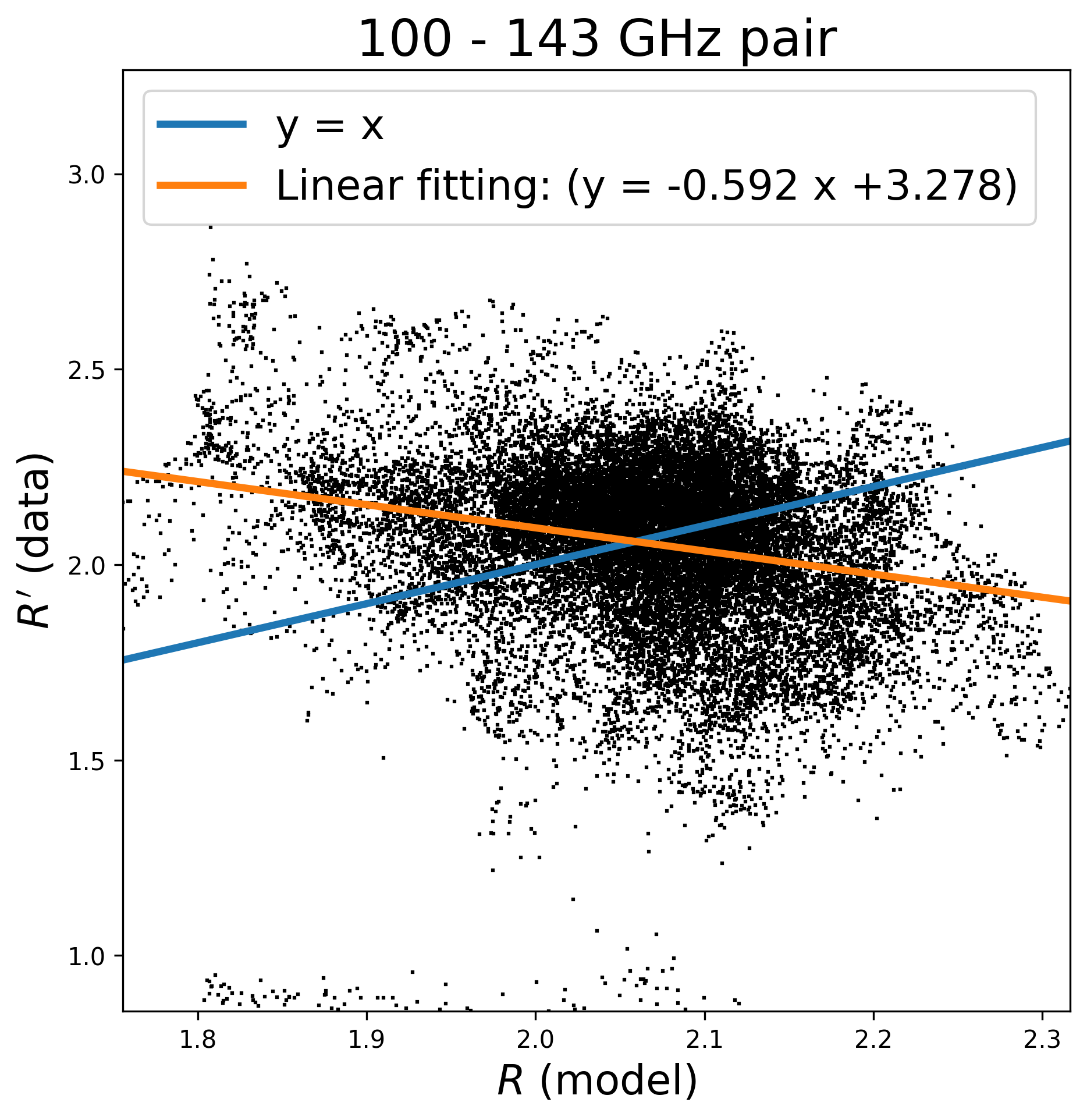}
\includegraphics[width=0.32\textwidth]{Scatter_model_2015_color_correction_yes_1_galactic_mask_80_smooth_degree_2_disk_degree_6_low_galac_mask_no_zodiacal_mask_no.png}
\includegraphics[width=0.32\textwidth]{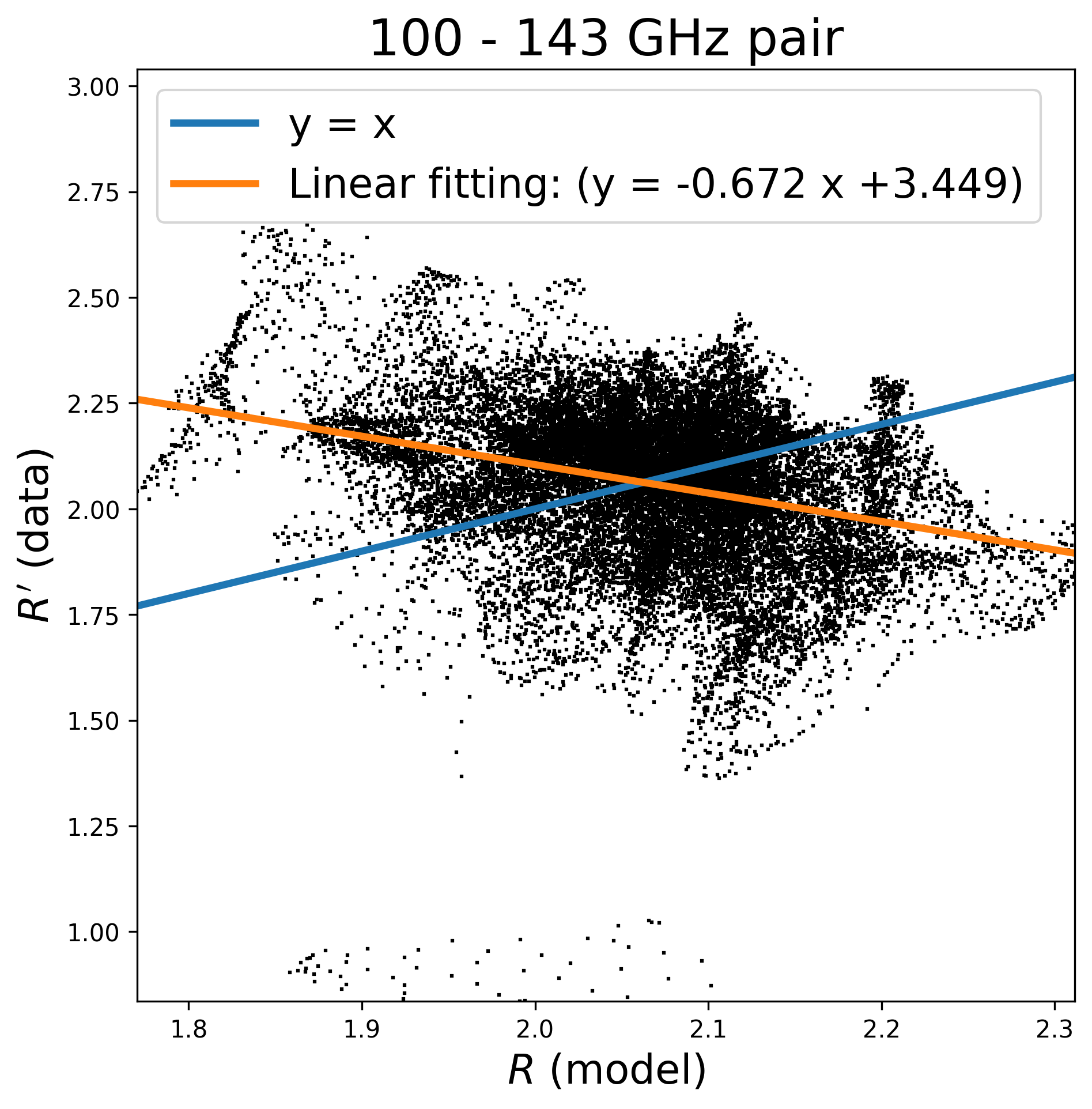}\\
\includegraphics[width=0.32\textwidth]{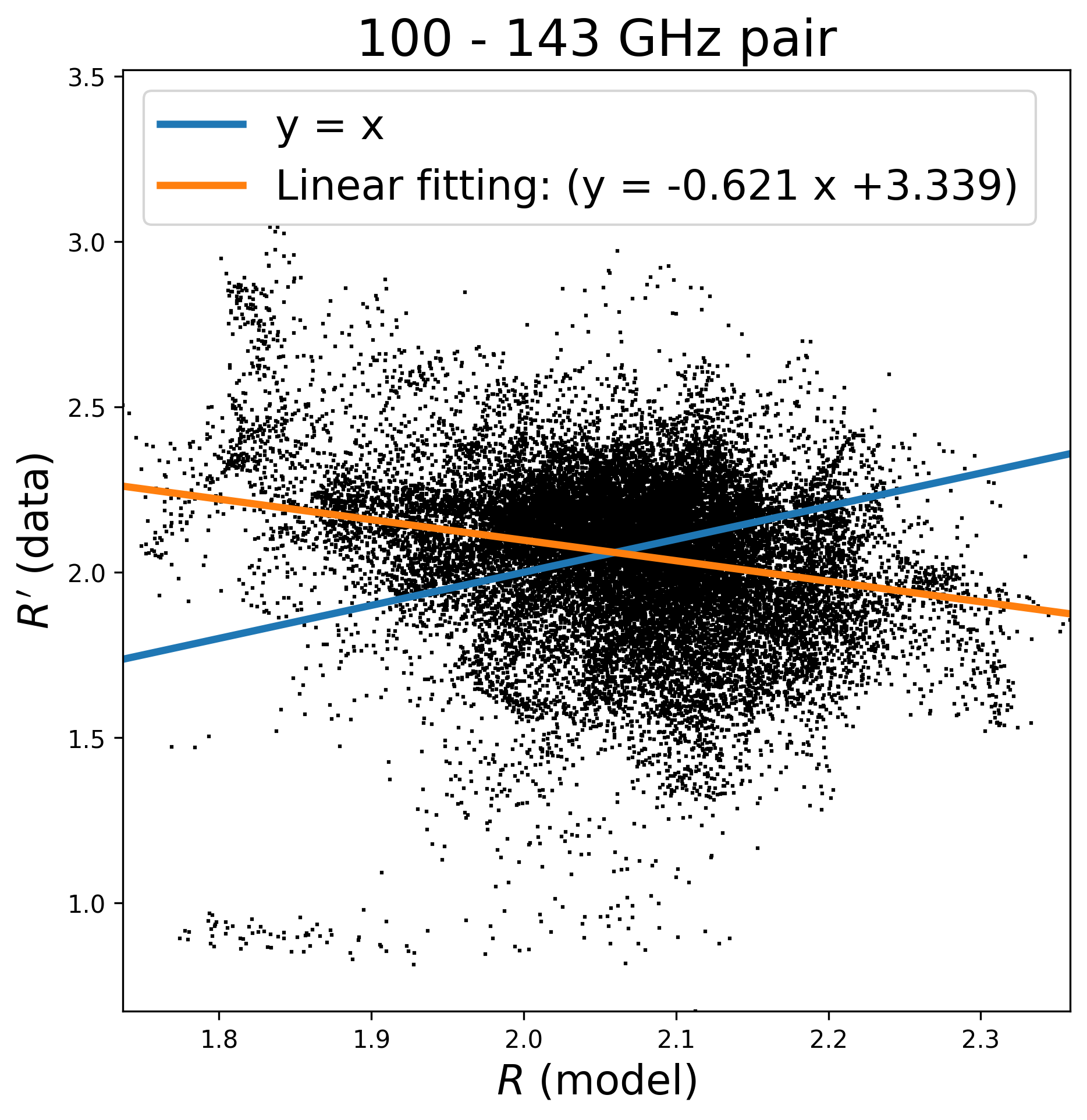}
\includegraphics[width=0.32\textwidth]{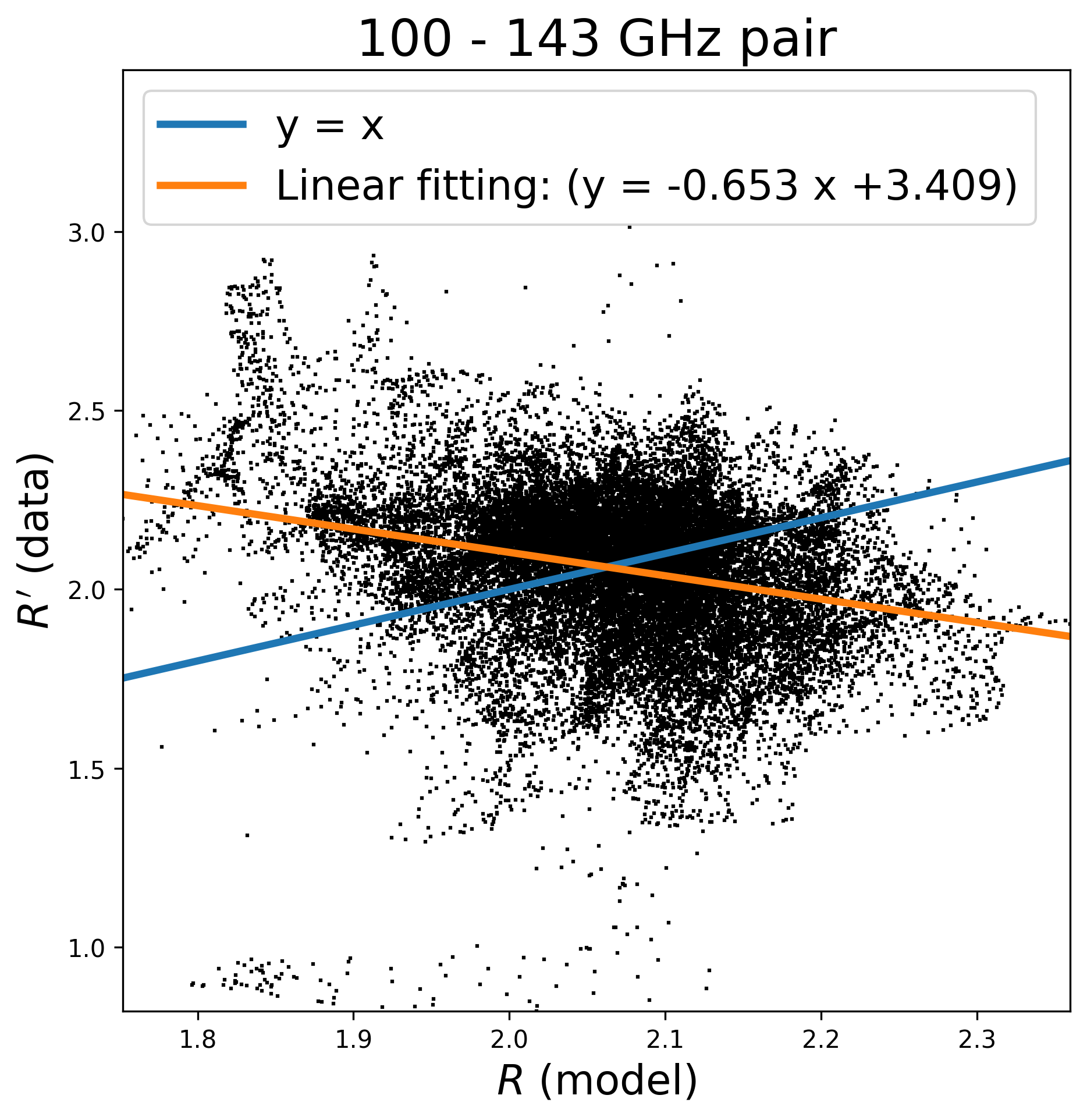}
\includegraphics[width=0.32\textwidth]{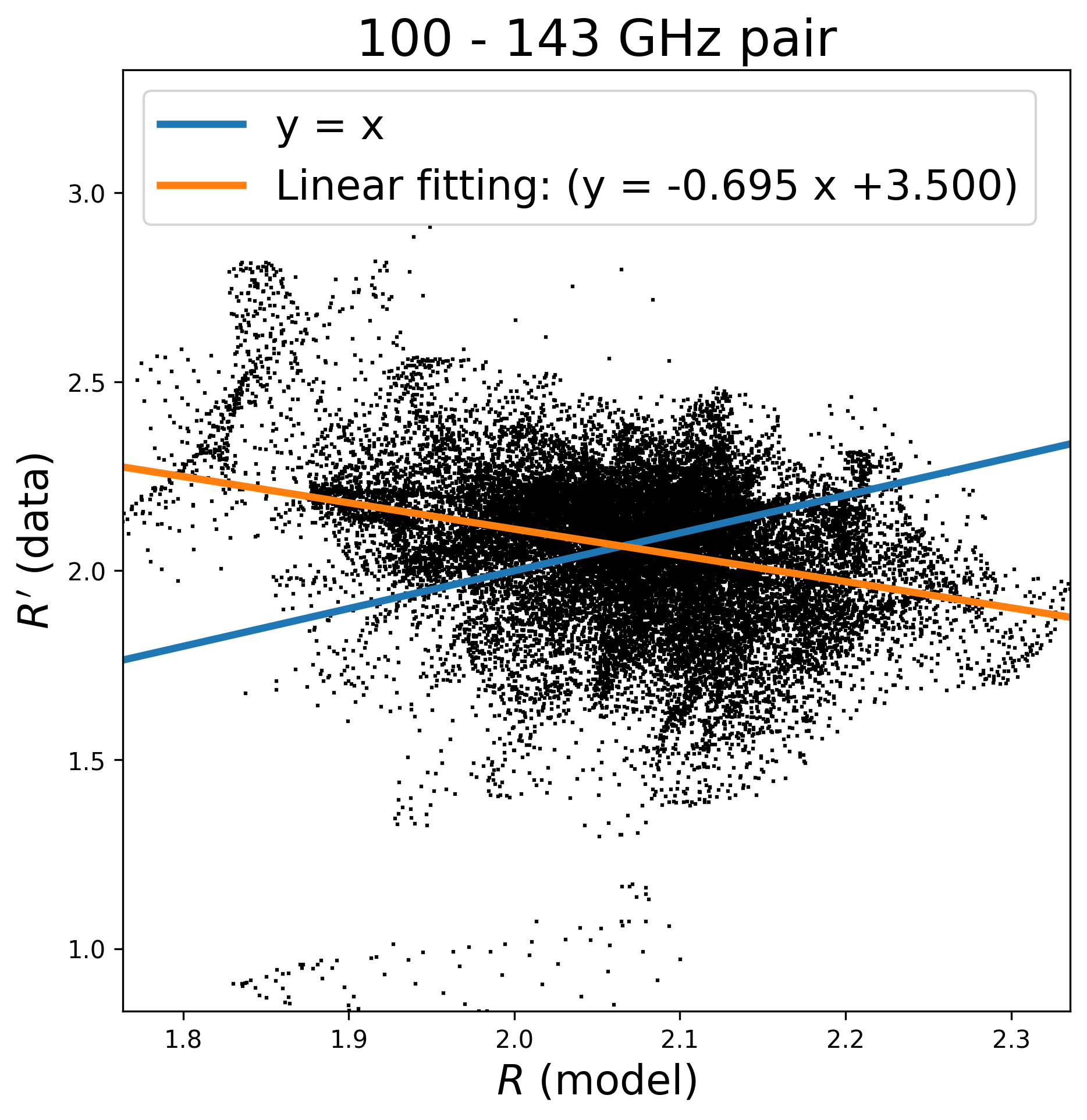}
\caption{Comparison of different smoothing angles and angular radii of mosaic disks for 143-100 GHz pair. 
$M_\mathrm{tot} = M_\mathrm{comp} \times M_{80}$. 
\textit{From left to right}: mosaic disks with angular radius of $5^\circ$, $6^\circ$, and $7^\circ$. 
\textit{From top to bottom}: smoothing angle of $1^\circ$, $2^\circ$, and $3^\circ$. }
\label{fig:scatter plots with different smoothing and disk angle 143-100}
\end{figure*}

\begin{figure*}[!htb]
\centering
\includegraphics[width=0.32\textwidth]{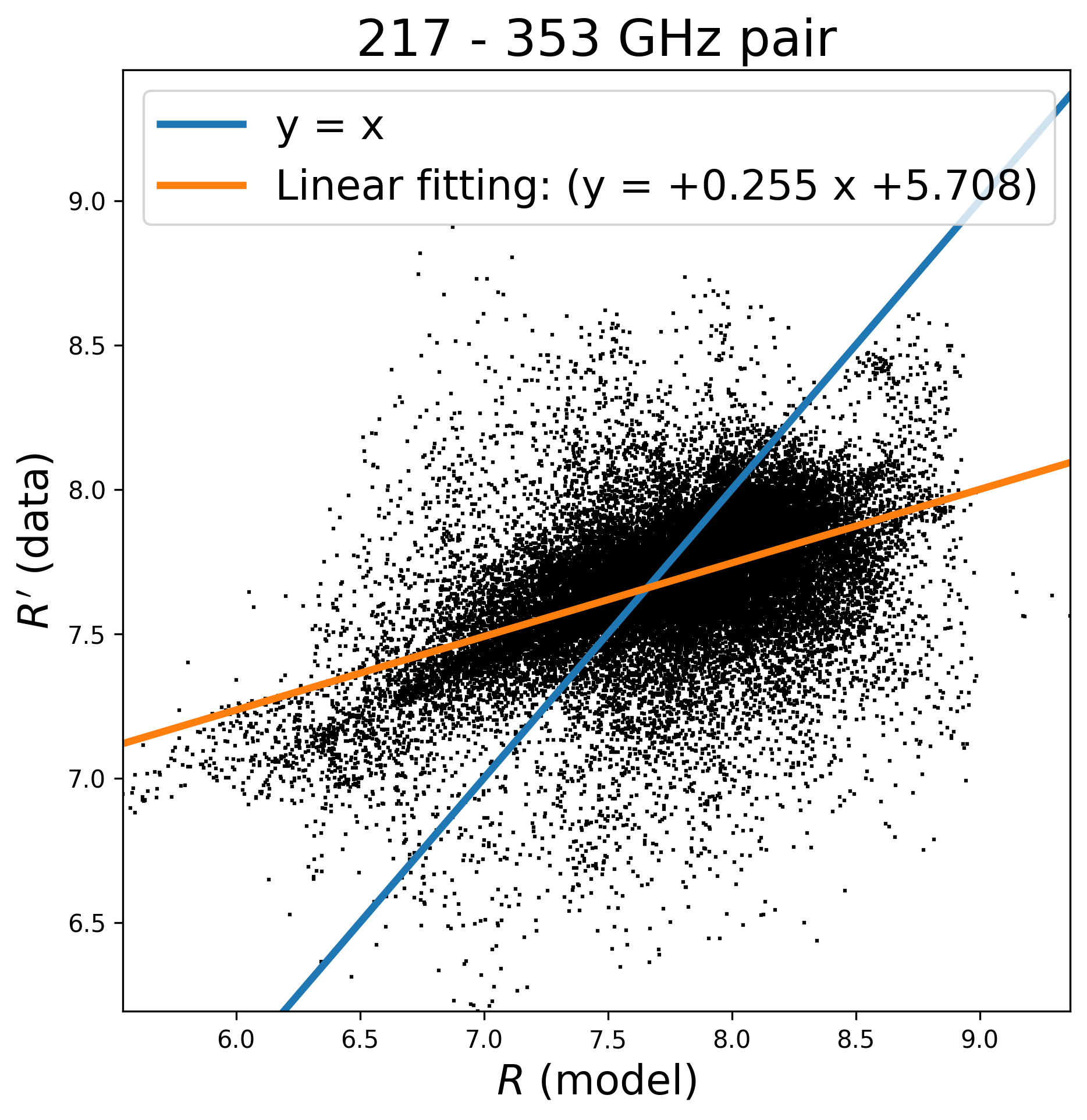}
\includegraphics[width=0.32\textwidth]{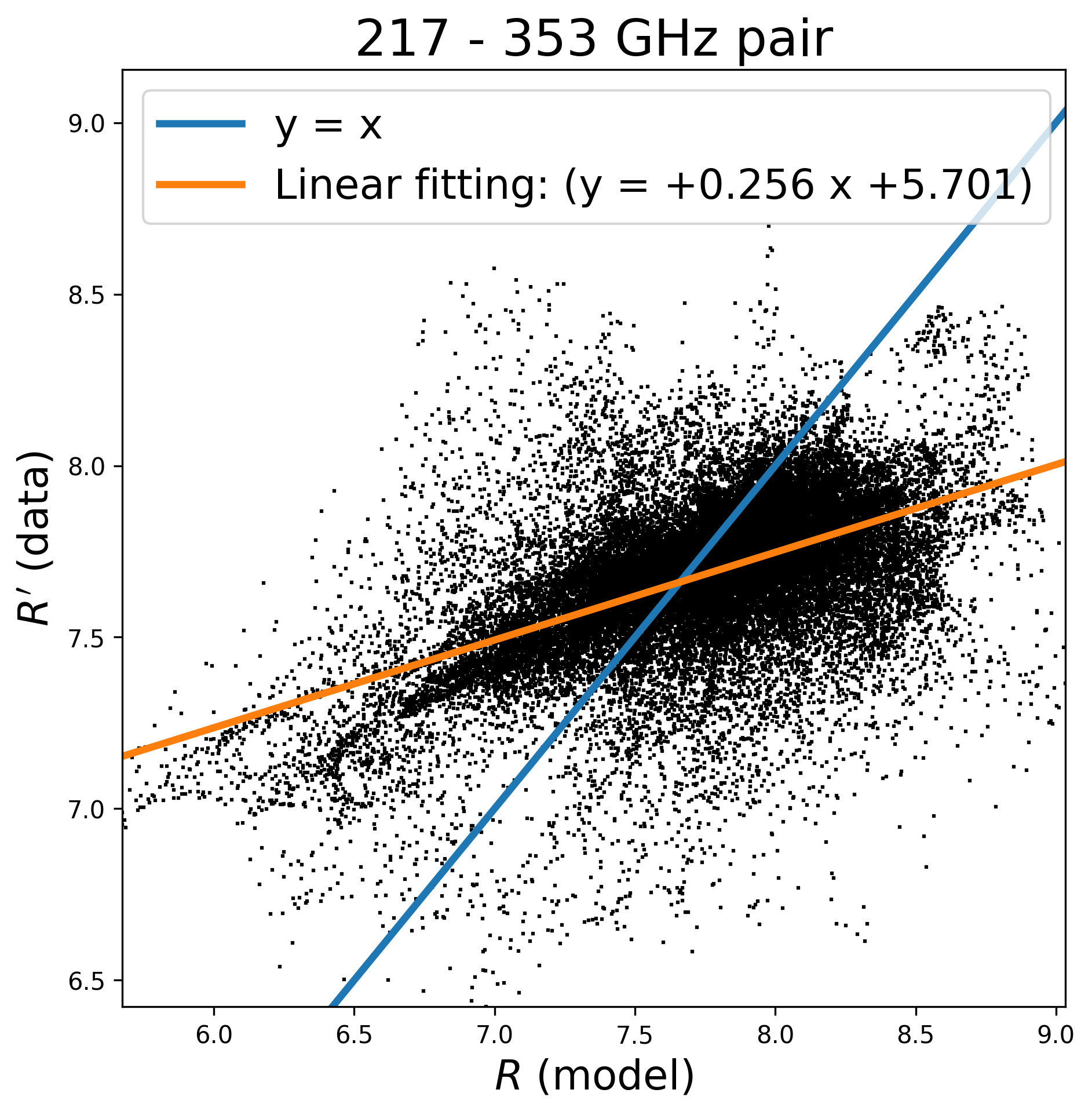}
\includegraphics[width=0.32\textwidth]{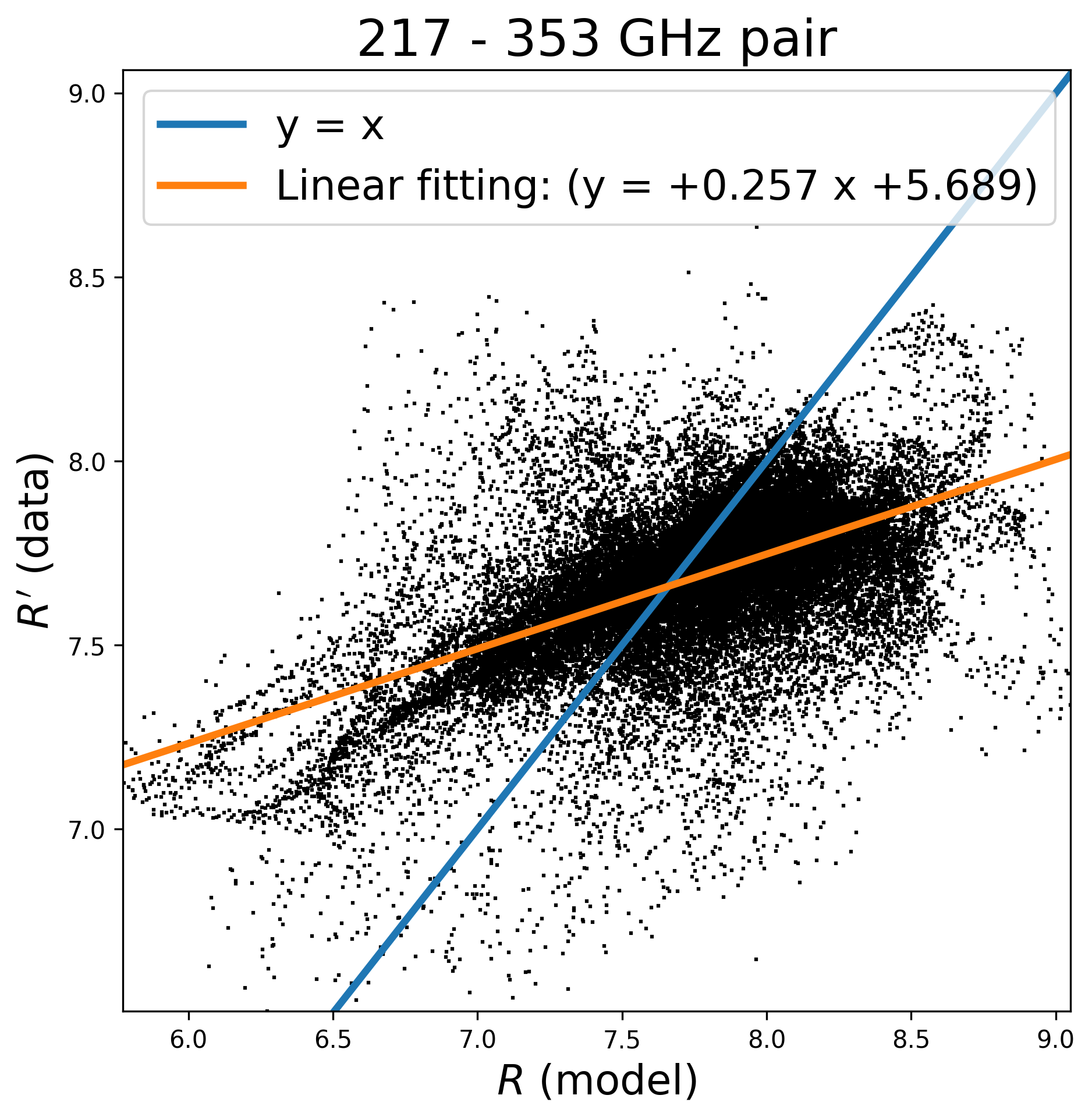}\\
\includegraphics[width=0.32\textwidth]{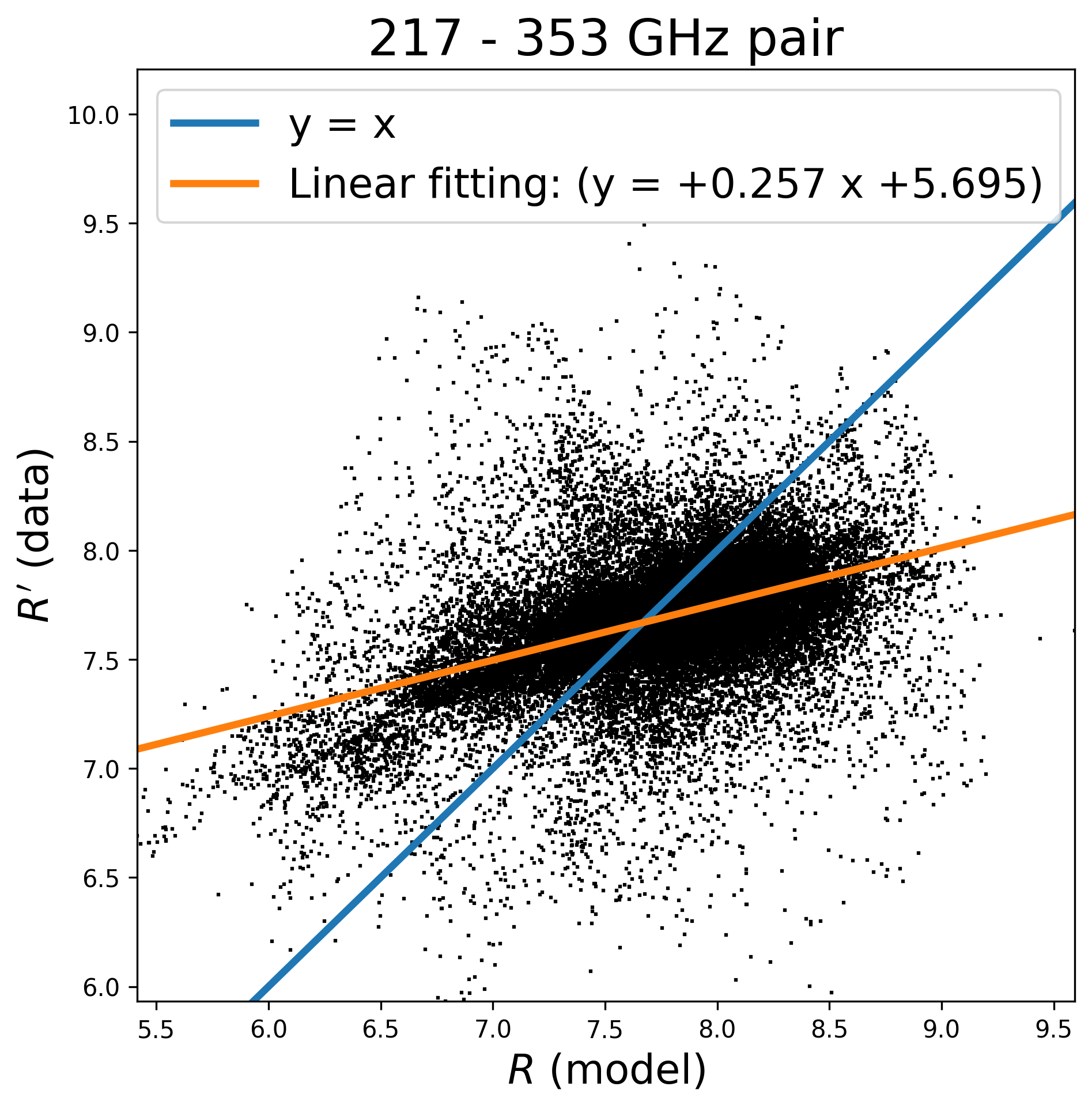}
\includegraphics[width=0.32\textwidth]{Scatter_model_2015_color_correction_yes_2_galactic_mask_80_smooth_degree_2_disk_degree_6_low_galac_mask_no_zodiacal_mask_no.png}
\includegraphics[width=0.32\textwidth]{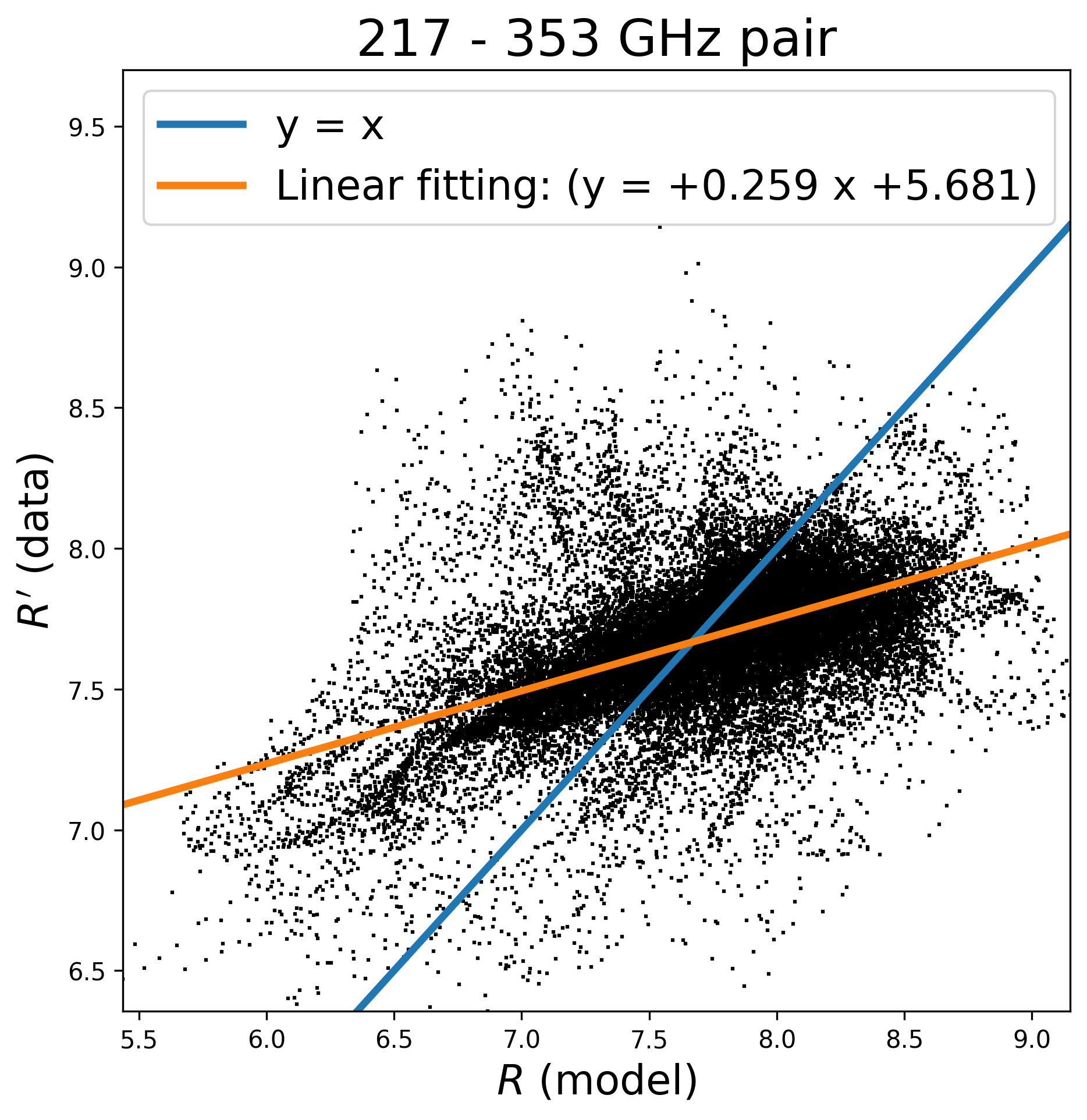}\\
\includegraphics[width=0.32\textwidth]{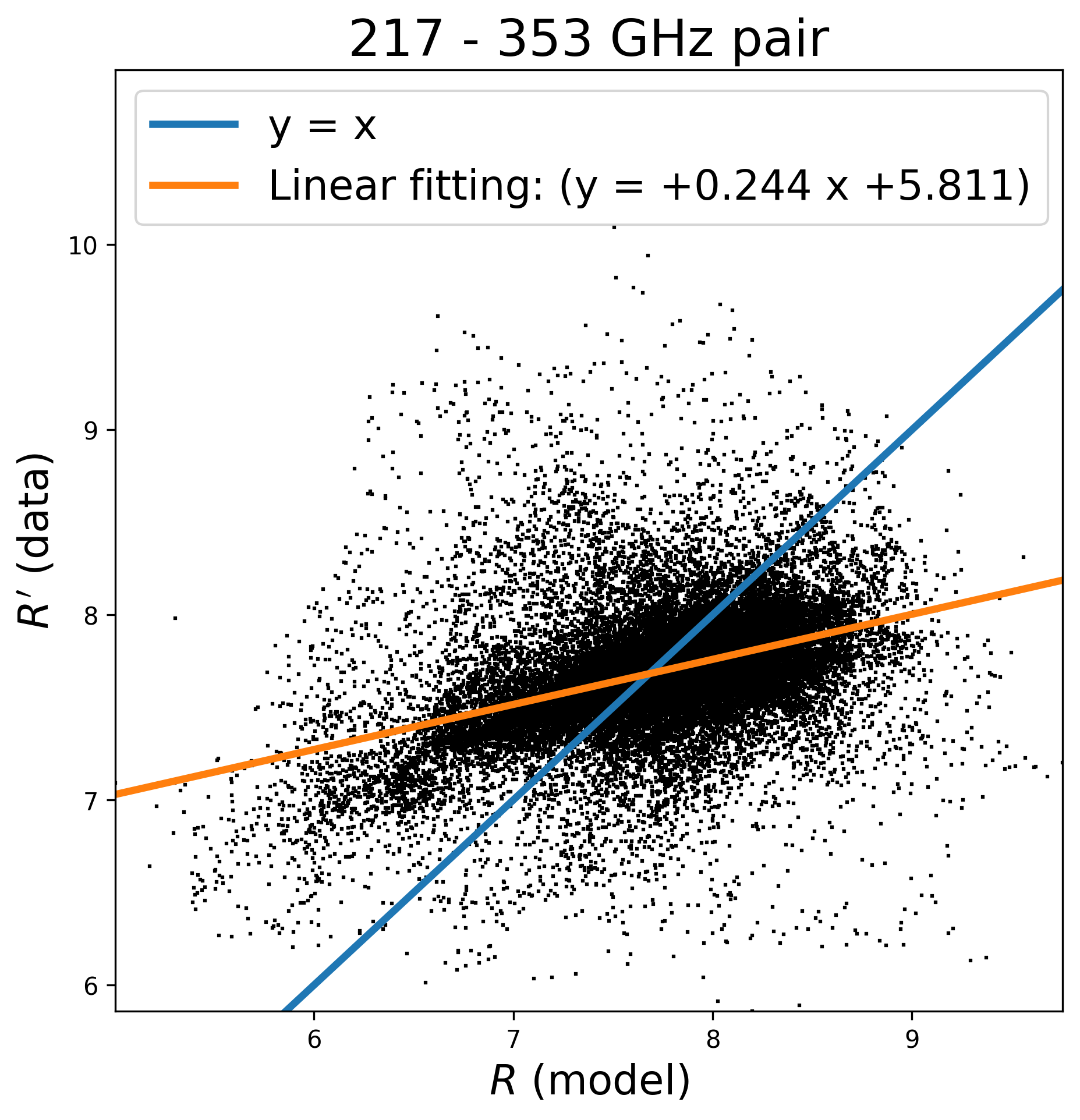}
\includegraphics[width=0.32\textwidth]{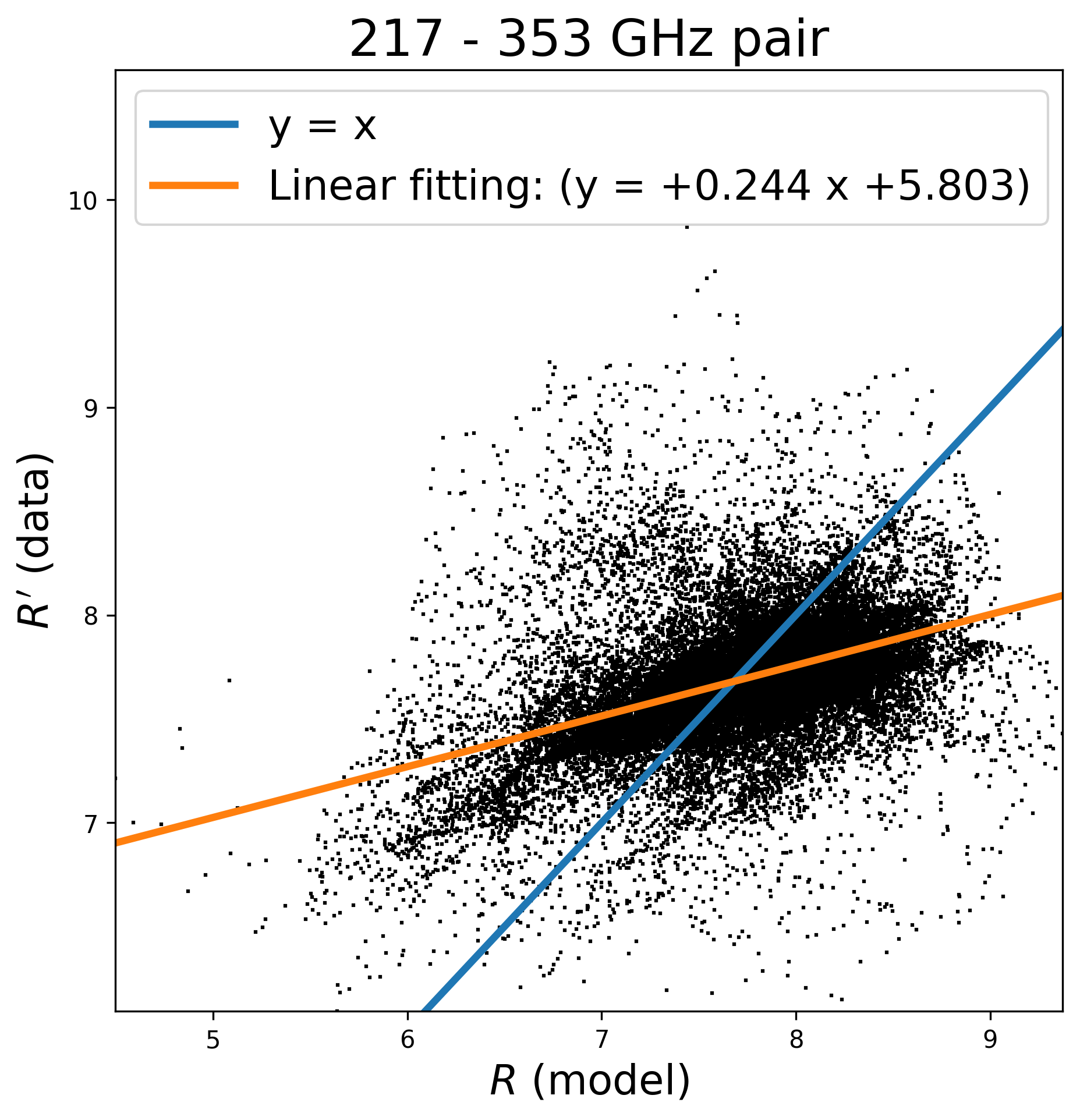}
\includegraphics[width=0.32\textwidth]{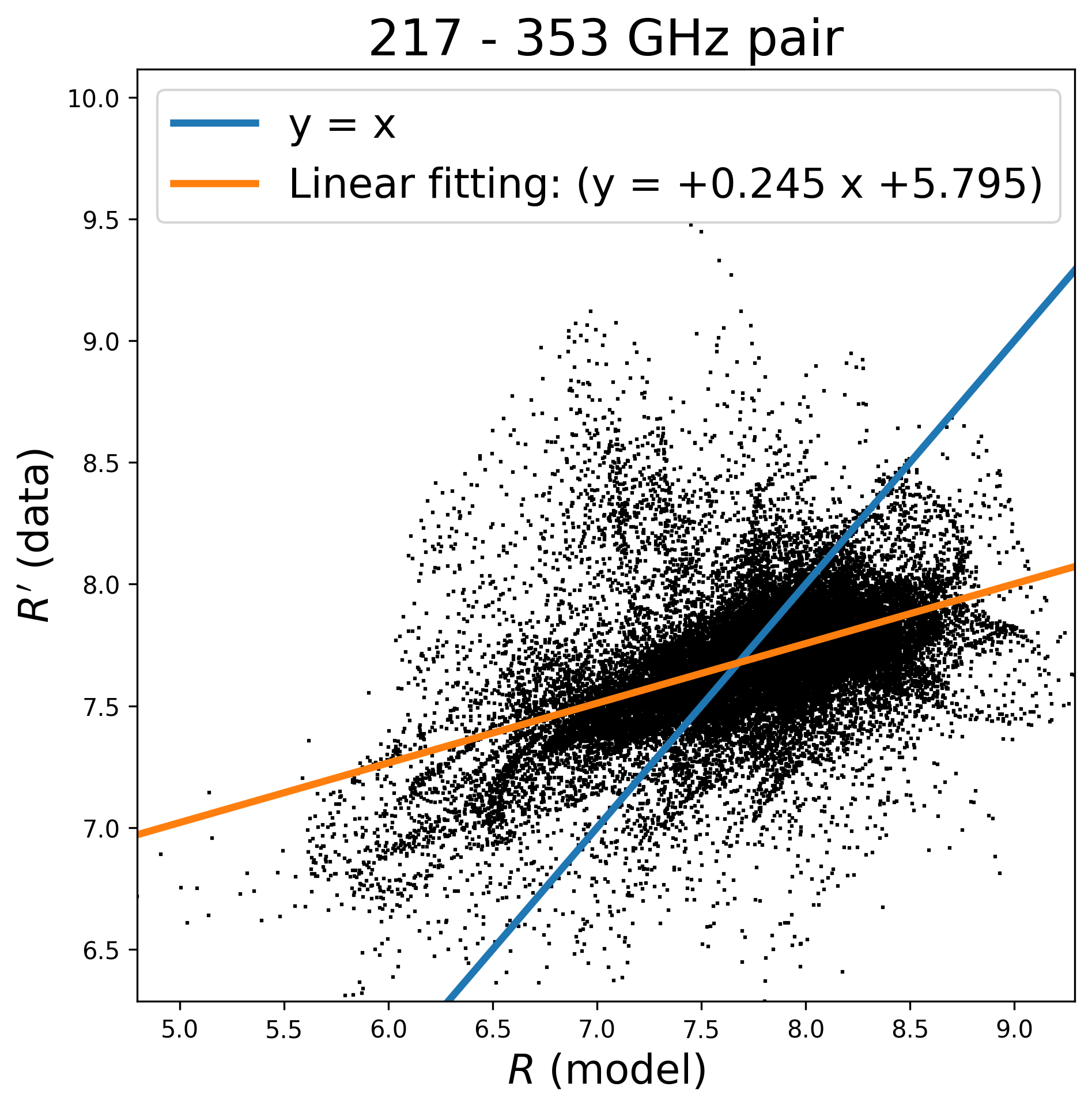}
\caption{Comparison of different smoothing angles and angular radii of mosaic disks for 217 - 353 GHz pair. 
$M_\mathrm{tot} = M_\mathrm{comp} \times M_{80}$. 
\textit{From left to right}: mosaic disks with angular radius of $5^\circ$, $6^\circ$, and $7^\circ$. 
\textit{From top to bottom}: smoothing angle of $1^\circ$, $2^\circ$, and $3^\circ$. }
\label{fig:scatter plots with different smoothing and disk angle 353-217}
\end{figure*}

\begin{figure*}[!htb]
\centering
\includegraphics[width=0.32\textwidth]{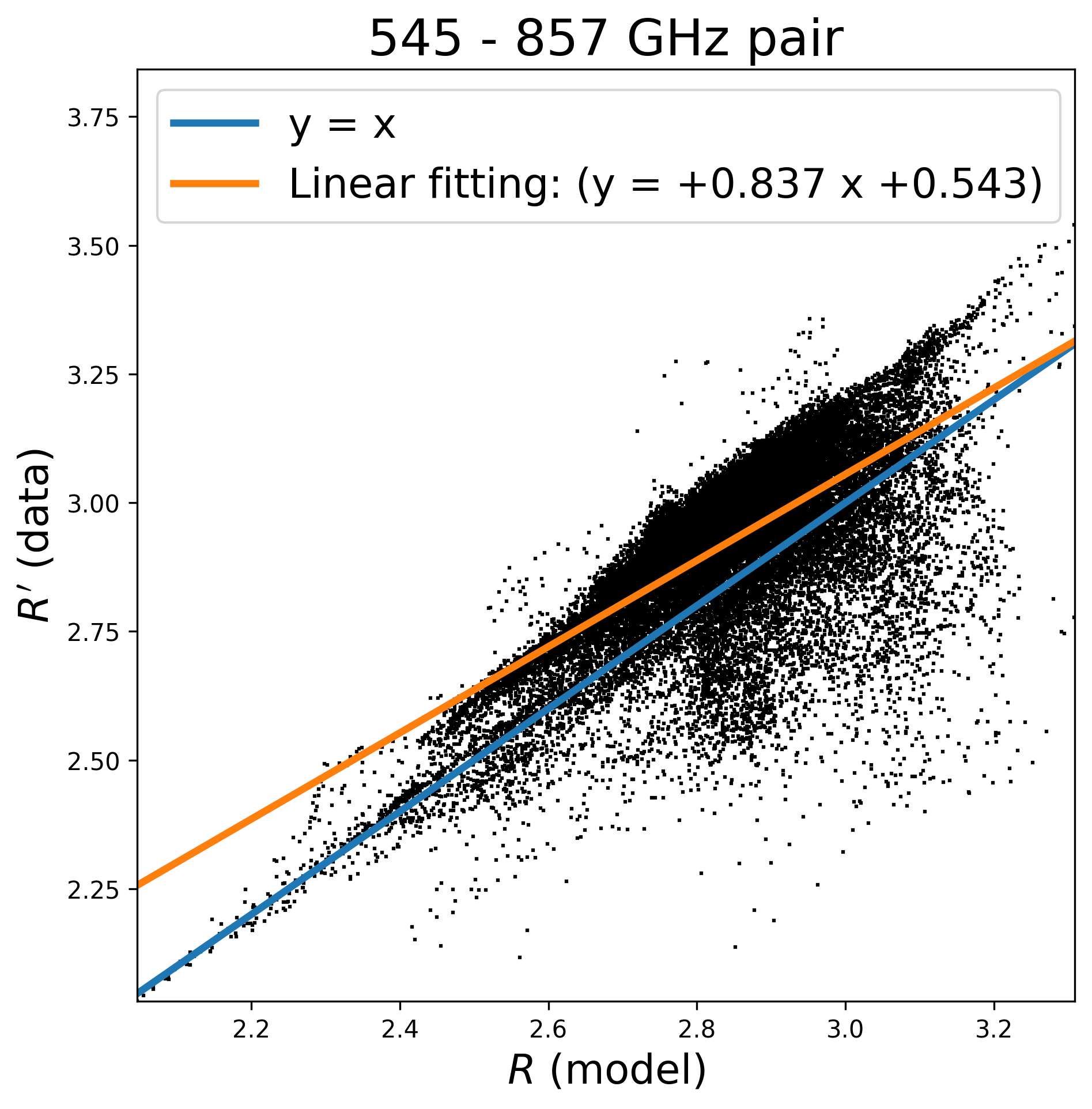}
\includegraphics[width=0.32\textwidth]{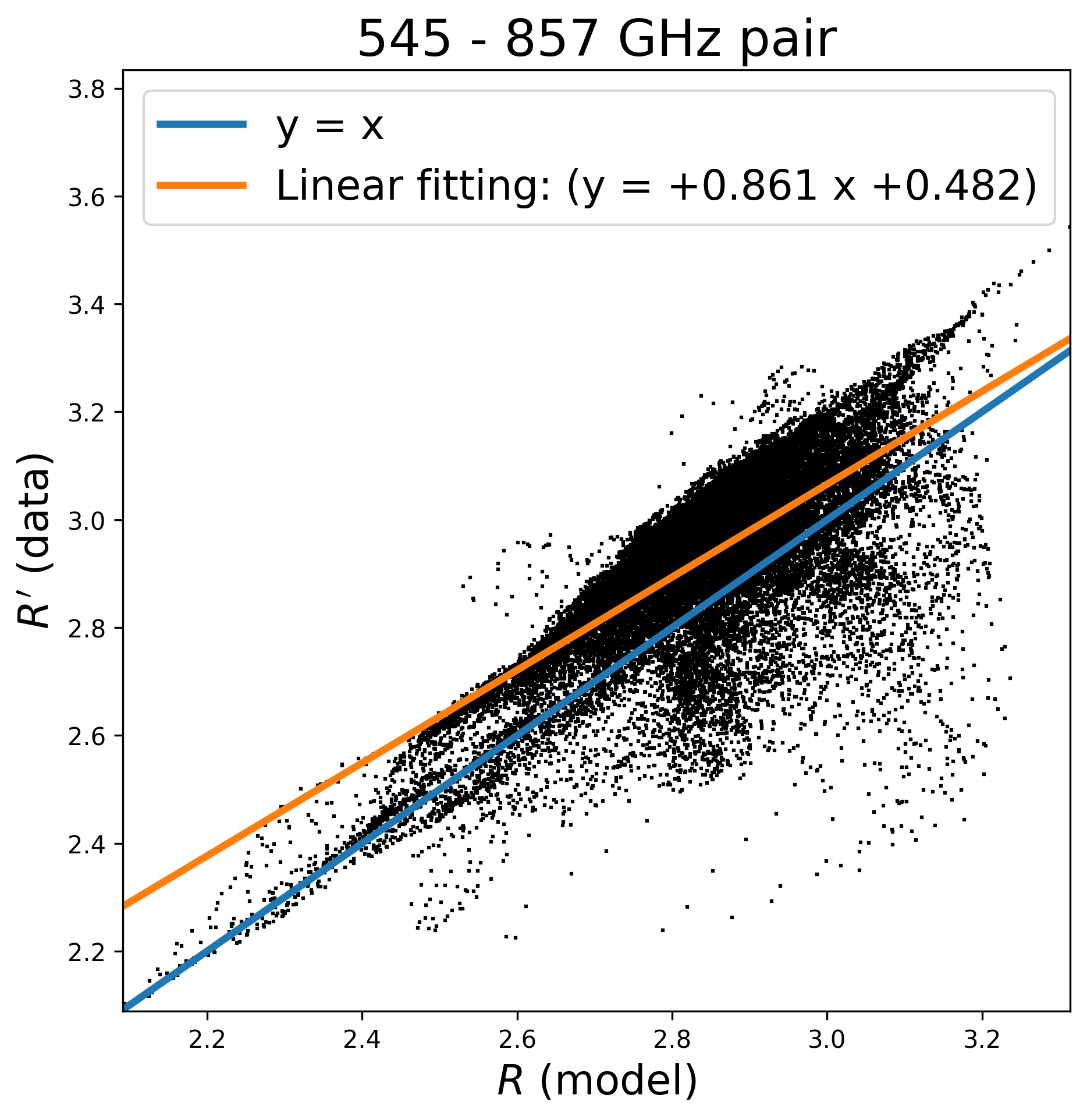}
\includegraphics[width=0.32\textwidth]{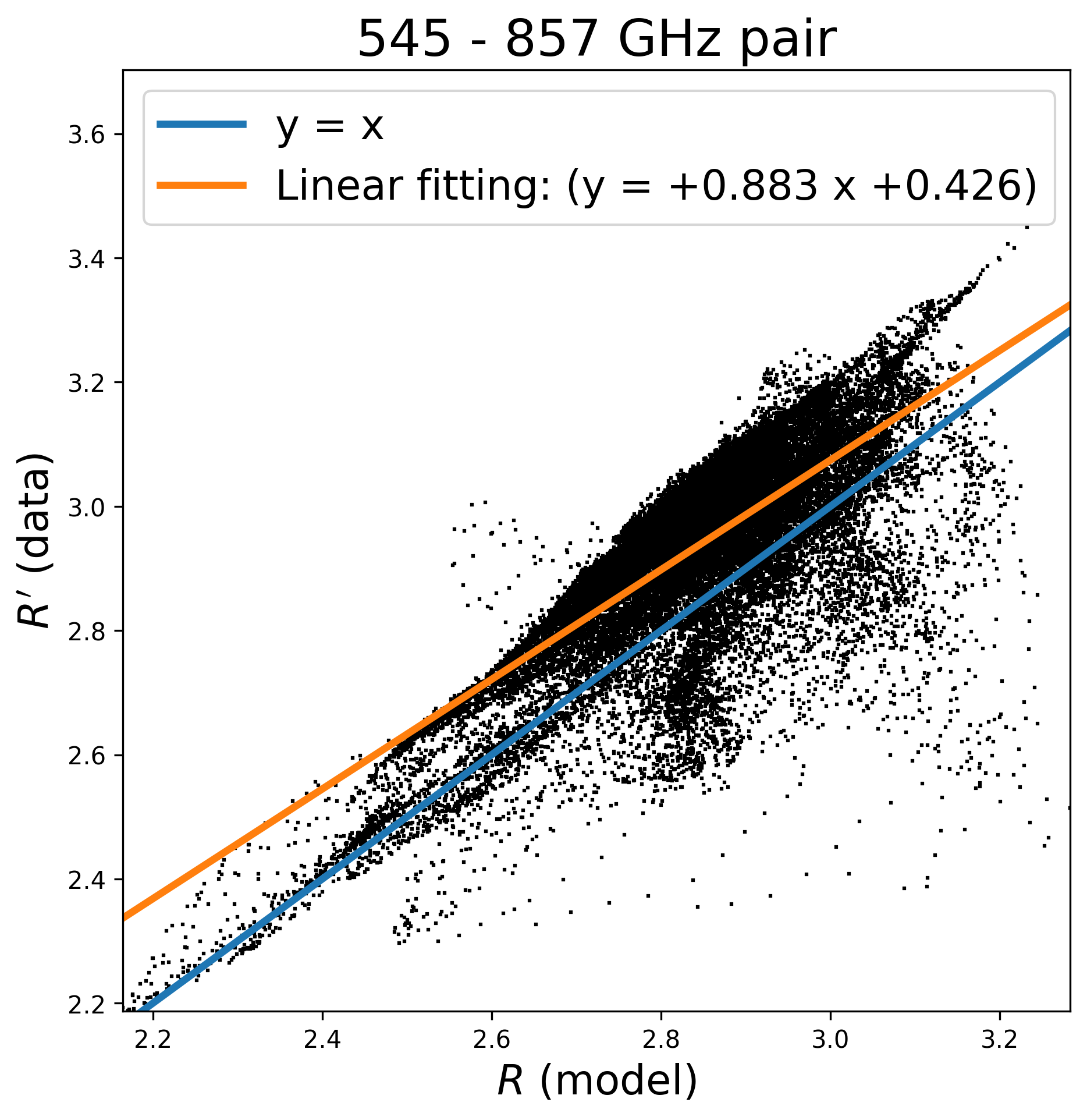}\\
\includegraphics[width=0.32\textwidth]{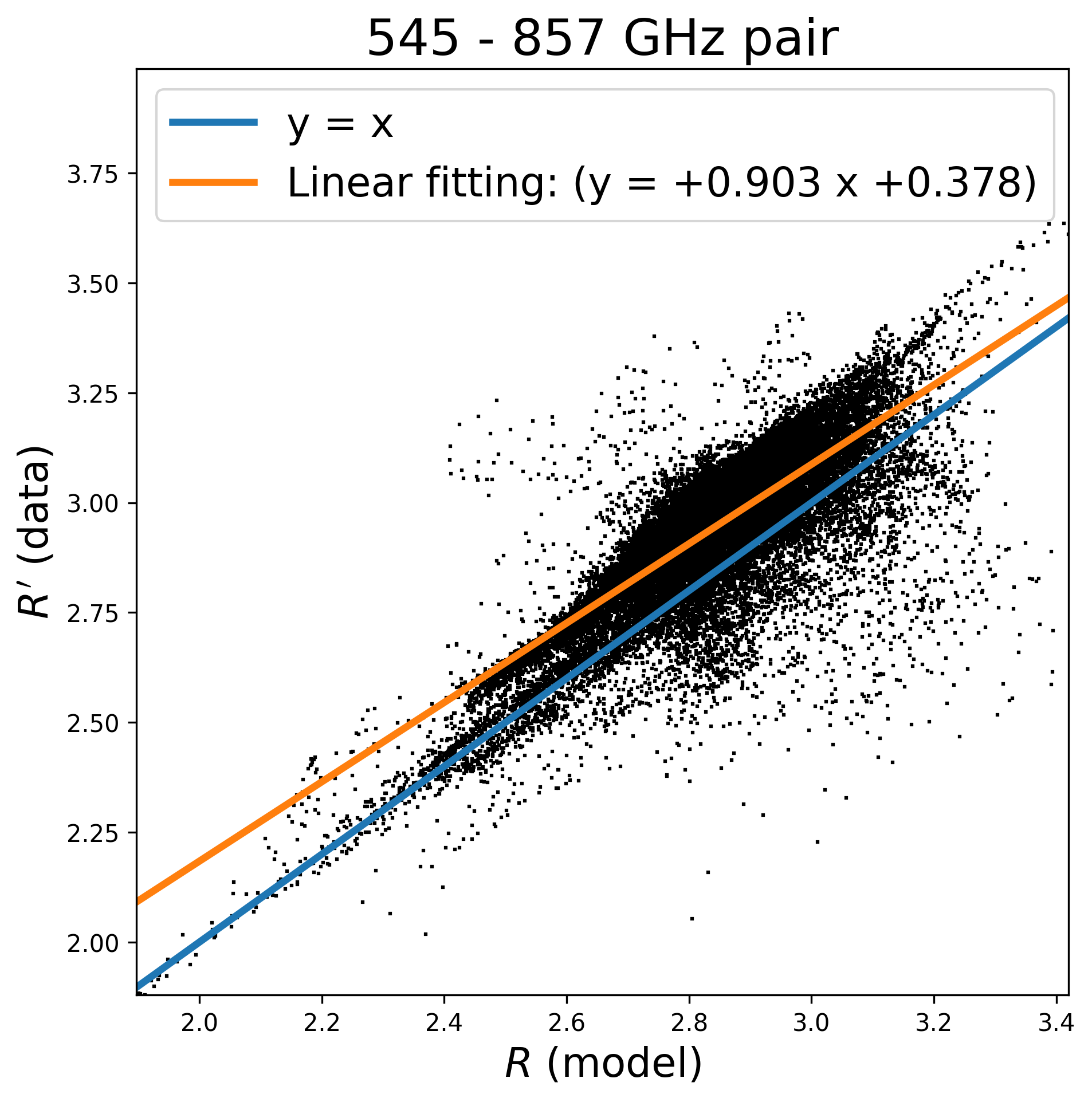}
\includegraphics[width=0.32\textwidth]{Scatter_model_2015_color_correction_yes_3_galactic_mask_80_smooth_degree_2_disk_degree_6_low_galac_mask_no_zodiacal_mask_no.png}
\includegraphics[width=0.32\textwidth]{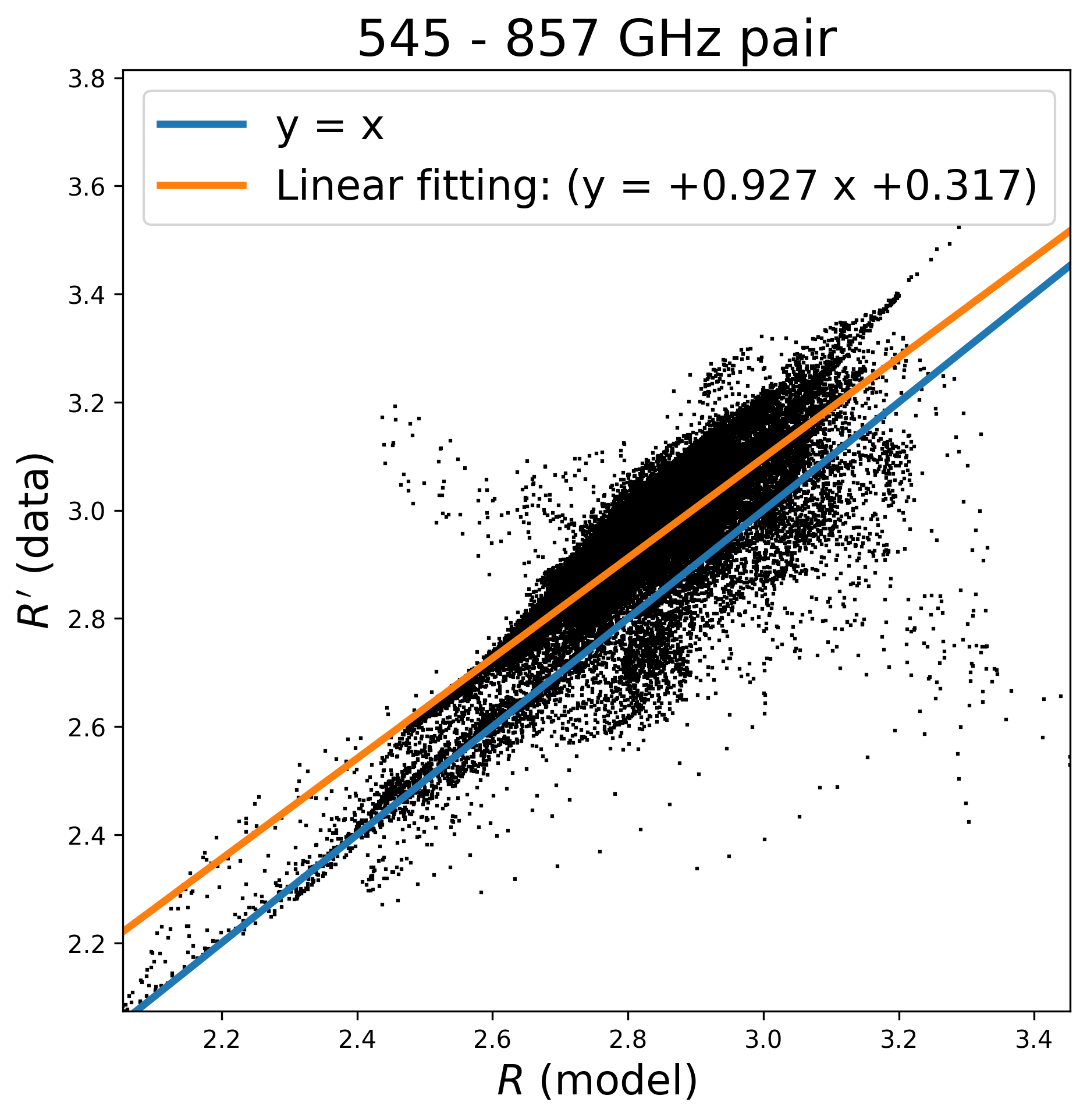}\\
\includegraphics[width=0.32\textwidth]{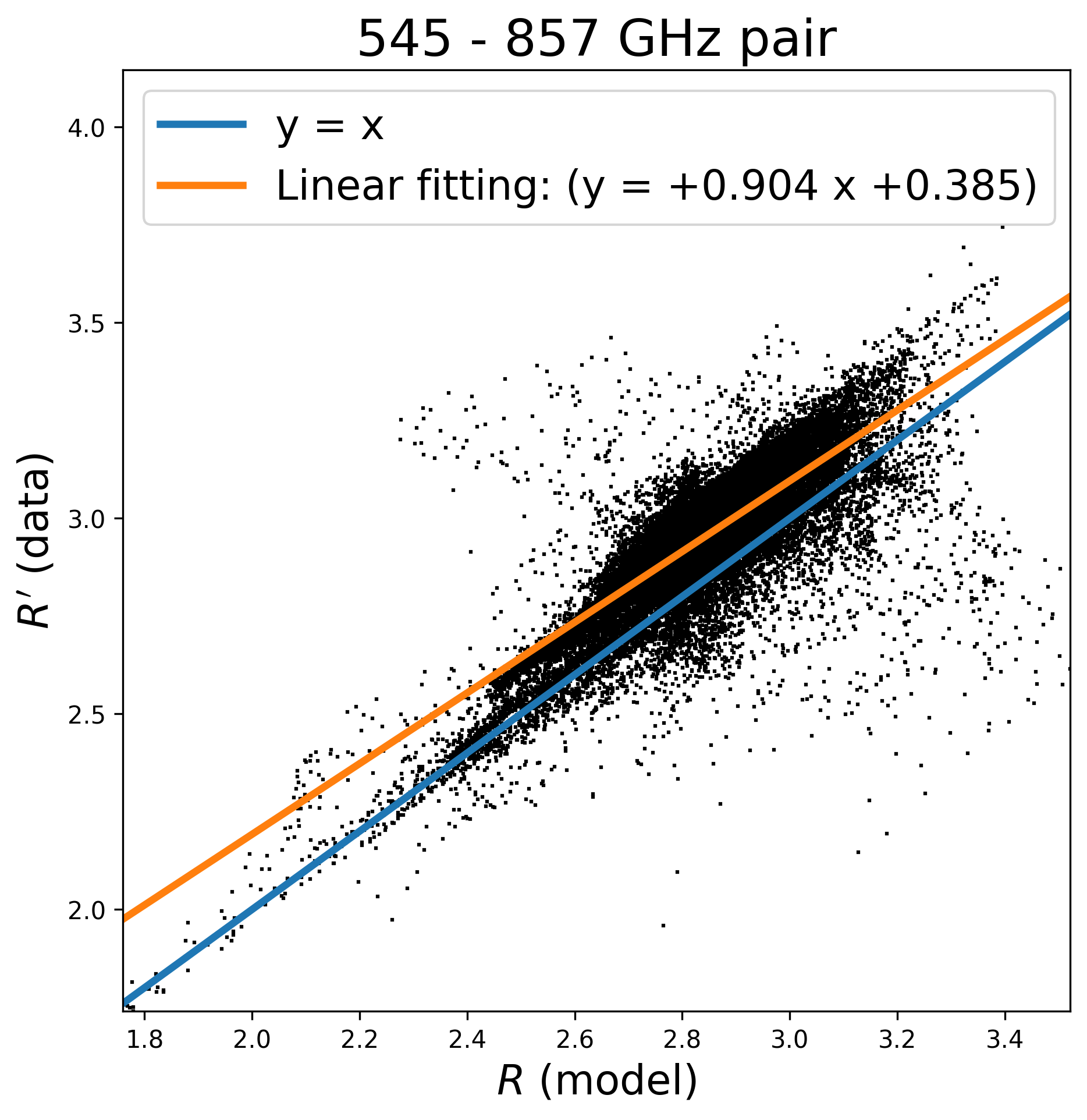}
\includegraphics[width=0.32\textwidth]{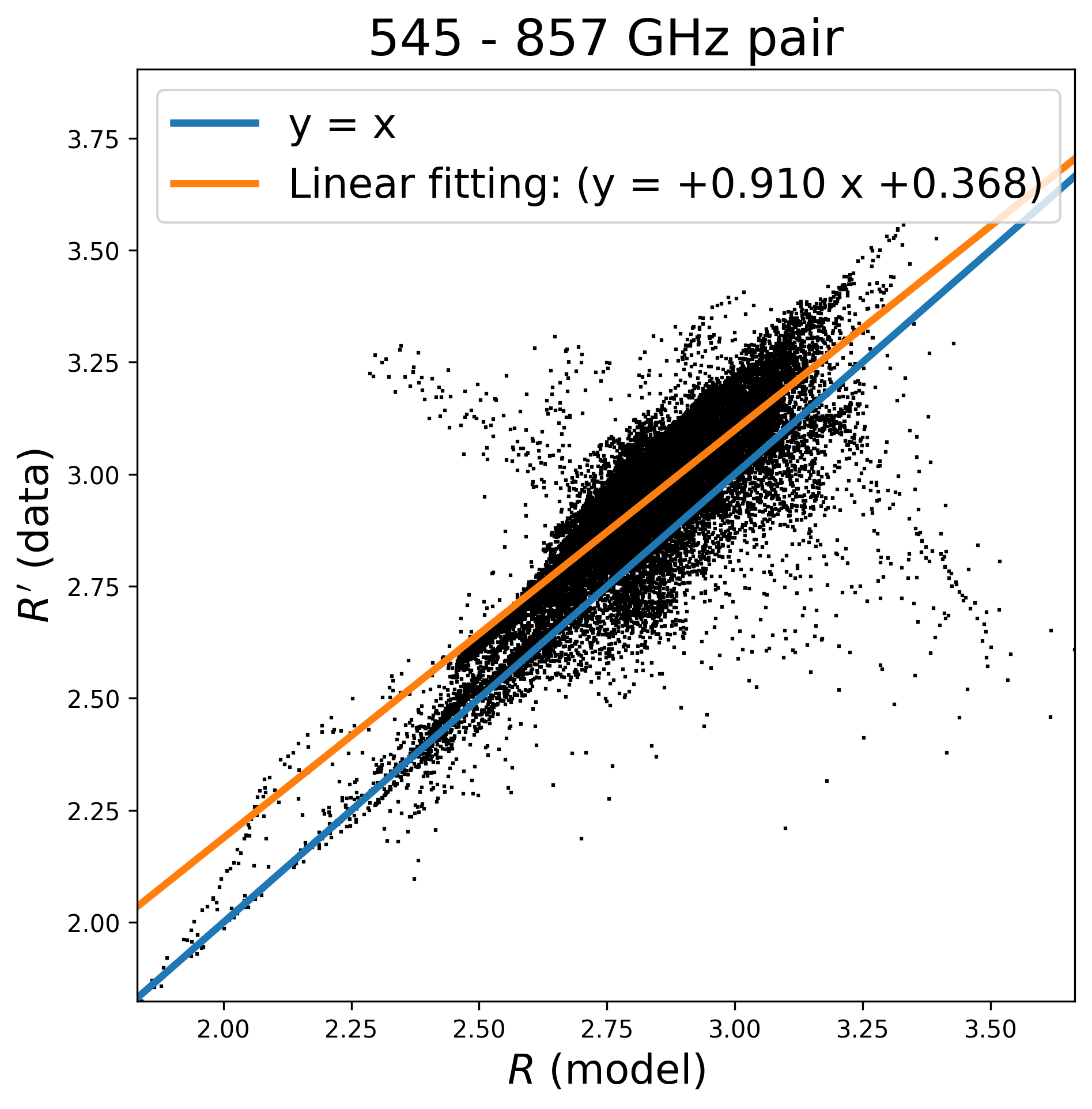}
\includegraphics[width=0.32\textwidth]{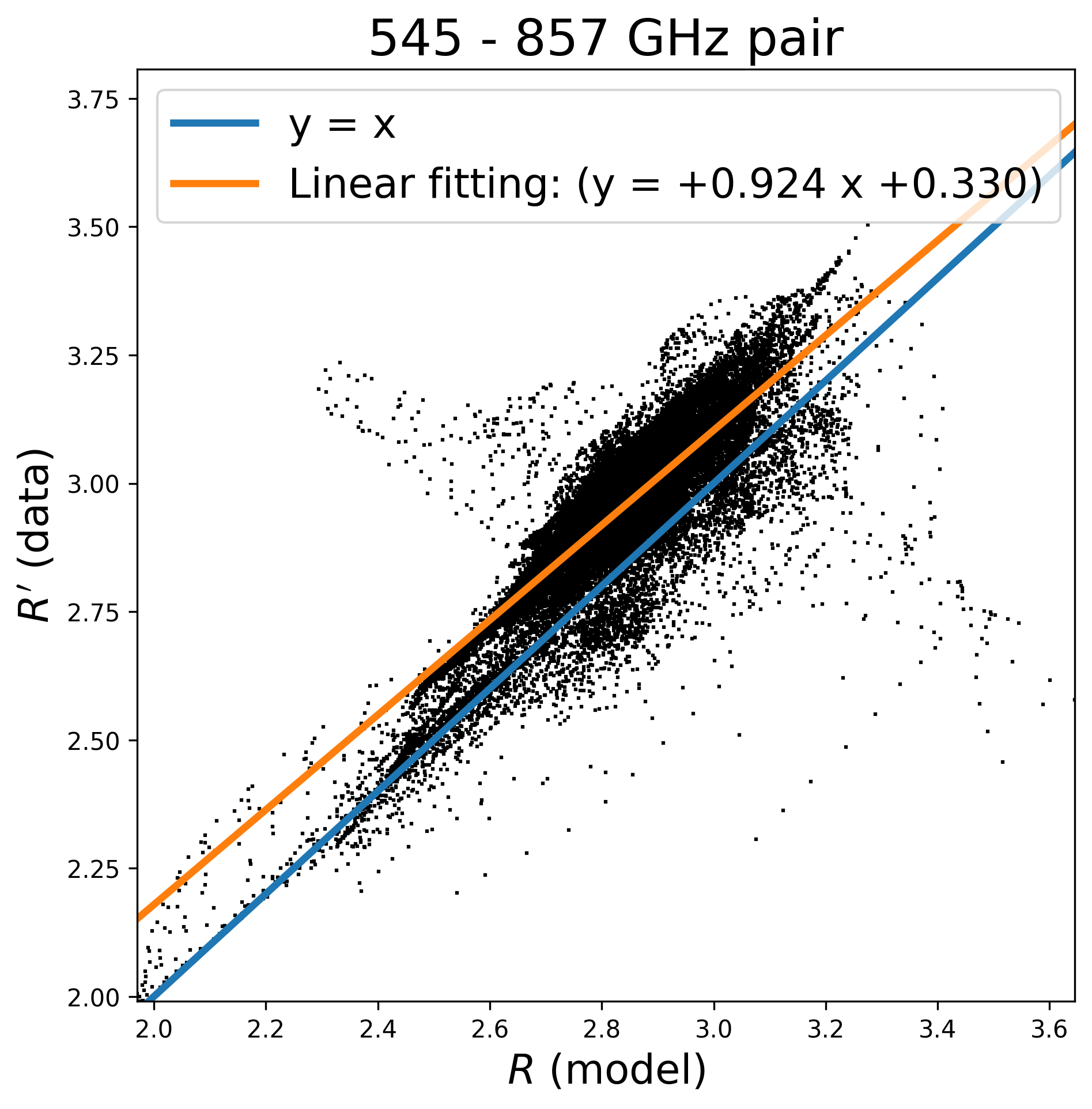}
\caption{Comparison of different smoothing angles and angular radii of mosaic disks for 545 - 857 GHz pair. 
$M_\mathrm{tot} = M_\mathrm{comp} \times M_{80}$. 
\textit{From left to right}: mosaic disks with angular radius of $5^\circ$, $6^\circ$, and $7^\circ$. 
\textit{From top to bottom}: smoothing angle of $1^\circ$, $2^\circ$, and $3^\circ$. }
\label{fig:scatter plots with different smoothing and disk angle 857-545}
\end{figure*}

\begin{figure*}[!htb]
\centering
\includegraphics[width=0.32\textwidth]{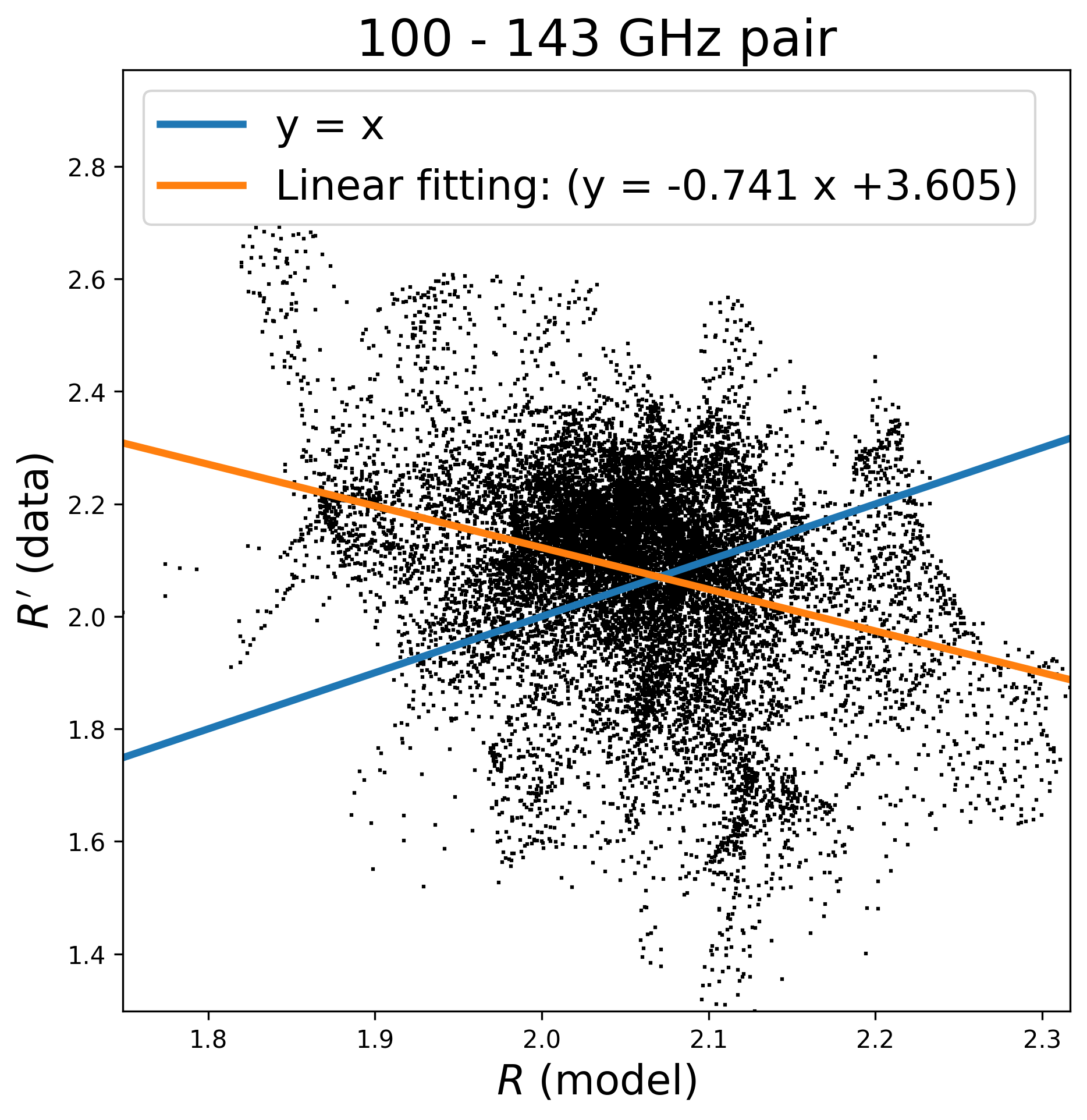}
\includegraphics[width=0.32\textwidth]{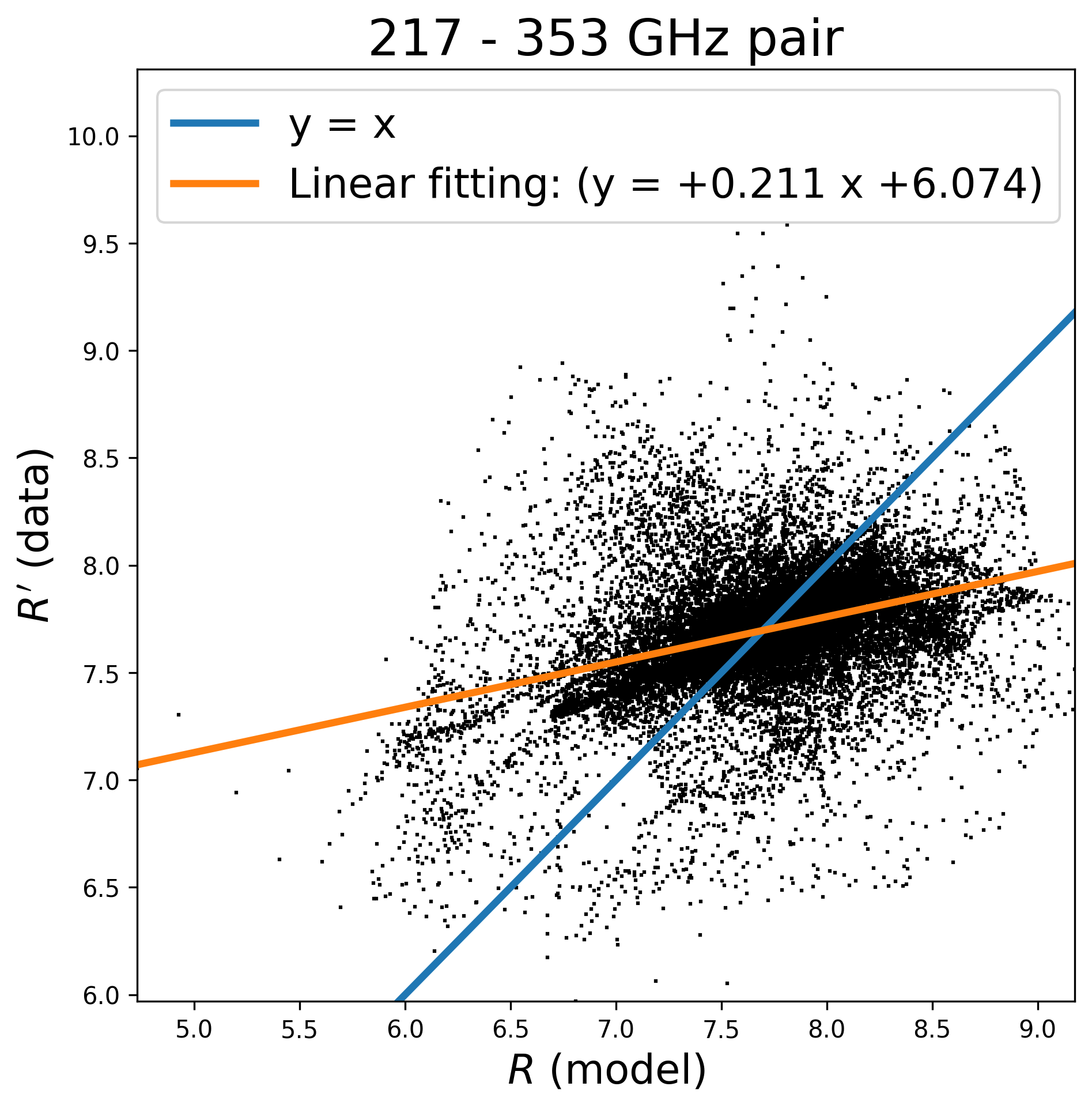}
\includegraphics[width=0.32\textwidth]{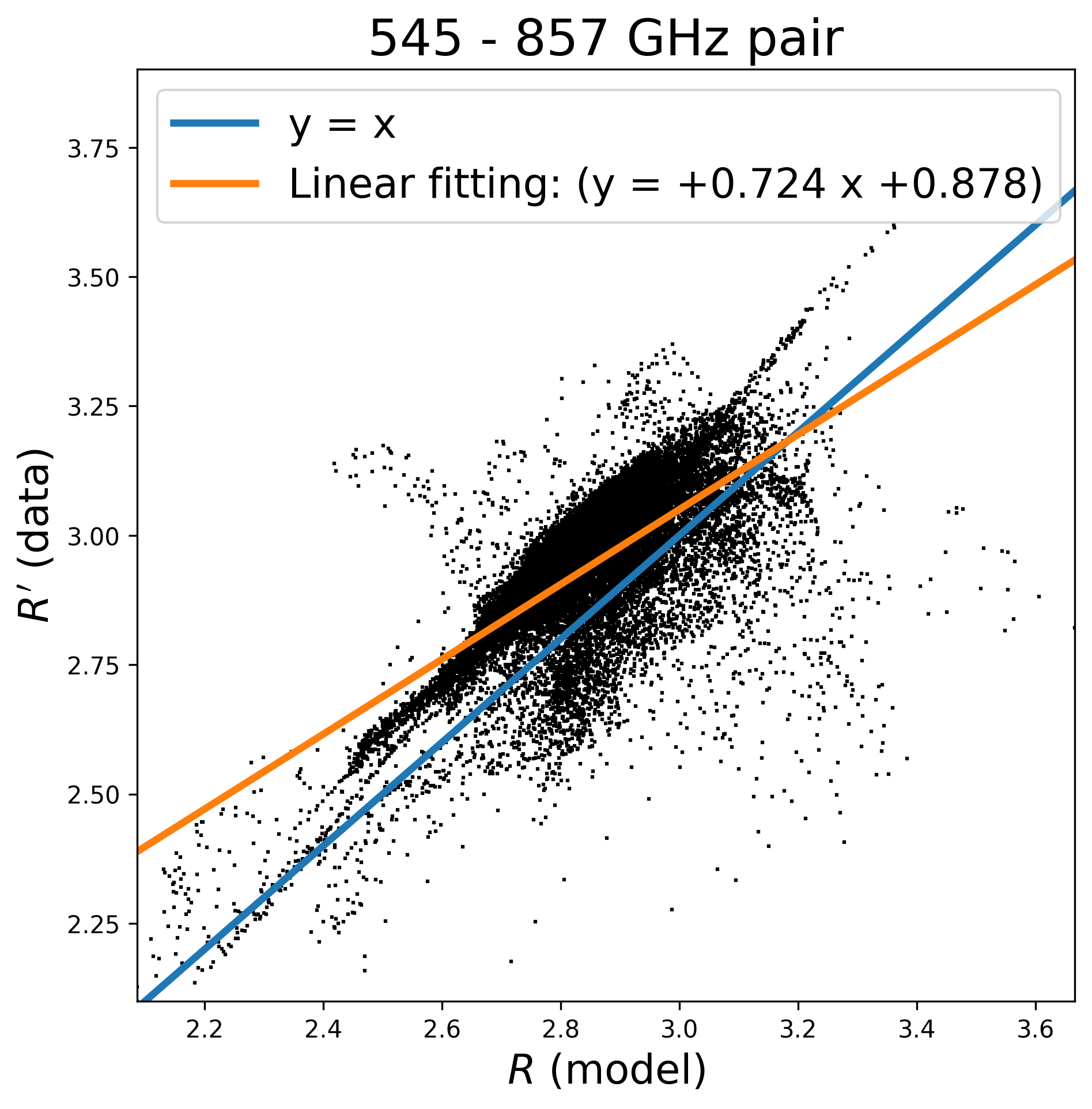}\\
\includegraphics[width=0.32\textwidth]{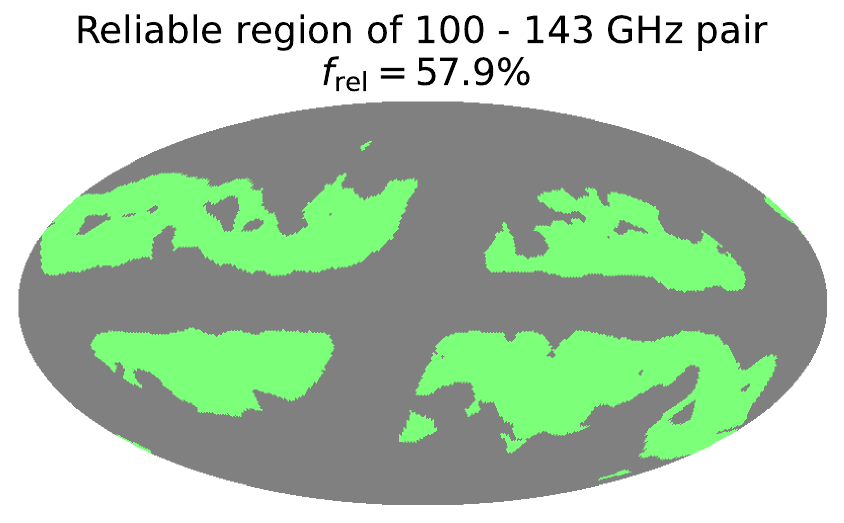}
\includegraphics[width=0.32\textwidth]{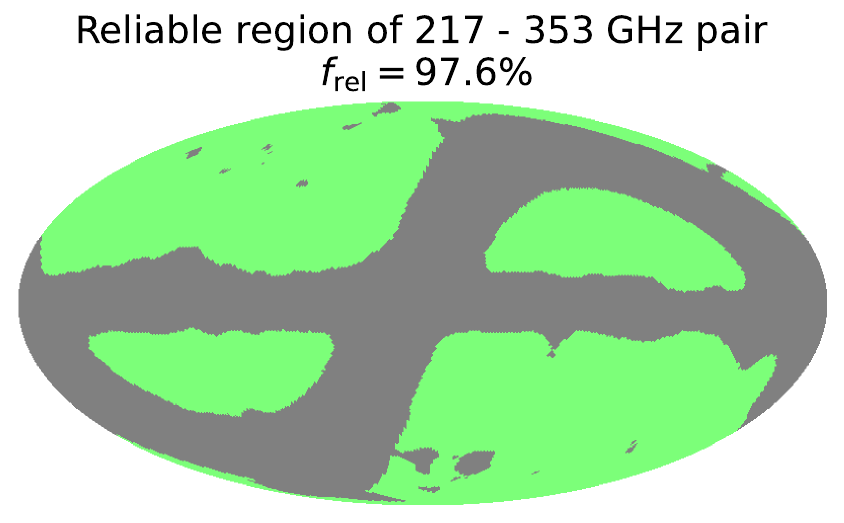}
\includegraphics[width=0.32\textwidth]{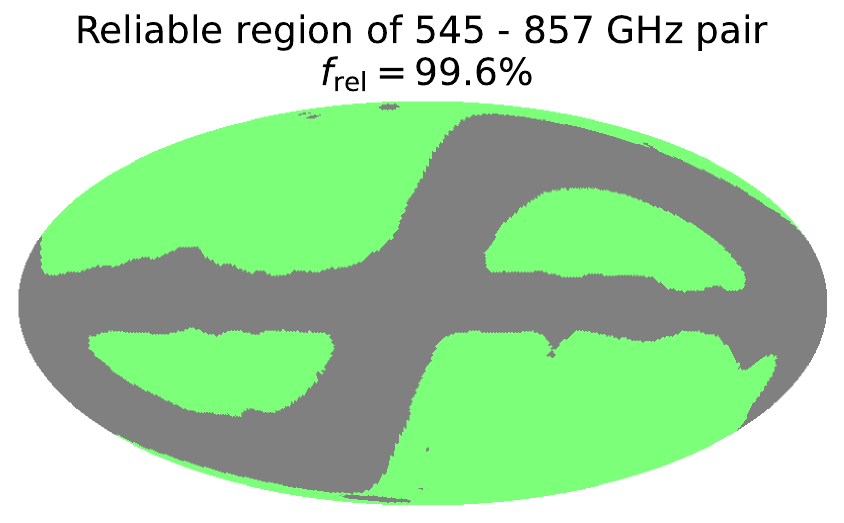}
\caption{Scatter plot of  $R'$ versus $R$, similar to Fig.~\ref{fig:dust_ratio_compare}, but excluding the regions with high zodiacal emission. 
\textit{From left to right}: 100-143, 217-353, and 545-857 GHz pairs. }
\label{fig:dust_ratio_compare_no zodiacal region}
\end{figure*}

\begin{figure*}[!htb]
\centering
\includegraphics[width=0.32\textwidth]{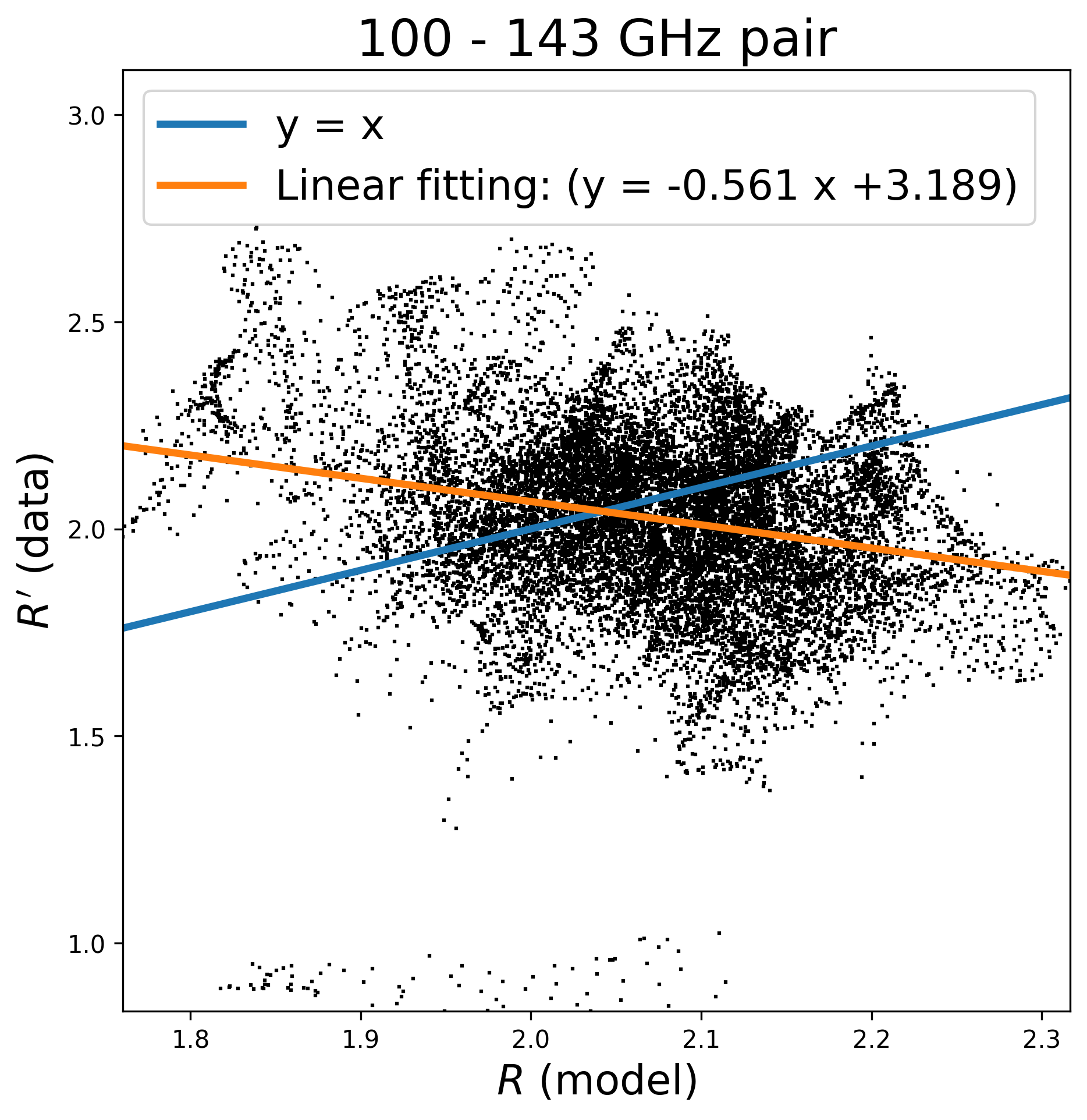}
\includegraphics[width=0.32\textwidth]{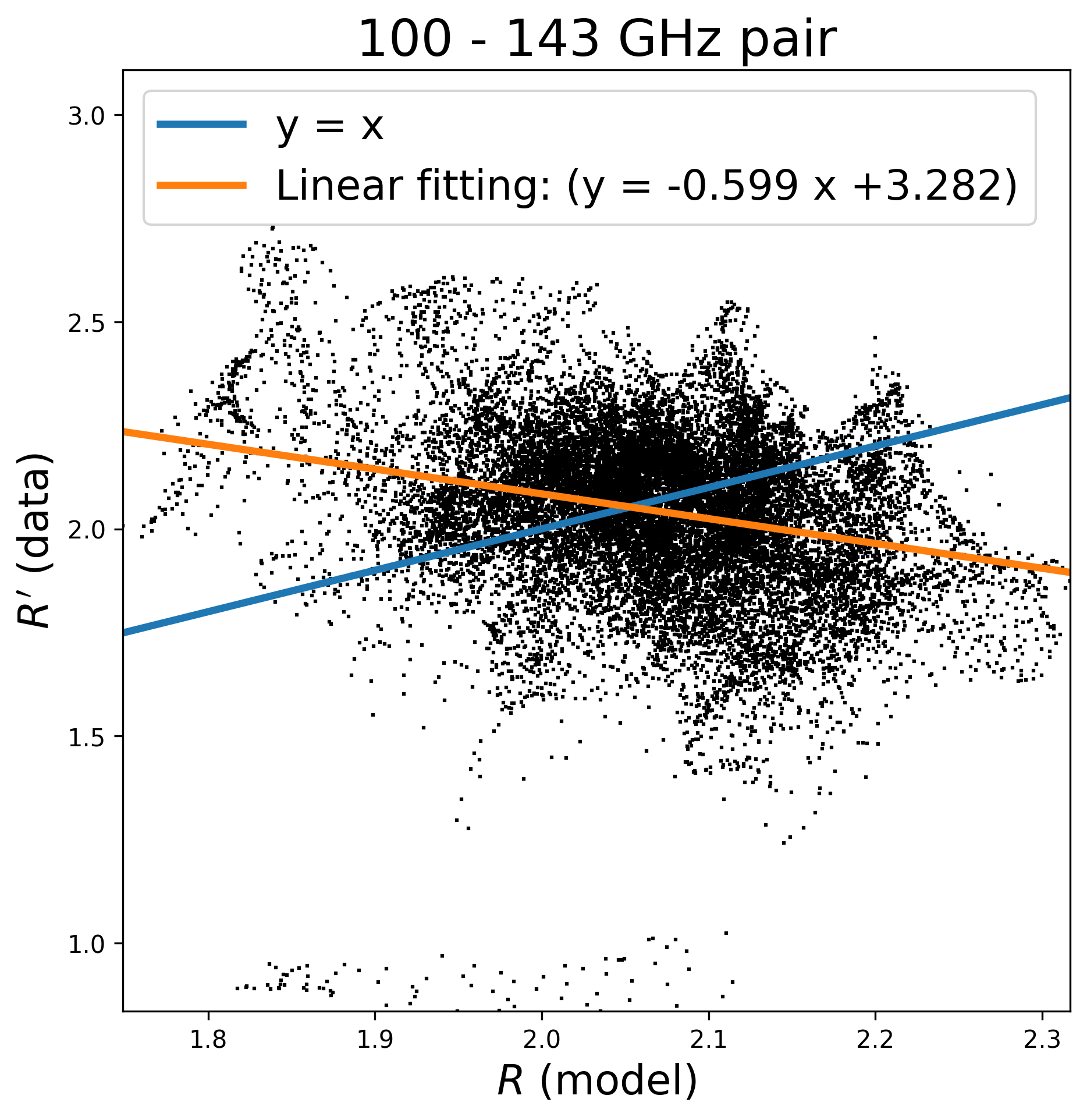}
\includegraphics[width=0.32\textwidth]{Scatter_model_2015_color_correction_yes_1_galactic_mask_80_smooth_degree_2_disk_degree_6_low_galac_mask_no_zodiacal_mask_no.png}\\
\includegraphics[width=0.32\textwidth]{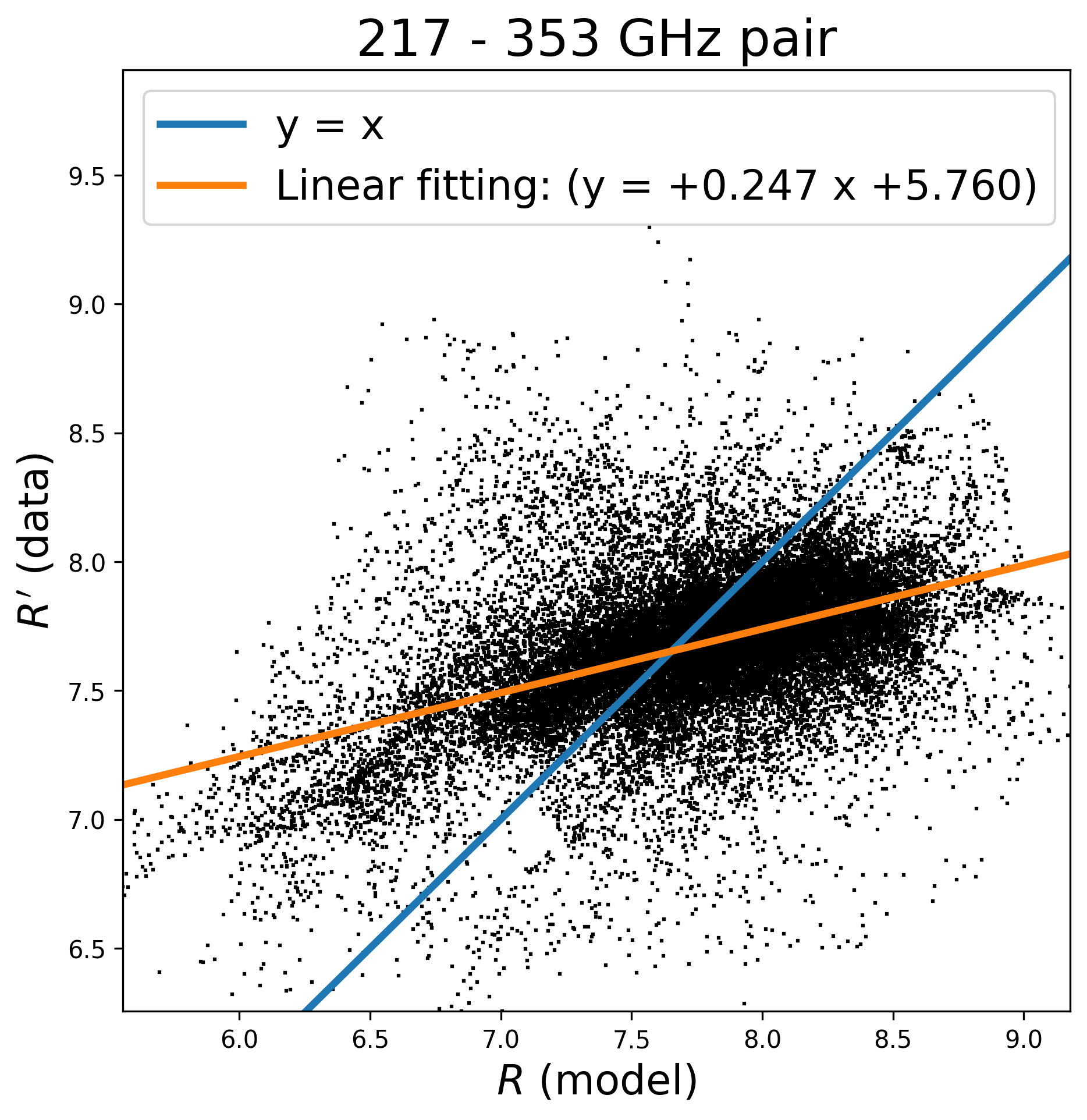}
\includegraphics[width=0.32\textwidth]{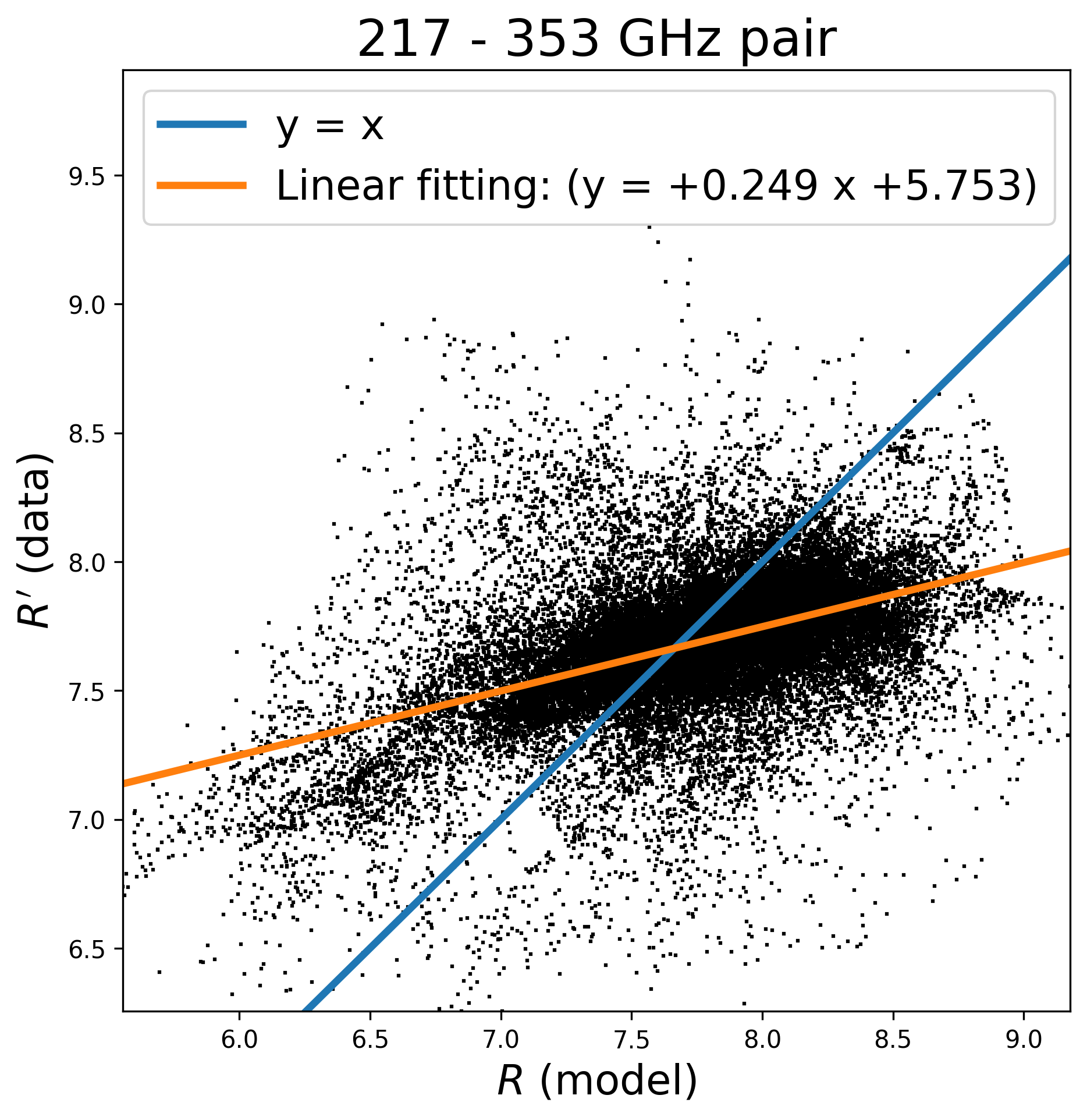}
\includegraphics[width=0.32\textwidth]{Scatter_model_2015_color_correction_yes_2_galactic_mask_80_smooth_degree_2_disk_degree_6_low_galac_mask_no_zodiacal_mask_no.png}\\
\includegraphics[width=0.32\textwidth]{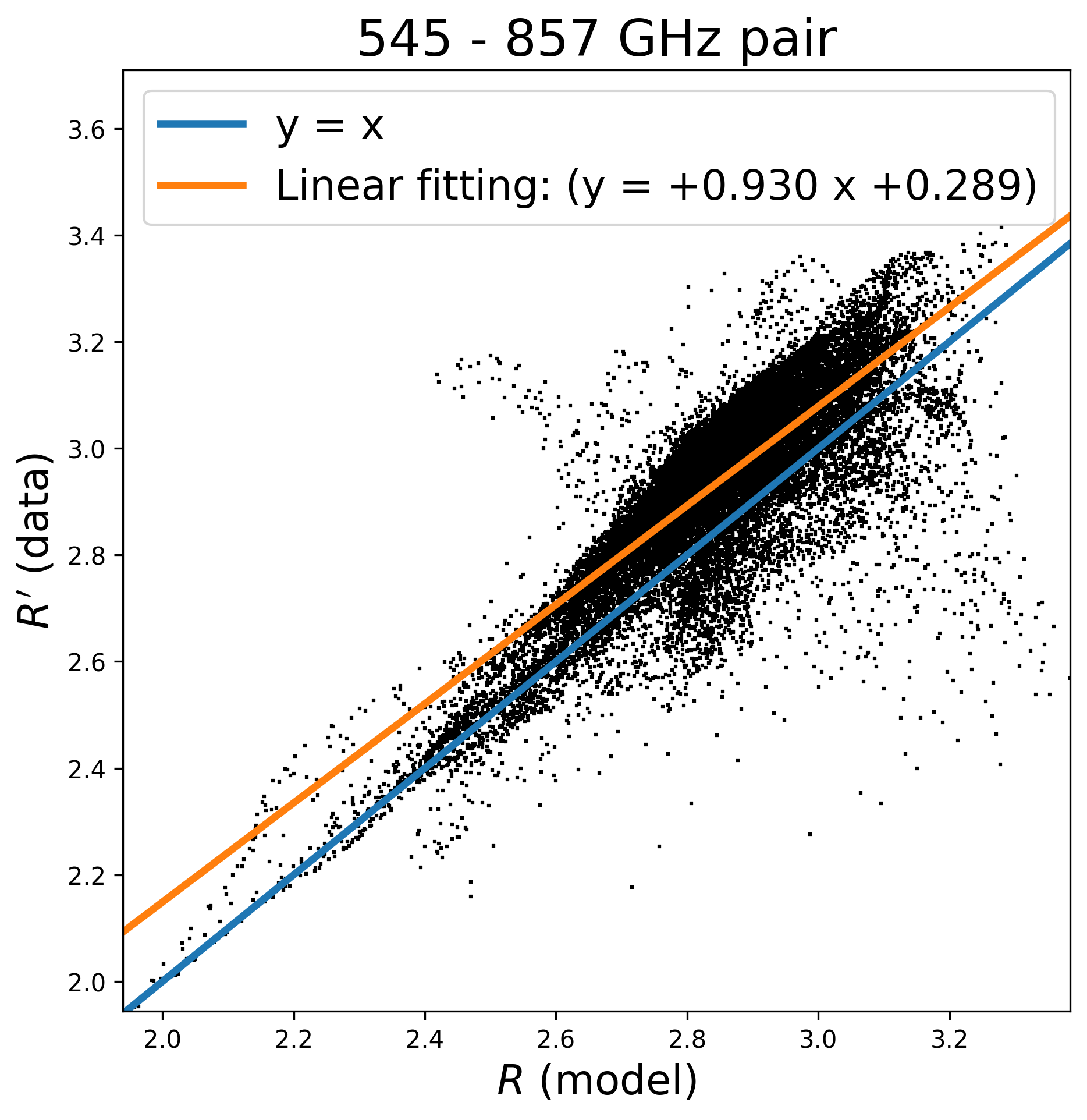}
\includegraphics[width=0.32\textwidth]{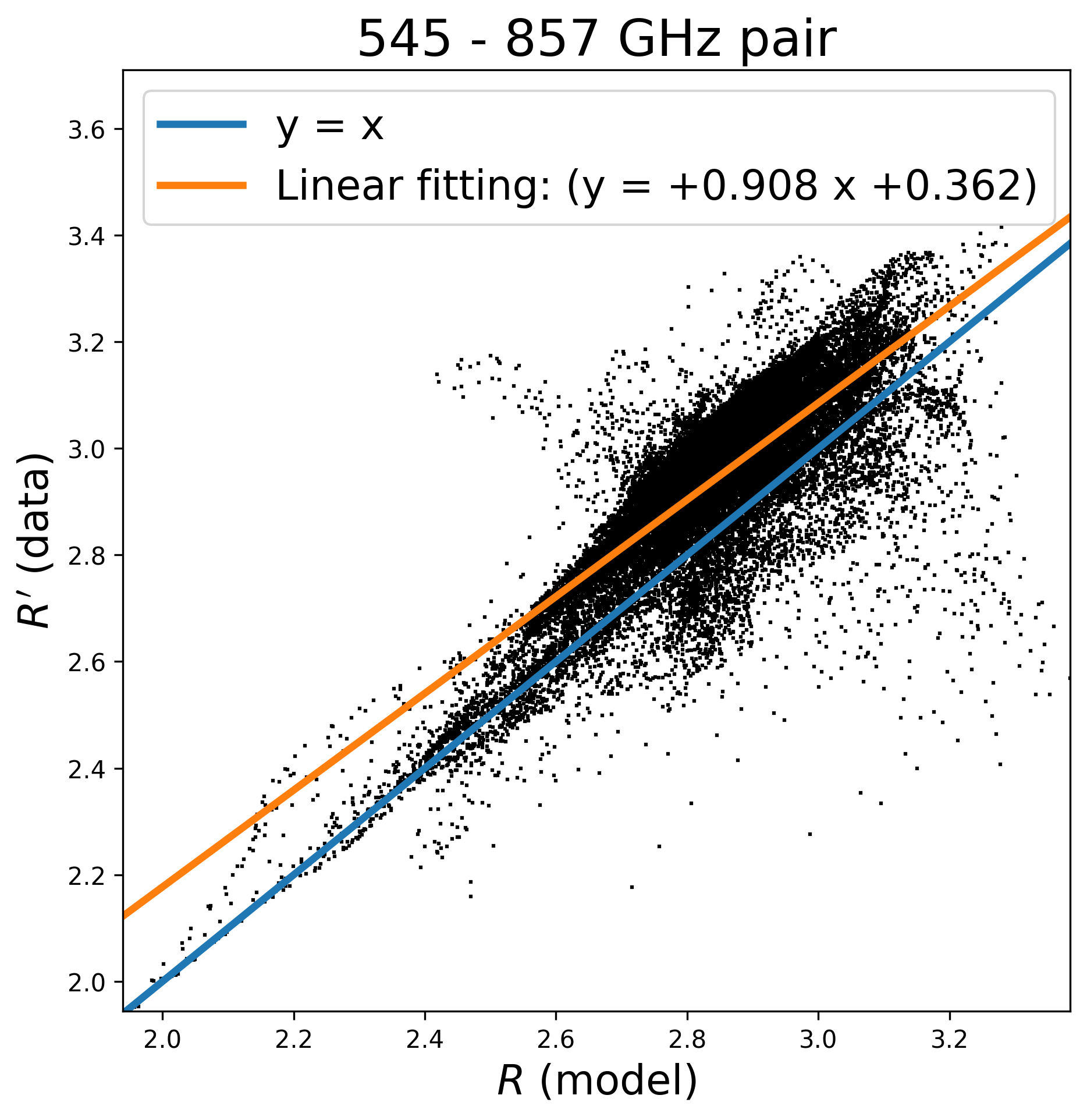}
\includegraphics[width=0.32\textwidth]{Scatter_model_2015_color_correction_yes_3_galactic_mask_80_smooth_degree_2_disk_degree_6_low_galac_mask_no_zodiacal_mask_no.png}
\caption{Comparison of different Galactic plane cut. 
\textit{From left to right}: 
$M_\mathrm{plane} = M_{60}$, $M_\mathrm{plane} = M_{70}$, and $M_\mathrm{plane} = M_{80}$. 
\textit{From top to bottom}: pairs of 100-143, 217-353, and 545-857 GHz, all with smoothing angle of $2^\circ$ and mosaic angular radius of $6^\circ$. }
\label{fig:scatter plots with different galac cut}
\end{figure*}

\begin{figure*}[htb]
\centering
\includegraphics[width=0.32\textwidth]{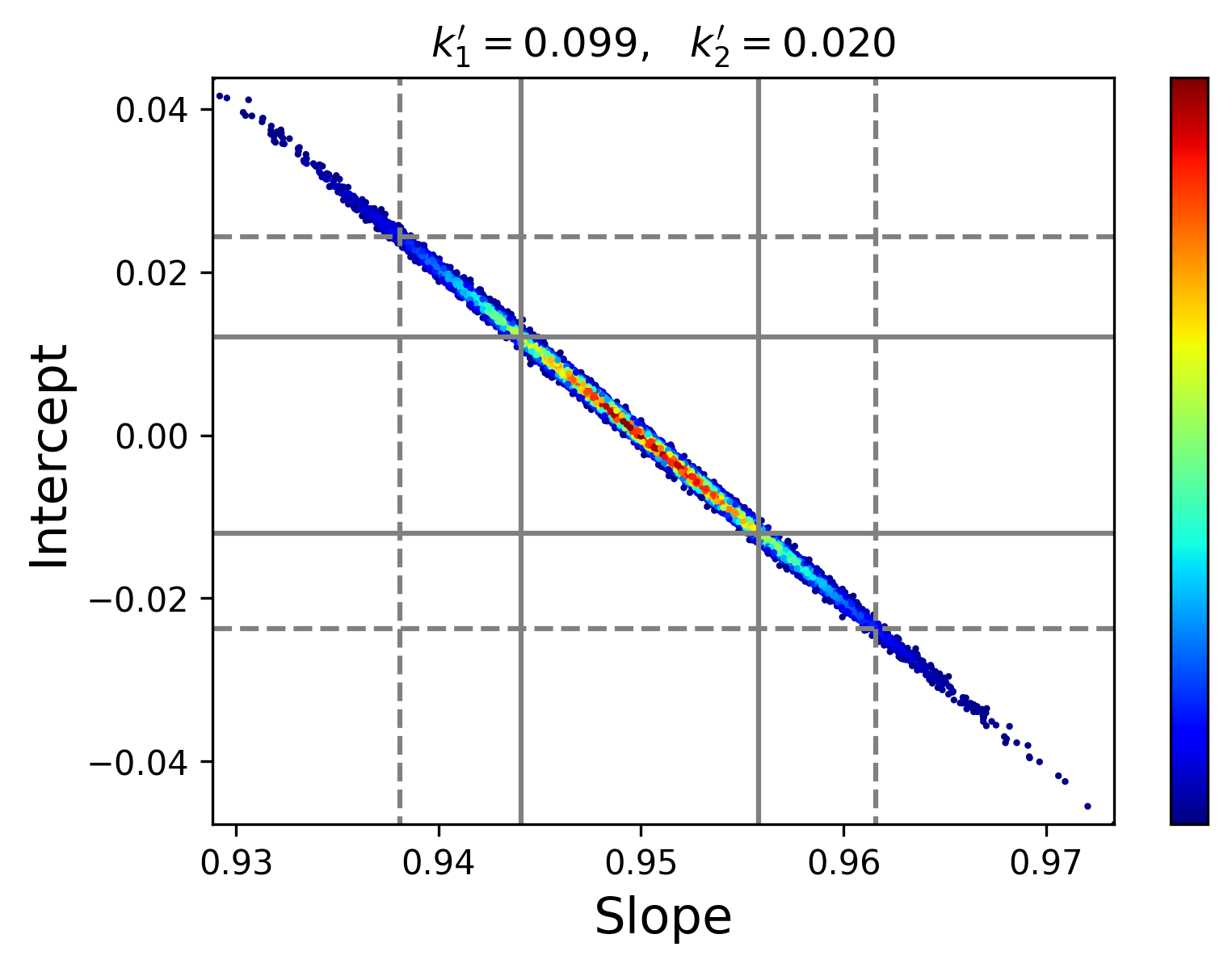}
\includegraphics[width=0.32\textwidth]{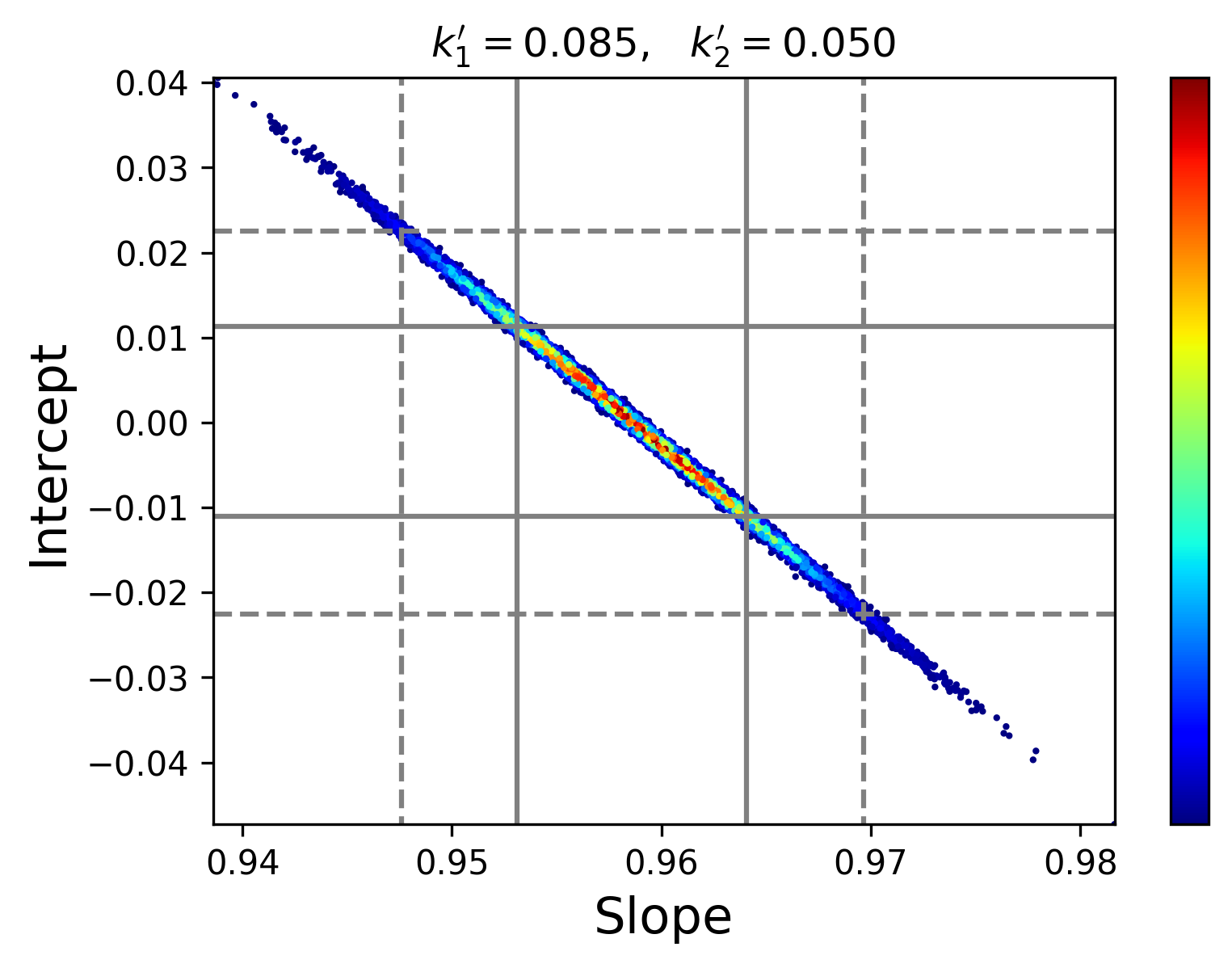}
\includegraphics[width=0.32\textwidth]{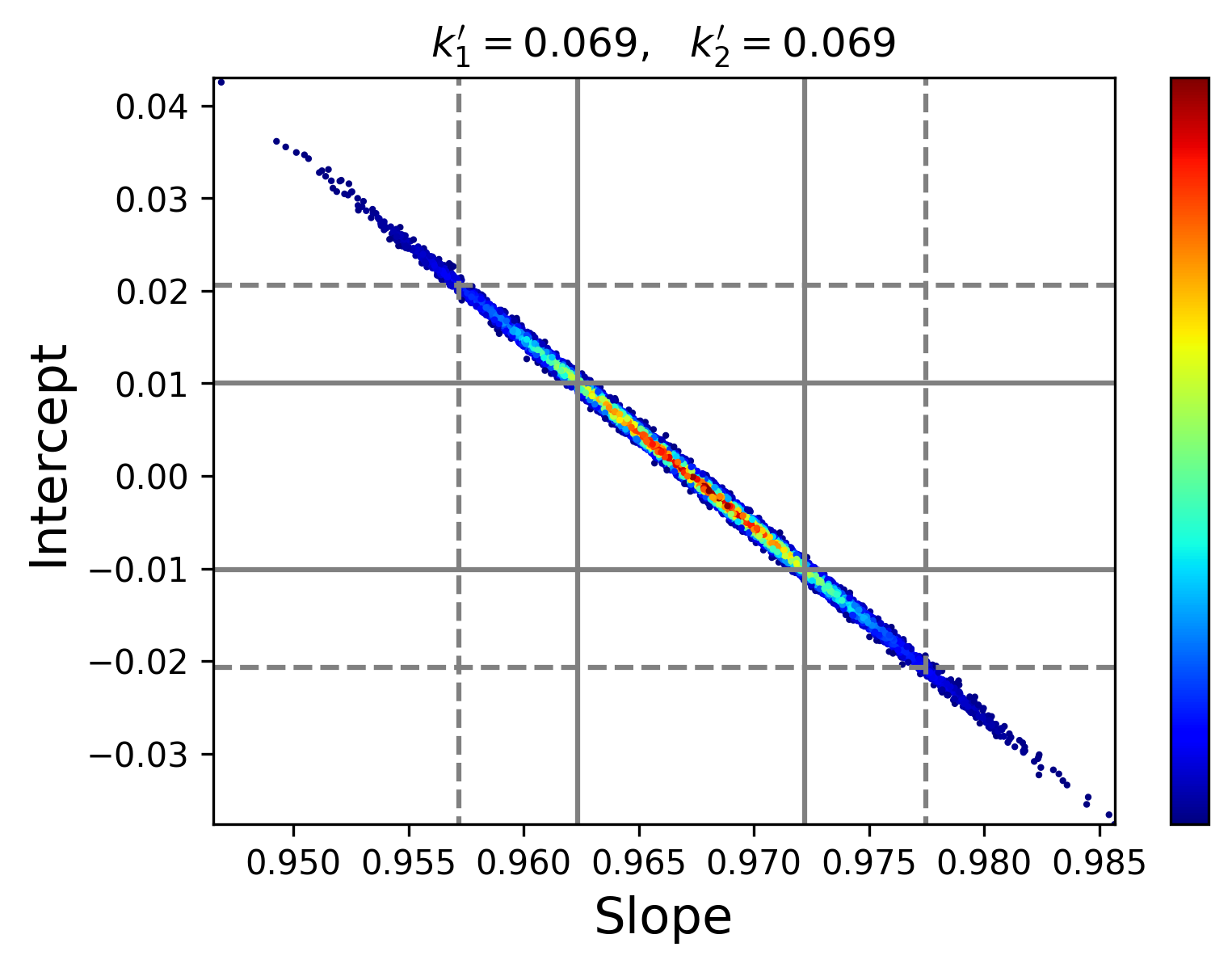}
\caption{Similar to Fig.~\ref{fig:simulation slope intercept with low galactic}, but with degree of freedom set to $N = 10$. }
\label{fig:simulation slope intercept with low galactic_freedom_10}
\end{figure*}

\section{Unit conversion and color correction}
\label{app:unit_conversion_colour_correction}
This section details unit conversions and color corrections involved in this study. 
\subsection{Rayleigh-Jeans unit}
In Rayleigh-Jeans unit system, the reading of one detector, $s_{\mathrm{RJ}}$, is expressed in terms of brightness temperature $T_{\mathrm{RJ}}$, with the calibration source being a blackbody in the Rayleigh-Jeans limit (long-wavelength limit, \citealp{2014A&A...571A...9P}): 
\begin{equation}
J_{0,\nu} = \frac{2 k_\mathrm{B} \nu^2}{c^2} T_\mathrm{RJ}
\equiv b'_\mathrm{RJ}(\nu) T_\mathrm{RJ}. 
\end{equation}
When observing a calibration source with a detector operating at an effective frequency $\nu_0$, the reading should be the brightness temperature of the calibration source at $\nu_0$, i.e., $T_{\mathrm{RJ}}$. 
It should be noted that for a blackbody, its brightness temperature does not vary with frequency: 
\begin{equation}
T_\mathrm{RJ} = A_\mathrm{RJ}\int J_{0,\nu} \mathcal{T}(\nu) \,\dd\nu
= T_\mathrm{RJ} A_\mathrm{RJ}\int b'_\mathrm{RJ}(\nu) \mathcal{T}(\nu) \,\dd\nu. 
\end{equation}
Thus
\begin{equation}
A_\mathrm{RJ} = \dfrac{1}{\displaystyle\int b'_\mathrm{RJ}(\nu) \mathcal{T}(\nu) \,\dd\nu}, 
\end{equation}
and the reading of this detector when observing any astronomical object, in Rayleigh-Jeans unit, is given by: 
\begin{equation}
s_\mathrm{RJ} = A_\mathrm{RJ} \int J_\nu \mathcal{T}(\nu) \,\dd\nu 
= \dfrac{\displaystyle\int J_\nu \mathcal{T}(\nu) \,\dd\nu}{\displaystyle\int b'_\mathrm{RJ}(\nu) \mathcal{T}(\nu) \,\dd\nu}. 
\end{equation}

\subsection{CMB unit}
In CMB unit system, the unit of emission intensity is the thermodynamics brightness differential temperature $T_\mathrm{CMB}$, with the CMB dipole adopted as the calibration source \citep{2014A&A...571A...9P}:
\begin{equation}
\begin{aligned}
J_{0,\nu} &= B_\nu(T_0+T_\mathrm{CMB})-B_\nu(T_0) 
\simeq \frac{\partial B_\nu(T_0)}{\partial T} T_\mathrm{CMB}\\
&= \frac{2k_B\nu^2}{c^2} \frac{x^2 \mathrm{e}^x}{(\mathrm{e}^x-1)^2} T_\mathrm{CMB}
\equiv b'_\mathrm{CMB}(\nu) T_\mathrm{CMB}. 
\end{aligned}
\end{equation}
Similarly, the calibration coefficient is: 
\begin{equation}
A_\mathrm{CMB} = \dfrac{1}{\displaystyle\int b'_\mathrm{CMB}(\nu) \mathcal{T}(\nu) \dd\nu}. 
\end{equation}
Then reading of the detector for the radiation from any astronomical object is: 
\begin{equation}
s_\mathrm{CMB} = \dfrac{\displaystyle\int J_\nu \mathcal{T}(\nu) \dd\nu}{\displaystyle\int b'_\mathrm{CMB}(\nu) \mathcal{T}(\nu) \dd\nu}. 
\end{equation}

\subsection{SI unit}
For SI unit (in unit of $\mathrm{W\,m^{-2}\,Hz^{-1}\,sr^{-1}}$ or  $\mathrm{MJy\,sr^{-1}}$), \textit{Planck} team adopts the IRAS convention \citep{2014A&A...571A...9P, 1988iras....1.....B},\footnote{\url{https://lambda.gsfc.nasa.gov/product/iras/docs/exp.sup/ch6/C3.html}} 
that is, the calibration source has the following spectral energy distribution: 
\begin{equation}
J_{0,\nu} = J_0 \left(\frac{\nu}{\nu_0}\right)^{-1}. 
\end{equation}
When observing this source via a detector with effective frequency $\nu_0$, the reading should be the flux intensity at $\nu_0$: 
\begin{equation}
J_0 = A_\mathrm{SI}\int J_0 \left(\frac{\nu}{\nu_0}\right)^{-1} \mathcal{T}(\nu) \,\dd\nu. 
\end{equation}
Thus in SI unit, reading of the detector for any astronomical object should be: 
\begin{equation}
s_\mathrm{SI} = \dfrac{\displaystyle\int J_\nu \mathcal{T}(\nu) \,\dd\nu}{\displaystyle\int \left(\dfrac{\nu}{\nu_0}\right)^{-1} \mathcal{T}(\nu) \,\dd\nu}. 
\end{equation}
where $\nu_0$ is 100, 143, 217, 353, 545, and 857 GHz.

\section{Removing non-dust components from HFI maps}
\label{app:removing other components}
This section outlines the detailed steps for subtracting CMB anisotropies and other Galactic foreground components from \textit{Planck} HFI maps.

\subsection{Subtracting CMB anisotropies}
CMB anisotropies dominate at 100, 143, and 217 GHz and also leave traces on maps with higher frequencies. 
We remove the SMICA constrained-realization CMB maps \citep{2020A&A...641A...4P} from \textit{Planck} HFI maps. 
For 100, 143, 217, and 353 GHz, CMB anisotropies are deducted from the single frequency maps directly since CMB anisotropies show a flat spectrum in CMB unit, whereas at 545 and 857 GHz they are transformed into unit of $\mathrm{MJy\,sr^{-1}}$.

\subsection{Subtracting free-free emission}
Free-free emission, also known as bremsstrahlung, is produced due to electron-electron and electron-ion scattering\footnote{Since electron has much smaller mass than ion, and so has much much larger acceleration, free-free emission primarily originates from free electrons. } in thermal plasma, depending on the emission measure (EM) of Galactic free electrons and the electron temperature, $T_e$ \citep{2016A&A...594A..10P, 2011piim.book.....D}. 
The Gaunt factor of free-free emission is: 
\begin{equation}
\label{equation_free-free_gaunt_factor}
g_\mathrm{ff} \simeq \ln\left\{\exp\left[5.960-\frac{\sqrt{3}}{\uppi}\log\left(\frac{\nu_9}{T_4^{3/2}}\right)\right] + \mathrm{e}\right\}, 
\end{equation}
where $\mathrm{e}$ is the natural logarithm base, $T_4 \equiv T_\mathrm{e}/{10^4\,\mathrm{K}}$, and $\nu_9 \equiv \nu/{10^9\,\mathrm{Hz}}$. 
Optical depth of the free electrons is: 
\begin{equation}
\label{equation_free-free_optical_depth}
\tau = 0.05468\, T_\mathrm{e}^{-3/2} \nu_9^{-2}\, \mathrm{EM}\, g_\mathrm{ff}, 
\end{equation}
and the flux intensity of free-free emission is: 
\begin{equation}
\label{equation_free-free_intensity}
s_\mathrm{ff} = T_\mathrm{e} \left(1-\mathrm{e}^{-\tau}\right) [\mathrm{K_{RJ}}]. 
\end{equation}

Free-free emission partially dominates the electromagnetic radiation at frequencies below 70 GHz (together with synchrotron) and goes down at higher frequencies \citep{2014A&A...571A..12P}. 
According to Eqs.~(\ref{equation_free-free_gaunt_factor})-(\ref{equation_free-free_intensity}), and based on the full-sky maps of EM and $T_\mathrm{e}$, we evaluate the intensity of free-free emission at \textit{Planck} HFI frequencies and then remove them.

\subsection{Subtracting synchrotron radiation}
Synchrotron radiation originates from relativistic high-energy electrons gyrating around the Galactic magnetic field, producing electromagnetic radiation that approximately follows a power-law spectrum with index $\beta_\mathrm{syn} \simeq -3.1$ \citep{2020A&A...641A...4P}. 
Thus for \textit{Planck} frequencies, synchrotron primarily operates at 30 and 44 GHz, and decay rapidly at higher frequencies. 
Nevertheless, for the sake of rigor, we still subtract the synchrotron radiation from \textit{Planck} HFI intensity maps. 
The reference emission map at 408 MHz and emission template for synchrotron are both obtained from COM\_CompMap\_Synchrotron-commander\_0256\_R2.00.fits in PR2.

\subsection{Subtracting CO line emission}
CO is the second most abundant molecular gas in the universe \citep{10.1039/9781839164798}. 
The tiny electric dipole moment of CO molecules allows them to emit a series of rotational transition lines.
In the CMB community, particular attention should be paid to the first three emission lines, namely CO(1-0) at 115.271 GHz, CO(2-1) at 230.538 GHz, and CO(3-2) at 345.796 GHz \citep{2014A&A...571A..13P}. 
The relative velocity $V$ between the gas cloud emitting CO lines and Earth results in Doppler effect: 
\begin{equation}
\nu = \nu_\mathrm{CO} \dfrac{\sqrt{1-\dfrac{V^2}{c^2}}}{1-\dfrac{V}{c}\cos\alpha}
\simeq \nu_\mathrm{CO} \left(1+\dfrac{v}{c}\right), 
\end{equation}
where $\nu_{\mathrm{CO}}$ and $\nu$ are the frequencies of the CO emission lines in the laboratory frame and those measured on Earth, respectively. 
$\alpha$ is the angle between the direction of the emission line and the direction of $V$ in the observer's frame. 
Here we denote the projection of $V$ along the line of sight (i.e., the velocity broadening) as $v \equiv V\cos\alpha$. 
Then the equation 
\begin{equation}
\Delta\nu = \dfrac{\nu_\mathrm{CO}}{c}\Delta v 
\end{equation}
converts velocity broadening into frequency broadening. 
\textit{Planck} team presents the maps of CO emission lines in terms of $\Delta T_{\mathrm{CO}}$ (in unit of $\mathrm{K_{RJ}\,km\,s^{-1}}$). 
Then
\begin{equation}
\label{Equation_intensity_CO}
\dfrac{\Delta T_\mathrm{CO}}{\Delta v} = \frac{\Delta T_\mathrm{CO}}{\Delta\nu} \frac{\nu_\mathrm{CO}}{c}
\end{equation}
represents the emission power of the CO emission lines in the band $\left[\nu_\mathrm{CO} - \dfrac{\Delta\nu}{2}, \nu_\mathrm{CO} + \dfrac{\Delta\nu}{2}\right)$ per unit frequency interval, per unit area, and per unit solid angle (in units of $\mathrm{K_{RJ}}$). 
Noting that the frequency broadening of the CO emission line is much smaller than the frequency resolution of the \textit{Planck} HFI detectors \citep{2014A&A...571A...9P}, Eq.~(\ref{Equation_intensity_CO}) can be approximated as a $\delta$-function. 
Thus, the reading for the CO emission line in unit of $\mathrm{K_{CMB}}$ (for 100, 143, 217, and 353 GHz) is given by: 
\begin{equation}
\label{Equation_CO_K_CMB}
\begin{aligned}
s_\mathrm{CMB} 
&= \dfrac{\displaystyle\int b'_\mathrm{RJ}(\nu)\left(\dfrac{\Delta T_\mathrm{CO}}{\Delta\nu} \dfrac{\nu_\mathrm{CO}}{c}\right)\mathcal{T}(\nu)\,\dd\nu}{\displaystyle\int b'_\mathrm{CMB}(\nu)\mathcal{T}(\nu)\,\dd\nu}\\
&= \Delta T_\mathrm{CO} \dfrac{b'_\mathrm{RJ}(\nu_\mathrm{CO}) \dfrac{\nu_\mathrm{CO}}{c}\mathcal{T}(\nu_\mathrm{CO})}{\displaystyle\int b'_\mathrm{CMB}(\nu)\mathcal{T}(\nu)\,\dd\nu}, 
\end{aligned}
\end{equation}
and that in unit of $\mathrm{MJy\,sr^{-1}}$ (for 545 and 857 GHz) is: 
\begin{equation}
\label{Equation_CO_SI}
s_\mathrm{SI} = \Delta T_\mathrm{CO} \dfrac{b'_\mathrm{RJ}(\nu_\mathrm{CO}) \dfrac{\nu_\mathrm{CO}}{c}\mathcal{T}(\nu_\mathrm{CO})}{\displaystyle\int \left(\dfrac{\nu}{\nu_0}\right)^{-1} \mathcal{T}(\nu) \,\dd\nu}. 
\end{equation}

In this study, we adopt the high-resolution map of CO(2-1) emission line provided in PR2. 
We assume that the intensity ratios of the CO(1-0), CO(2-1), and CO(3-2) emission lines are constant across the full sky:
CO(2-1) / CO(1-0) = 0.595 and CO(3-2) / CO(1-0) = 0.297 \citep{2014A&A...571A..13P}. 
Based on these ratios, one can derive the maps of the CO(1-0) and CO(3-2) emission lines in units of $\mathrm{K_{RJ}\,km\,s^{-1}}$. 
According to Eqs.~(\ref{Equation_CO_K_CMB})-(\ref{Equation_CO_SI}), one can further obtain the total CO intensity components of these three emission lines in each HFI frequency channel.

\subsection{Subtracting 94/100 GHz line emission}
Beyond the CO line emission, PR2 release also considers additional molecular emission lines such as HCN, CN, HCO+, and CS, collectively referred to as the ``94/100 GHz line emission'' \citep{2016A&A...594A..10P} and abbreviated as ``xline'' in the published foreground separation maps. 
In this study, we directly subtract the 94/100 GHz intensity map provided in PR2 (in unit of $\mu\mathrm{K_{CMB}}$) from the HFI map at 100 GHz.

\section{Supplement figures}
\label{app:supplement figures}
This section presents supplementary figures used for further cross-validation. 

Fig.~\ref{fig:scatter plots with different smoothing and disk angle 143-100} (for the 100 - 143 GHz pair), Fig.~\ref{fig:scatter plots with different smoothing and disk angle 353-217} (for the 217 - 353 GHz pair), and Fig.~\ref{fig:scatter plots with different smoothing and disk angle 857-545} (for the 545 - 857 GHz pair) illustrate the scatter plots of $R'$ versus $R$ corresponding to different smoothing angles ($1^\circ$, $2^\circ$, and $3^\circ$) and mosaic disk radii ($5^\circ$, $6^\circ$, and $7^\circ$). 
Fig.~\ref{fig:dust_ratio_compare_no zodiacal region} shows the $R'$-to-$R$ scatter plots over the regions excluding zodiacal light. 
Fig.~\ref{fig:scatter plots with different galac cut} examines the impact of selecting different galactic plane masks on the scatter plots. 
The results of the above figures show no significant differences compared to Fig.~\ref{fig:dust_ratio_compare}. 
Fig.~\ref{fig:simulation slope intercept with low galactic_freedom_10} shows the simulated slopes and intercept of $R'$-$R$ regression lines, with degree of freedom set to $N = 10$, with no obvious departures compared to Fig.~\ref{fig:simulation slope intercept with low galactic}.



\FloatBarrier
\bibliography{A_file_hfi_ratio}

\end{document}